\documentclass[12pt]{article}
\usepackage{eqsection,subeqnarray,indent,amsfonts,amssymb,amsmath}
\usepackage{bm}    
\usepackage{cite}  
\usepackage{pstricks,pst-node,pst-text,pst-3d,pst-plot}  
\usepackage{epic,eepic}
\usepackage{rotating}  
\usepackage{graphicx}
\usepackage{url} 
\usepackage{hyperref}

\usepackage{amsthm}
\theoremstyle{plain}

\newtheorem{algthm}{Algorithm}

\theoremstyle{definition}

\theoremstyle{remark}
\newtheorem{remark}{Remark}[section]

\begin{document}

\bibliographystyle{plain}

\date{April 4, 2011}

\title{\vspace*{-3.6cm}\hspace*{-0.9cm}
     \hbox{Dynamic Critical Behavior of the Chayes--Machta Algorithm} \\
       for the Random-Cluster Model \\[5mm]
       \large\bf I.~Two Dimensions}

\author{
  {\small Timothy M.~Garoni}                       \\[-0.2cm]
  {\small\it Department of Mathematics and Statistics}        \\[-0.2cm]
  {\small\it University of Melbourne}        \\[-0.2cm]
  {\small\it Vic.~3010, AUSTRALIA}                  \\[-0.2cm]
  {\small\tt T.GARONI@MS.UNIMELB.EDU.AU}              \\[2mm]
  {\small Giovanni Ossola}                       \\[-0.2cm]
  {\small\it Department of Physics}                  \\[-0.2cm]
  {\small\it New York City College of Technology}                  \\[-0.2cm]
  {\small\it 300 Jay Street}                  \\[-0.2cm]
  {\small\it Brooklyn, NY 11201, USA}                  \\[-0.2cm]
  {\small\tt GOSSOLA@CITYTECH.CUNY.EDU}              \\[2mm]
  {\small Marco Polin}                       \\[-0.2cm]
  {\small\it Department of Applied Mathematics and Theoretical Physics}                  \\[-0.2cm]
  {\small\it University of Cambridge}                  \\[-0.2cm]
  {\small\it Wilberforce Road}                  \\[-0.2cm]
  {\small\it Cambridge CB3 0WA, UK}              \\[-0.2cm]
  {\small\tt M.POLIN@DAMTP.CAM.AC.UK}              \\[2mm]
  {\small Alan D.~Sokal\thanks{Also at Department of Mathematics,
           University College London, London WC1E 6BT, United Kingdom.}}  \\[-0.2cm]
  {\small\it Department of Physics}            \\[-0.2cm]
  {\small\it New York University}              \\[-0.2cm]
  {\small\it 4 Washington Place}               \\[-0.2cm]
  {\small\it New York, NY 10003, USA}           \\[-0.2cm]
  {\small\tt SOKAL@NYU.EDU}                    \\[-0.2cm]
  {\protect\makebox[5in]{\quad}}  
  \\[-4mm]
}

\maketitle
\thispagestyle{empty}   

\vspace*{-1cm}

\begin{abstract}
We study, via Monte Carlo simulation, the dynamic critical behavior
of the Chayes--Machta dynamics for the Fortuin--Kasteleyn random-cluster
model, which generalizes the Swendsen--Wang dynamics for the $q$-state Potts
ferromagnet to non-integer $q \ge 1$.
We consider spatial dimension $d=2$ and $1.25\le q \le 4$ in steps of $0.25$,
on lattices up to $1024^2$,
and obtain estimates for the dynamic critical exponent $z_{\rm CM}$.
We present evidence that when $1 \le q \lesssim 1.95$
the Ossola--Sokal conjecture $z_{\rm CM} \ge \beta/\nu$ is violated,
though we also present plausible fits compatible with this conjecture.
We show that the Li--Sokal bound $z_{\rm CM} \ge \alpha/\nu$
is close to being sharp over the entire range $1 \le q \le 4$,
but is probably non-sharp by a power.
As a byproduct of our work, we also obtain evidence concerning
the corrections to scaling in static observables.
\end{abstract}


\clearpage

\tableofcontents

\clearpage

%
 
\newcommand{\ontop}[2]{\genfrac{}{}{0pt}{}{#1}{#2}}
\newcommand{\supp}{\text{supp}}
\newcommand{\var}{\text{var}}
\newcommand{\cov}{\text{cov}}
\newcommand{\varN}{C_H}
\newcommand{\colorset}{\{1,2,\dots,m\}^V}
\newcommand{\spinset}{\{1,2,\dots,q\}^V}
\renewcommand{\color}{\boldsymbol{\sigma}}
\newcommand{\bond}{\boldsymbol{n}}
\newcommand{\edgeset}{A}
\newcommand{\pcolor}{P_{\rm color}}
\newcommand{\pbond}{P_{\rm bond}}
\newcommand{\palpha}{P_{\alpha}}
\newcommand{\jointmeasure}{\mu_{G,{\bf W}}}
\newcommand{\configurationspace}{\{0,1\}^E}
\newcommand{\lmin}{L_{\text{min}}}
\newcommand{\activeE}{E_A^{(\boldsymbol{\sigma})}}
\newcommand{\activebonds}{\Omega_A^{(\boldsymbol{\sigma})}}
\newcommand{\inactiveE}{E_I^{(\boldsymbol{\sigma})}}
\newcommand{\inactivebonds}{\Omega_I^{(\boldsymbol{\sigma})}}
\newcommand{\activeW}{\widehat{W}}
\newcommand{\obs }{\ensuremath{{\cal O}}}
\newcommand{\E }{\ensuremath{{\cal E'}}}
\newcommand{\F }{\ensuremath{{\cal F'}}}
\newcommand{\N }{\ensuremath{{\cal N}}}
\newcommand{\corner}{\ensuremath{v_L}}
\newcommand{\Y }{\ensuremath{{\cal T}}}
\newcommand{\y }{\ensuremath{\tau(\corner)}}
\newcommand{\Snought}{\ensuremath{{\cal S}_0}}
\newcommand{\Sm}{\ensuremath{{\cal S}_m}}
\newcommand{\Stwo}{\ensuremath{{\cal S}_2}}
\newcommand{\Sfour}{\ensuremath{{\cal S}_4}}
\newcommand{\Cone}{\ensuremath{{\cal C}_1}}
\newcommand{\Ci}{\ensuremath{{\cal C}_i}}
\newcommand{\Ctwo}{\ensuremath{{\cal C}_2}}
\newcommand{\Cthree}{\ensuremath{{\cal C}_3}}
\newcommand{\Cfour}{\ensuremath{{\cal C}_4}}
\newcommand{\Cfive}{\ensuremath{{\cal C}_5}}
\newcommand{\Csumtwo}{\ensuremath{{\widetilde{\cal C}}_2}}
\newcommand{\Csumthree}{\ensuremath{{\widetilde{\cal C}}_3}}
\newcommand{\Csumfour}{\ensuremath{{\widetilde{\cal C}}_4}}
\newcommand{\Csumfive}{\ensuremath{{\widetilde{\cal C}}_5}}
\newcommand{\Z}{{\mathbb{Z}}}
\newcommand{\torus}{\mathbb{Z}_L^d}
\newcommand{\m}{m} 
\newcommand{\dm}{\delta m}
\newcommand{\gtapprox}{\gtrsim}
\newcommand{\ltapprox}{\lesssim}
\newcommand{\reff}[1]{(\protect\ref{#1})}
\newcommand{\bsigma}{{\boldsymbol{\sigma}}}

\section{Introduction}
\label{Introduction}

Since nontrivial models in statistical mechanics are rarely exactly solvable,
Monte Carlo (MC) simulations have become a standard tool for obtaining
information on phase diagrams and critical exponents
\cite{Binder86,Binder87,Binder95}.
Unfortunately, MC simulations typically suffer from severe
{\em critical slowing-down}\/ \cite{Hohenberg_77,SokalLectures},
so that the computational efficiency tends rapidly to zero
as the critical point is approached.
More precisely, the autocorrelation (relaxation) time $\tau$ diverges
in the critical limit, most often like $\tau\sim\xi^z$,
where $\xi$ is the spatial correlation length.
The dynamic critical exponent $z$ depends on both
the model being investigated and the MC algorithm being used.
For local algorithms one typically has $z\approx2$.  

An important advance was made in 1987 with the invention of the
Swendsen--Wang (SW) cluster algorithm \cite{SwendsenWang87}
for simulating the ferromagnetic $q$-state Potts model
\cite{Potts52,Wu82,Wu84} at integer $q\ge2$.
The SW algorithm is based on passing back and forth
between the Potts spin representation
and the Fortuin--Kasteleyn (FK) bond representation
\cite{KasteleynFortuin69,FortuinKasteleyn72,Fortuin72a,Fortuin72b,EdwardsSokal88,Grimmett_06}.
More precisely, one introduces a joint probability distribution
\cite{EdwardsSokal88} of spin and bond variables, whose marginal on the spins
(integrating out the bonds) is the Potts spin model
and whose marginal on the bonds (integrating out the spins)
is the FK random-cluster model \cite{Grimmett_06};
one then updates this joint distribution
by alternately applying the two conditional distributions
(see Section~\ref{subsec.SW} below for details).

Since a local move in one set of variables can have highly
nonlocal effects in the other, it is not surprising that
the SW algorithm might have less critical slowing-down
than the conventional local algorithms.
And this is in fact the case:
although the SW algorithm does not eliminate critical slowing-down,
it does radically reduce it compared to local algorithms.
Much effort has therefore been devoted, for both theoretical and
practical reasons, to understanding the dynamic critical behavior of
the SW algorithm as a function of the spatial dimension $d$ and the
number $q$ of Potts spin states.
The best information on the dynamic critical exponent $z_{\rm SW}$
prior to the present work is summarized in Table~\ref{z_sw_table}.
Unfortunately, it is very difficult to develop a physical understanding
from the small number of non-trivial ``data points'' at our disposal:
second-order non-mean-field transitions occur only for
$(d,q)=(2,2),(2,3),(2,4),(3,2)$ and $(4,2)$.\footnote{
  A continuous (second-order) transition occurs also in the
  Ising ($q=2$) model in dimensions $d>4$,
  but here the static behavior is mean-field.
  One expects the dynamic critical exponents likewise to be
  dimension-independent for $d \ge 4$
  (with possible multiplicative logarithmic corrections at $d=4$).
}

%
%
\begin{table}[t]
\begin{center}
\begin{tabular}{|c||c|c|c|c|} \hline
        & \multicolumn{4}{|c|}{Estimates of $z_{\rm SW}$} \\ \hline
        &  $q=1$  &   $q=2$   &   $q=3$   &   $q=4$   \\ \hline\hline
$d=1$   &   0     &     0     &     0     &    0                     \\
$d=2$   &   0     & $0.222 \pm 0.007$  & $0.514 \pm 0.006$ &
   $1$ ($\times \log^{??}$) \\
$d=3$   &   0     & $0.46 \pm 0.03$            & ---     & ---     \\
$d=4$   &   0     &  $1$ ($\times \log^{??}$)  & ---     & ---     \\
\hline
\end{tabular}
\end{center}
\caption{
   Best estimates of the dynamic critical exponent $z$
   for the Swendsen--Wang algorithm prior to the present work.
   Estimates are taken from \protect\cite{Salas_Sokal_Ising_v1}
   for $d=2$, $q=2$;
   \cite{Salas_Sokal_Potts3} for $d=2$, $q=3$;
   \cite{Salas_Sokal_AT,Salas_Sokal_Potts4} for $d=2$, $q=4$;
   \protect\cite{OssolaSokal04} for $d=3$, $q=2$;
   and \protect\cite{Klein_89,Ray_89,CoddingtonBaillie92,PerskyBenAvKanterDomany96}
   for $d=4$, $q=2$.
   Error bars are one standard deviation,
   and include statistical errors only.
}
\label{z_sw_table}
\end{table}

A further advance was made in 1998 by
Chayes and Machta (CM) \cite{ChayesMachta98},
who devised a cluster algorithm for simulating the
FK random-cluster model at any {\em real}\/ value $q \ge 1$.
The idea behind the CM algorithm is very similar to that of SW,
but now one starts with the random-cluster (bond) measure
and introduces auxiliary color (spin) variables on the sites,
so that we again have a joint model of spins and bonds,
analogous to the one employed in the SW algorithm.
Although the marginal measure on the spins
no longer has any obvious physical interpretation when $q$ is noninteger,
the joint measure can still be used to construct an efficient cluster
algorithm, by alternately applying the conditional distributions exactly
as in standard SW.
The CM algorithm thus generalizes the SW algorithm
and in fact reduces to (a slight variant of) it when $q$ is an integer:
see Section \ref{subsec.CM} for details.
Indeed, the CM algorithm can be thought of
as a ``natural'' interpolation of the SW algorithm
to noninteger $q$ (though unfortunately only for $q \ge 1$).
Thus, by using the CM algorithm we can study the dynamic critical behavior
of the SW--CM dynamic universality class as a function of the
{\em continuous}\/ variable $q$ throughout the range
$1\le q\le q_c(\mathcal{L})$, where $q_c(\mathcal{L})$
is the maximum $q$ for which the transition is
second-order on the given lattice $\mathcal{L}$.\footnote{
   We stress that $q_c(\mathcal{L})$ is \emph{not}\/ necessarily
   the same for all lattices of a given dimension $d$;
   the first-order or second-order nature of the transition
   is a {\em non-universal}\/ question.
   See \cite{BloteDengQianSokal_inprep} for further discussion.
   For the standard two-dimensional lattices (square, triangular and hexagonal)
   we have $q_c(\mathcal{L}) = 4$.
}
This vastly enhances our ability to make theoretical sense
of the numerical results.

In the present paper we perform high-precision MC simulations
of the random-cluster model on the square lattice ($d=2$),
for $1.25\le q \le 4$ in steps of $0.25$, using the CM algorithm.
Our main goal is to gain better insight into the
SW--CM dynamic universality class in two dimensions
by studying the behavior as a function of the continuous variable~$q$.
We estimate numerically several dynamic critical exponents
for each value of $q$,
and we attempt to understand their behavior as a function of $q$.
As a byproduct we also obtain new information on corrections to scaling
in the {\em static}\/ quantities.
In the companion paper \cite{cm3_fullpaper}
we carry out an analogous study for the simple-cubic lattice ($d=3$)
at $q=1.5,1.8,2.2$;
and in a forthcoming paper \cite{cmKn_fullpaper}
we will analyze the case of the complete graph (Curie--Weiss model).\footnote{
   A brief summary of the preliminary results in this trio of papers
   has appeared in letter form \cite{DengGaroniMachtaOssolaPolinSokal07}.
}

One advantage of considering $d=2$ is of course is that we have available
a remarkable amount of information concerning the static behavior of the
random-cluster model as a function of $q$.
This information includes the exact location~\cite{Baxter82,Beffara_10} of
the transition point on the square lattice,
namely $p_c=\sqrt{q}/(1+\sqrt{q})$,
and the knowledge that the transition is second-order for $0 \le q \le 4$
and first-order for $q > 4$.
Furthermore, the leading and next-to-leading thermal and magnetic
critical exponents are known exactly (though non-rigorously)
for the entire range $0 \le q \le 4$
\cite{Nienhuis84,Nienhuis_DL11,denNijs_83}:
\begin{eqnarray}
   y_{T1}  & = &  {3g-6 \over g}  \\[2mm]
   y_{T2}  & = &  {4g-16 \over g}  \\[2mm]
   y_{H1}  & = &  {(g+2)(g+6) \over 8g}  \\[2mm]
   y_{H2}  & = &  {(g-2)(g+10) \over 8g}
\end{eqnarray}
where $g$ is the Coulomb-gas coupling defined by
$q=4\cos^2(\pi g/4)$ and $2\le g \le 4$.\footnote{
   See \cite[Appendix~A.1]{Salas_Sokal_Potts4}
   for further discussion and references.
   Note that there is a typographical error in equation~(A.10)
   of \cite{Salas_Sokal_Potts4}, which should read
   $\Delta_{r,s}=([2(s-r)+s x]^2 -x^2)/(8[(2+x)]$.
   We remark that the same formulae for $4 \le g \le 6$
   give the exponents $y_{T1},y_{T2},y_{H1},y_{H2}$
   of the tricritical Potts model \cite{Nienhuis84,Nienhuis_DL11}.
}
For the standard critical exponents this implies
\begin{eqnarray}
   1/\nu     & = &  y_{T1}  \;=\;  {3g-6 \over g}
        \label{eq.nu.exact}  \\[2mm]
   \Delta_1  & = &  -y_{T2}  \;=\;  {16-4g \over g}
         \label{eq.Delta1.exact} \\[2mm]
   d_F       & = &  y_{H1}  \;=\;  {(g+2)(g+6) \over 8g}  \\[2mm]
   \alpha/\nu & = & 2y_{T1} - d  \;=\;  \frac{4g-12}{g}
         \label{eq.alphaovernu.exact}  \\[2mm]
   \beta/\nu  & = & d - y_{H1}  \;=\; \frac{(g-2)(6-g)}{8g}
         \label{eq.betaovernu.exact}  \\[2mm]
   \gamma/\nu & = & 2y_{H1} - d  \;=\; \frac{12+g^2}{4g}
         \label{eq.gammaovernu.exact}
\end{eqnarray}
(Here $\Delta_1$ is the leading correction-to-scaling exponent,
 defined e.g.\ by corrections $\sim L^{-\Delta_1}$
 in finite volume at criticality;
 and $d_F$ is the cluster fractal dimension.)
The numerical values of these critical exponents
for the values of $q$ employed in our simulations
are collected for reference in Table~\ref{table_exact_exponents}.

A key theoretical result concerning the SW dynamics is the Li--Sokal bound
\cite{LiSokal89}, which states that 
\begin{equation}
\tau_{\text{int},\N}, \, \tau_{\text{exp},\N}
   \;\ge\; \text{const} \times C_H
\label{SW_Li--Sokal}
\end{equation}
where $\tau_{\text{int},\N}$ and $\tau_{\text{exp},\N}$
are, respectively, the integrated and exponential autocorrelation times
for the observable $\N =$ number of occupied bonds,
and $C_H$ is the specific heat.
(For the precise definitions of $\tau_{\text{int}}$ and
$\tau_{\text{exp}}$, see Section \ref{autocorrelation times}.)
It follows from this lower bound that
the SW algorithm {\em cannot}\/ completely eliminate critical slowing-down
if the specific heat is divergent at criticality.
In particular, (\ref{SW_Li--Sokal}) implies the lower bound
\begin{equation}
   z_{\text{int},\N}, \, z_{\text{exp}}  \;\ge\;  \alpha/\nu
\label{SW_Li--Sokal_z}
\end{equation}
for the dynamic critical exponents of the SW algorithm,
where $\alpha/\nu$ is the static exponent for the specific heat.
Now, simple ``Fortuin--Kasteleyn identities'' show that $\var(\N)/V$
is a specific-heat-like quantity and provides a
natural continuation of the notion of specific heat to random-cluster models.
It turns out that the Li--Sokal bound (\ref{SW_Li--Sokal})
with this definition of specific heat can be easily extended
from the SW to the CM algorithm at arbitrary real $q \ge 1$,
as we show in \cite[Appendix~A]{cm3_fullpaper}.

The physical mechanism underlying the Li--Sokal proof is the slow evolution
of $\N$, which is an ``energy-like'' observable.
Previous studies in two dimensions for $q=2,3,4$
have shown empirically that the Li--Sokal bound is very
close to being sharp for all three values of $q$
\cite[Section~6]{Salas_Sokal_Ising_v1}
\cite{Salas_Sokal_Potts3,Salas_Sokal_AT,Salas_Sokal_Potts4};
and we will show here that this is the case also for noninteger $q$
throughout the range $1 \le q \le 4$
(see Sections~\ref{subsec.zintE} and \ref{subsec.li-sokal}).
For the three-dimensional Ising model, by contrast,
the Li--Sokal bound is very far from sharp \cite{OssolaSokal04}.
Therefore there must be another mechanism,
beyond the one captured in the Li--Sokal proof, that is principally
responsible for the critical slowing-down in three dimensions.
A few years ago, Ossola and Sokal \cite{OssolaSokal04} suggested that
this as-yet-not-understood mechanism causing slowness might perhaps be
somehow related to the typical size of the largest cluster.
More specifically, they conjectured that perhaps the SW dynamics
for any Potts ferromagnet in any dimension satisfies
\begin{equation}
\tau_{\text{SW}} \;\stackrel{?}{\ge}\; \text{const} \times \frac{L^d}{C_1}
\label{Ossola-Sokal_conjecture}
\end{equation}
where $C_1$ is the expected number of sites in the largest cluster.
If true, this would imply the critical-exponent inequality
\begin{equation}
  z_{\text{SW}} \;\stackrel{?}{\ge}\; \beta/\nu
   \;,
\label{Ossola-Sokal_conjecture_z}
\end{equation}
where $\beta/\nu$ is the static exponent for the magnetization.
Indeed, Coddington and Baillie \cite{CoddingtonBaillie92}
had earlier suggested that the inequality \reff{Ossola-Sokal_conjecture_z}
holds as an {\em equality}\/ for the Ising models in dimensions $d=2,3,4$.
The numerical results reported by Ossola and Sokal \cite{OssolaSokal04}
for the three-dimensional Ising model are consistent with this
conjectured equality, though they also present plausible fits consistent with
$z_{\text{SW}} < \beta/\nu$ (i.e.\ violation of their conjectured inequality).
Finally, an analytical treatment \cite{PerskyBenAvKanterDomany96} 
of the SW dynamics for the Ising model on the complete graph
suggests that $z_{\text{SW}}=\beta/\nu$ in this case also
(namely, $z_{\text{SW}}=\beta/\nu = 1$).

Alas, by considering the Ossola--Sokal conjecture
within the more general framework of the Chayes--Machta dynamics,
one can see at a glance that it is probably false!
To start with, for $q=1$ on any lattice,
the Swendsen--Wang--Chayes--Machta algorithm
reduces to independent sampling for independent bond percolation,
so that $z_{\text{CM}}= 0$;
but in general $\beta/\nu > 0$ for percolation,
which shows that the Ossola--Sokal conjecture \reff{Ossola-Sokal_conjecture_z}
is false for $q=1$.
Moreover, it is reasonable to believe that $z_{\text{CM}}$
is a continuous function of $q$:
if so, then for $q$ slightly greater than 1
one has either $z_{\text{CM}}$ identically zero
or else $z_{\text{CM}}$ very close to zero;
and in either case one would have $z_{\text{CM}} < \beta/\nu$,
so that the Ossola--Sokal conjecture
would be violated also for $q$ in some interval above 1.\footnote{
   It is of course conceivable that $z_{\text{CM}}$ is discontinuous at $q=1$:
   for instance, one might have $z_{\text{CM}} = \beta/\nu$ exactly
   for all $q$ near 1, but with an amplitude that vanishes as $q \downarrow 1$.
   In Section~\ref{subsec.ossola-sokal}
   we will test this scenario against our data.
}
Our numerical results in this paper
(see Section~\ref{subsec.zintE})
suggest that in $d=2$
the Ossola--Sokal conjecture fails for $1 \le q \ltapprox 1.95$;
indeed, for $1 \le q \ltapprox 1.6$
we apparently have $z_{\text{CM}} = 0$ exactly,
while of course $\beta/\nu > 0$.
(But see Section~\ref{subsec.ossola-sokal} for an alternative fit
that is compatible with the Ossola--Sokal conjecture.)
Likewise, the numerical results for $d=3$
to be presented in~\cite{cm3_fullpaper} for $q=1.5$ and $q=1.8$
strongly suggest that the Ossola--Sokal conjecture is
violated for these values of $q$ and hence presumably for all $q \ltapprox 2$,
but that it holds as a strict inequality $z_{\text{CM}} > \beta/\nu$
when $q=2.2$ and hence presumably for all $q$
in the range $2 \ltapprox q \le q_c(\Z^3)$.
Indeed, in $d=3$ it is found that $z_{\text{CM}}$ is a strongly increasing
function of $q$, while $\beta/\nu$ varies very slowly with $q$;
the two curves presumably cross somewhere near $q = 2$.
Finally, in our forthcoming paper \cite{cmKn_fullpaper}
we will extend the method of \cite{PerskyBenAvKanterDomany96}
to cover the CM dynamics on the complete graph
for all $q$ in the range $1\le q \le 2$.
We find that there are two markedly different behaviors depending on $q$:
for $1 \le q<2$ we have $z_{\text{CM}} =0 < \beta/\nu = 1$,
while for $q=2$ we have $z_{\text{CM}} =\beta/\nu=1$.
In particular, the Ossola--Sokal conjecture (\ref{Ossola-Sokal_conjecture_z})
is violated when $1 \le q<2$ for the CM dynamics on the complete graph.

It therefore seems that the mechanism causing slowness of SW--CM
in three dimensions is {\em not}\/ related in any simple way
to the size of the largest cluster, contrary to the intuition
of Ossola and Sokal.\footnote{
   Or at least, no such relation seems to hold for {\em all}\/ $q$.
   It is still conceivable that such a relation might hold
   for $q=2$ only, or for all $q \ge 2$.
}
And so we are back to square one:
no one seems to have the slightest idea
what is the physical mechanism dominating the
critical slowing-down of the SW--CM dynamics in dimensions $d \ge 3$.


The present paper is organized as follows:
In Section~\ref{algorithm} we briefly review the Swendsen--Wang algorithm
and then present a simple explanation of the Chayes--Machta algorithm.
In Section~\ref{observables and autocorrelation times} we define the
observables and autocorrelation times that we measured in our simulations,
while in Section~\ref{statistics} we discuss our methods of
statistical data analysis.
In Section~\ref{simulations} we summarize the characteristics
of our MC simulations.
In Section~\ref{sec.data_static} we analyze our numerical data
for the static observables, with an emphasis on detecting
corrections to scaling and estimating their exponents and amplitudes.
In Section~\ref{sec.data_dynamic} we analyze our numerical data
for the dynamic quantities (i.e.\ autocorrelation times),
with an emphasis on estimating the dynamic critical exponent $z_{\rm CM}$
as a function of $q$;
in particular we test the Ossola--Sokal conjecture
and the sharpness of the Li--Sokal bound.
Finally, in Section~\ref{discussion} we briefly discuss our results
and the prospects for future work.

\section{The Swendsen--Wang and Chayes--Machta algorithms}
\label{algorithm}

In this section we describe a family of algorithms
for simulating the FK random-cluster model at any real $q \ge 1$,
which generalize slightly the original algorithms introduced by
Chayes and Machta~\cite{ChayesMachta98}.
We begin (Section~\ref{subsec.SW}) by reviewing the
Swendsen--Wang \cite{SwendsenWang87} algorithm for simulating the
ferromagnetic $q$-state Potts model at integer $q\ge2$,
with emphasis on the perspective afforded by the
joint probability distribution of spins and bonds
(sometimes known as the Edwards--Sokal coupling~\cite{EdwardsSokal88}).
We then present (Section~\ref{subsec.CM}) our version of the
Chayes--Machta algorithm and a proof of its validity.
For an alternative (and rather more general) presentation
of the Chayes--Machta algorithm,
see \cite{cm3_fullpaper}.

\subsection{Swendsen--Wang algorithm}   \label{subsec.SW}

Let $G = (V,E)$ be a finite graph with vertex set $V$ and edge set $E$,
and let $q$ be an integer $\ge 2$.
The $q$-state Potts model on $G$
with nearest-neighbor couplings $\boldsymbol{\beta}=\{\beta_e\}_{e\in E}$
is defined by the Gibbs measure 
\begin{subeqnarray}
   \pi_{\boldsymbol{\beta},q}(\color) 
   & \propto &
   \exp\left(\sum_{e \in E} \beta_e \, \delta_e(\color) \right)
            \\[1mm]
   & \propto &
   \prod_{e\in E} \, [(1-p_e) \,+\, p_e \delta_e(\color)]  \;,
 \label{Potts measure}
\end{subeqnarray}
for $\color\in \spinset$,
where $p_e=1-e^{-\beta_e}$ for each $e\in E$ and
\begin{equation}
   \delta_e(\color) \;=\;  \delta_{\sigma_x , \sigma_y}
    \text{  for  } e=\langle xy \rangle \;.
\end{equation}

One rather natural way to elucidate the Swendsen--Wang (SW) algorithm
is via the Edwards--Sokal coupling~\cite{EdwardsSokal88} of the Potts and 
random-cluster models.
Recall first that the random-cluster model \cite{Grimmett_06},
introduced by Fortuin and Kasteleyn
\cite{KasteleynFortuin69,FortuinKasteleyn72,Fortuin72a,Fortuin72b},
is a correlated bond-percolation model
defined on a finite graph $G = (V,E)$ for parameter $q>0$
(not necessarily an integer)
and edge probabilities $\mathbf{p}=\{p_e\}_{e\in E}$,
by the probability measure
\begin{equation}
   \phi_{\mathbf{p},q}(\mathbf{n})
   \;\propto\;
   q^{k(\mathbf{n})}\, B_{\mathbf{p}}(\mathbf{n})
   \;,
 \label{RC measure}
\end{equation}
for $\bond\in\configurationspace$, where
\begin{equation}
   B_{\mathbf{p}}(\bond)  
   \;=\;
   \prod_{e\in E \colon\, n_e = 1} p_e \prod_{e\in E \colon\, n_e = 0} (1-p_e)
\end{equation}
and $k(\bond)$ is the number of connected components (``clusters'')
in the spanning subgraph whose edges are those with $n_e = 1$
(we call these ``components of $\bond$'' for short).
The Potts and random-cluster models are intimately related,
in the sense that correlation functions of the Potts spins can be
expressed in terms of connectivity functions of the random-cluster model.
The Edwards--Sokal coupling is a joint measure on
the space of spin and bond configurations, defined by
\begin{equation}
   \mu_{\mathbf{p},q}(\bond,\color)
   \;\propto\;
   \prod_{e \in E} \,
     [(1-p_e)\delta_{n_e,0} \,+\, p_e\,\delta_{n_e,1} \delta_e(\color)]
   \;.
 \label{ES_measure}
\end{equation}
It is easy to show \cite{EdwardsSokal88,SokalLectures}
that the marginal measure $\sum_{\bond}\mu_{\mathbf{p},q}(\bond,\cdot)$
on the spins is the Potts measure (\ref{Potts measure}),
and the marginal measure $\sum_{\color}\mu_{\mathbf{p},q}(\cdot,\color)$
on the bonds is the random-cluster measure (\ref{RC measure}).
Explicit forms for the conditional measures are also easily computed:
\begin{eqnarray}
   \mu_{\mathbf{p},q}(\color|\bond) 
   & = &
   \Delta(\bond,\color)\, q^{-k(\bond)}
        \label{ES conditioned on bonds} \\[4mm]
   \mu_{\mathbf{p},q}(\bond|\color)
   & = &
   \Delta(\bond,\color)\, B_{\mathbf{p}}(\bond)
\label{ES conditioned on spins}
\end{eqnarray}
where
\begin{equation}
   \Delta(\bond,\color)
   \;=\;
   \begin{cases}
       1  & \text{ if $\delta_e(\color)=1$ for all $e$ with $n_e=1$} \\
       0  & \text{ otherwise}
   \end{cases}
 \label{delta definition}
\end{equation}
In words, $\Delta(\bond,\color)$ is the indicator for the event
that ``we draw occupied bonds only between vertices with the same spin value''.

The SW algorithm simulates the Edwards--Sokal measure \reff{ES_measure},
and hence both the Potts and random-cluster models,
by alternately updating the spins conditioned on the bonds
using (\ref{ES conditioned on bonds})
and the bonds conditioned on the spins using (\ref{ES conditioned on spins}).
In words, the spin-updating rule (\ref{ES conditioned on bonds})
states that, independently for each cluster, we choose a new spin value
uniformly at random from the $q$ possible choices and assign
this spin value to all vertices in the cluster;
while the bond-updating rule (\ref{ES conditioned on spins})
states that on each subgraph induced by the vertices of a given spin value
we update the bonds via independent bond percolation.

\subsection{Chayes--Machta algorithm}   \label{subsec.CM}

Consider the random-cluster model defined on a finite graph $G=(V,E)$ by (\ref{RC measure}).
Recall that to arrive at the SW
algorithm for the Potts model, one introduces auxiliary bond
variables and then considers the joint model of Potts spins on the vertices
and auxiliary bond variables on the edges. The SW algorithm updates
this joint model by alternately applying its two conditional measures
(bond variables given spins and spin variables given bonds). 
Here we do the reverse:
starting from the random-cluster model (\ref{RC measure}),
we introduce auxiliary color variables on the vertices of $G$
and then consider the joint model of our original bond variables
and the new auxiliary color variables;
we will then update this joint model
by alternately applying its conditional measures.
To this end, let us introduce color variables on the sites taking values in
some finite set $S$, i.e. $\color\in S^V$.
For each $\alpha \in S$, we choose a number $q_\alpha \ge 0$, and write $\mathbf{q}=\{q_{\alpha}\}_{\alpha\in S}$.
We can now introduce the coupled probability measure
\begin{equation}
   \mu_{\mathbf{p},\mathbf{q}}(\bond,\color)
   \;\propto\;
   \Delta(\bond,\color) \:  B_{\mathbf{p}}(\bond) \,
     \prod_{\alpha \in S} q_\alpha^{k_\alpha(\color,\bond)}  \;.
 \label{joint measure}
\end{equation}
Here $\Delta(\bond,\color)$ is again given by (\ref{delta definition})
but now with with $\color\in S^V$,
i.e.\ it is the indicator of the event
``we draw occupied bonds only between vertices of the same color'';
and we denote by $k_\alpha(\color,\bond)$ the number of components of $\bond$
that are colored $\alpha$ by $\color$.
Note that $k_\alpha(\color,\bond)$ is well-defined
whenever $\Delta(\bond,\color)\neq0$.
{}From the definition (\ref{joint measure}) we can observe the following:

\begin{itemize}
   \item[1)]
The marginal measure $\sum_{\color} \mu_{\mathbf{p},\mathbf{q}}(\cdot,\color)$ 
on the bond variables is the random-cluster measure with parameter $q= \sum_{\alpha \in S} q_\alpha$.
   \item[2)]
The conditional measure of $\color$ given $\bond$ is as follows:
Independently for
each component of $\bond$, randomly choose a value $\alpha \in S$ with
probability $q_\alpha/q$ and impose it on all vertices in that component.
   \item[3)]
Every coloring $\color$ defines a partition of the vertex set into $V=\bigcup_{\alpha\in S}V_{\alpha}$, where $V_{\alpha}$ is the 
set of all sites $v\in V$ colored $\alpha$ by $\color$. 
The conditional measure of $\bond$ given $\color$ is then simply
the product over $\alpha \in S$ of random-cluster measures with parameter
$q_\alpha$ on the induced subgraphs $G[V_{\alpha}]$. All bonds not lying within a single
$G[V_{\alpha}]$ are forced to be unoccupied.
\end{itemize}

Now suppose that for each $\alpha \in S$ and each induced subgraph
$H \subseteq G$, we have a valid Monte Carlo algorithm for updating the
random-cluster model with parameter $q_\alpha$ on $H$. 
We can then apply the following algorithm to simulate the joint model
(\ref{joint measure}), and therefore, via Observation 1, the
random-cluster model with parameter $q=\sum_{\alpha\in S}q_{\alpha}$:

\begin{algthm}[Random-cluster algorithm] $\,$
  \begin{enumerate}
  \item Given a bond configuration, choose a new color
  configuration using the conditional measure described in Observation 2.
  \item Given the new color configuration, for each $\alpha\in S$
  update the random-cluster model on $G[V_{\alpha}]$
  with parameter $q_{\alpha}$, using the given algorithm.
  \end{enumerate}
  \label{RC algorithm}
\end{algthm}
Note that for any $\alpha \in S$, one valid Monte Carlo update is to do nothing
(i.e.\ perform the identity operation),
since detailed balance is satisfied trivially.
Furthermore, if $q_\alpha$ happens to equal 1, one can use a Bernoulli update,
i.e.\ erase the bonds on $G[V_{\alpha}]$ and choose
new ones via independent bond percolation.  But if one has some other
way of updating a random-cluster measure with parameter $q_\alpha$,
then that is fine too.
A number of special cases of Algorithm~\ref{RC algorithm} are now clear:
\begin{enumerate}
\item If $q$ is an integer, $S = \{1,\dots,q\}$, $q_{\alpha} = 1$
   for all $\alpha$, and one uses Bernoulli updates for every $\alpha$,
   then Algorithm~\ref{RC algorithm} is simply the standard SW algorithm.
\item If $q$ is an integer, $S = \{1,\dots,q\}$, $q_{\alpha} = 1$
   for all $\alpha$, and one uses Bernoulli updates for $1 \le \alpha \le k$
   and do-nothing updates for $k+1 \le \alpha \le q$
   (where $k$ is any integer satisfying $1 \le k \le q$),
   then Algorithm~\ref{RC algorithm} is a variant of the SW algorithm
   in which only the first $k$ colors are ``active''.
\item If $q \ge 1$, $S = \{active,inactive\}$, $q_{active} = 1$, $q_{inactive} = q-1$,
  and one uses a Bernoulli update for $\alpha=active$ and a do-nothing
  update for $\alpha=inactive$, then Algorithm~\ref{RC algorithm} is the
  original Chayes--Machta algorithm presented in \cite{ChayesMachta98}.
\item If $q \ge k \ge 1$ with $k$ an integer, $S=\{1,\dots,k,inactive\}$, $q_{\alpha} = 1$ for
  $1 \le \alpha \le k$, $q_{inactive} = q-k$, and one uses a Bernoulli update
  for $1 \le \alpha \le k$ and a do-nothing update for $\alpha=inactive$,
  this is the generalized algorithm presented in
  \cite[pp.~482--483]{ChayesMachta98}.
\end{enumerate}

Note that since each of these cases involves performing Bernoulli updates,
they all require $q\ge 1$. 
For a given value of $q\ge1$, Case 4 provides a family of algorithms,
indexed by the integer $k$, for simulating the random-cluster model.
We call this ``the Chayes--Machta algorithm with $k$ active colors''.
Any choice of $1\le k \le \lfloor q\rfloor$ is legitimate,
though it seems reasonable to expect that $k=\lfloor q\rfloor$
should be the most efficient. 
Indeed, our numerical simulations suggest that in practice
the autocorrelation time is approximately proportional to $1/k$
(see Section~\ref{subsec.dependence_on_k} below).

\section{Observables and autocorrelation times}
\label{observables and autocorrelation times}

In this section we recall the definitions of the various autocorrelation times
(Section~\ref{autocorrelation times})
and then list the observables that we measured in our simulations
(Section~\ref{observables}).

\subsection{Autocorrelation functions and autocorrelation times}
\label{autocorrelation times}

Consider an observable $\obs$ in the random-cluster model,
i.e.\ a real-valued function of $\bond\in\{0,1\}^E$.
Then a realization of the Chayes--Machta (CM) Markov chain
gives rise to a time series $\obs(t)$,
where each unit of time corresponds to one step of the CM algorithm.
The autocovariance function of $\obs$ is defined to be
\begin{equation}
C_{\obs \obs}(t) \;=\;  \langle \obs(0)\obs(t)\rangle -\langle\obs\rangle^2,
\end{equation}
where the expectation is taken in equilibrium.
The normalized autocorrelation function of $\obs$ is then
\begin{equation}
\rho_{\obs \obs}(t) \;=\; \frac{C_{\obs \obs}(t)}{C_{\obs \obs}(0)}.
\end{equation}
{}From $\rho_{\obs \obs}(t)$ we define the integrated autocorrelation time as
\begin{equation}
\tau_{{\text{int}},\obs} \;=\; \frac{1}{2}\,\sum_{t=-\infty}^{\infty}\,\rho_{\obs\obs}(t)
\label{tauint definition}
\end{equation}
and the exponential autocorrelation time as
\begin{equation}
\tau_{{\text{exp}},\obs} \;=\; \limsup_{|t|\to\infty}
\frac{-|t|}{\log\,\rho_{\N \N}(t)}.
\label{tauexp definition}
\end{equation}
Finally, the exponential autocorrelation time of the system is defined
as
\begin{equation}
\tau_{\text{exp}} \;=\; \sup_{\obs}\tau_{{\text{exp}},\obs},
\end{equation}
where the supremum is taken over all observables $\obs$. This
autocorrelation time thus measures the decay rate of the slowest mode of
the system. All observables that are not orthogonal to this slowest
mode satisfy $\tau_{{\text{exp}},\obs}=\tau_{\text{exp}}$.

It is important to remember that there is not just one autocorrelation
time, but many: namely $\tau_{\text{exp}}$ as well as
$\tau_{{\text{int}},\obs}$ for each $\obs$. In all but the most
trivial Markov chains these autocorrelation times are {\em not}
equal. Correspondingly, there are many dynamic critical exponents:
namely $z_{\text{exp}}$ as well as $z_{\text{int},\obs}$ for each
$\obs$. These exponents {\em may} in some cases be equal, but they
need not be; this is a detailed dynamical question, and the answer
will vary from algorithm to algorithm and model to model.

More information on the principles of Markov-chain Monte Carlo
and the relations between these autocorrelation times
can be found in \cite{SokalLectures}.

\subsection{Observables to be measured}
\label{observables}

We now take the graph $G$ to be the $d$-dimensional simple-hypercubic lattice
of linear size $L$ with periodic boundary conditions;
we denote this graph by $\torus$ in order to emphasize that
addition of coordinates is always taken mod $L$.
When the precise form of the graph $G$ is unimportant to our definitions,
we denote the vertex set and edge set as simply $V(G)$ and $E(G)$, respectively.

For $x_1, \ldots, x_m \in V(G)$, not necessarily all distinct,
we denote by $\gamma_{x_1 \cdots x_m}$ the indicator for the event
that the vertices $x_1,\ldots,x_m$ are all in the same cluster, i.e. 
\begin{equation}
\gamma_{x_1 \cdots x_m}(\bond)  \;=\;
\begin{cases}
1 & 
\text{if 
$x_1\leftrightarrow x_2 \leftrightarrow x_3 \leftrightarrow \dots \leftrightarrow x_m$
in configuration $\bond$} \\
0 & \text{otherwise}
\end{cases}
\end{equation}
and $x \leftrightarrow y$ denotes that $x$ is connected to $y$
by at least one path of occupied bonds.
In particular, $\tau(x,y) = \langle \gamma_{x y}\rangle$
is the 	``two-point connectivity function'',
i.e.\ the probability that $x$ is connected to $y$.
On $\torus$ we write $\tau(x) = \tau(x,0)$,
and translational invariance gives us $\tau(x,y)=\tau(x-y)$
for all $x,y\in\torus$.

We measured the following observables in our MC simulations:
\begin{itemize}
\item The number of occupied bonds
\begin{equation}
\N  \;=\;  \sum_{e\in E(G)}n_e
\end{equation}

\item The nearest-neighbor connectivity (which is an energy-like observable \cite{Salas_Sokal_Potts3})
\begin{equation}
\E  \;=\;  \sum_{\langle xy\rangle\in E(G)}\gamma_{xy}
\end{equation}

\item The cluster-size moments
\begin{subeqnarray}
   \Sm
   & = &
     \sum_{C\in K(\bond)}|C|^m   \\
   & = & \!\!
     \sum_{x_1,\ldots, x_m \in V(G)} \gamma_{x_1 \cdots x_m}(\bond) \;,
\label{S cluster sum}
\end{subeqnarray}
where the size $|C|$ of a cluster means the number of vertices.
In this work we measured $\Sm$ for $m=2,4,6,8$;

\item The size $\Ci$ of the $i$th largest cluster. In this work we
measured $\Ci$ for $i=1,2,3$;

\item An observable used to compute the Fourier transform of $\tau(x,y)$
   evaluated at the smallest non-zero momentum $(2\pi/L,0,\ldots,0)$:
\begin{subeqnarray}
   \F
   & = &
   \frac{1}{d}\sum_{j=1}^d \sum_{x,y\in\torus}
        \gamma_{x y} \,\,e^{2\pi i(x_j-y_j)/L}
        \\
   & = &
   \frac{1}{d}\sum_{j=1}^d \sum_{C \in K(\bond)}
      \left|\sum_{x\in C} e^{2\pi i x_j/L}\right|^2
\end{subeqnarray}
\end{itemize}

 From these observables we computed the following quantities:
\begin{itemize}
\item The bond density
\begin{equation}
N  \;=\; \frac{\langle\cal{N}\rangle}{B}
   \;,
\end{equation}
where $B = |E|$ is the total number of edges in the graph $G$
(i.e.\ $B = dL^d$ for $G = \torus$);
\item The connectivity density
\begin{equation}
E'  \;=\;  \frac{\langle\cal{E}'\rangle}{B}
\end{equation}
\item The specific heat. There are a number of sensible definitions
  for specific heat, and we considered the following
  (see Remark~\ref{remark_C_H} below):
\begin{eqnarray}
   C_H^{(1)}
   & = &
   \frac{1}{B} \var(\mathcal{N})
       \label{CH1} \\[2mm]
   C_H^{(2)}
   & = &
   \frac{d}{p^2} \, [C_H^{(1)} - (1-p) \, N]
       \label{CH2} \\[2mm]
   C_H^{(3)}
   & = &
   \frac{q^2}{(q-1)^2} \, C_H^{(2)}
      \label{CH3}
\end{eqnarray}
In this paper we used $C_H = C_H^{(2)}$
in agreement with \cite{Salas_Sokal_Ising_v1,Salas_Sokal_Potts3}.
\item The expected size of the $i$th largest cluster,
\begin{equation}
C_i  \;=\; \langle \Ci \rangle
\end{equation}
We expect that in the critical region $C_i \sim L^{d-\beta/\nu}$ as
$L\to\infty$, for any fixed $i$;

\item The susceptibility (see Remark~\ref{remark_chi} below)
\begin{equation}
\chi \;=\; \frac{\langle \Stwo\rangle}{L^d}
\end{equation}
In the random-cluster model, $\chi$ is the mean size of the
cluster containing any specified point.

\item The Fourier transform of the two-point connectivity function
$\tau(x)$, evaluated at the smallest nonzero momentum, $(2\pi/L,0,...0)$:
\begin{equation} 
F  \;=\;
 {\widetilde \tau}(p)\Big\vert_{p=(2\pi/L,0,...0)}
  \;=\; \frac{\langle\cal{F}'\rangle}{L^d}
\end{equation}

\item The finite-size second-moment correlation length
 (see Remark~\ref{correlation length remark} below):
\begin{equation}
\xi  \;=\; \frac{1}{2\sin(\pi/L)} \left(\frac{\chi}{F}-1\right)^{1/2}
\label{correlation length definition}
\end{equation}
\end{itemize}

\bigskip

\begin{remark}
   \label{remark_C_H}
The definition $C_H^{(1)}$ of the specific heat
seems quite natural from the perspective of the random-cluster model,
in which we view the bond variables as fundamental;
in particular, it arises in a very direct way in
the proof of the Li--Sokal bound for the CM algorithm
\cite[Appendix]{cm3_fullpaper}.
The definitions $C_H^{(2)}$ and $C_H^{(3)}$, by contrast,
are designed to reduce to more familiar expressions in terms of
the energy of the Potts spin model when $q$ is an integer.
Specifically, if $q$ is an integer we have
\begin{eqnarray}
   C_H^{(2)}
   & = &
\frac{1}{V} \, \var \Biggl( \sum_{\langle xy \rangle\in E}
\delta_{\sigma_x,\sigma_y}  \Biggr) 
     \\[2mm]
   C_H^{(3)}
   & = &
\frac{1}{V} \, \var \Biggl( \sum_{\langle xy \rangle\in E}
\bsigma_x \cdot \bsigma_y  \Biggr)
\end{eqnarray}
where $\sigma_x\in\{1,2,\dots,q\}$
and $\bsigma_x \in \mathbb{R}^{q-1}$ is a Potts spin in the
hypertetrahedral representation, so that
\begin{equation}
\bsigma_x \cdot \bsigma_y
  \;=\;
\frac{q\,\delta_{\sigma_x,\sigma_y} -1}{q-1}  \;.
\label{hypertetrahedral  identity}
\end{equation}
\end{remark}

\bigskip

\begin{remark}
   \label{remark_chi}
We note that for $q\in\{2,3,\ldots\}$, the Fortuin--Kasteleyn identity 
$\langle \bsigma_x \cdot \bsigma_y \rangle
 = \langle\gamma_{x y}\rangle$ shows that 
$\langle{\cal{M}}^2\rangle=\langle\Stwo\rangle$,
where $\mathcal{M}^2$ is the squared magnetization, and hence $\chi$ is simply
equal to the magnetic susceptibility.
\end{remark}

\bigskip

\begin{remark}
\label{correlation length remark}
In terms of ${\widetilde{\tau}}(p)$ the definition (\ref{correlation length definition}) has the general form 
\begin{equation}
\xi \;=\; \frac{1}{2\sin(\pi/L)} \left(\frac{{\widetilde \tau}(0,\ldots,0)}{\widetilde{\tau}(2\pi/L,0,\ldots,0)}-1\right)^{1/2}
\label{correlation function definition of correlation length}
\end{equation}
and applies equally well to the integer $q$ case where we interpret
$\tau(x)$ as the spin-spin correlation function.
A very nice account of the origin of the definition
(\ref{correlation function definition of correlation length})
of the finite-size correlation length, relating it to the thermodynamic
second-moment correlation length is given in \cite[Part~III]{AmitMayor05};
see also \cite{CaraccioloEdwardsPelissettoSokal93,CooperFreedmanPreston82}.

The prime on $\F$ is employed to distinguish it from another
estimator for $F$, often denoted ${\cal F}$,
which can be defined in terms of Potts spins when $q\in\{2,3,\ldots\}$:
see e.g.\ \cite{Salas_Sokal_Potts3}.
We note that $\F$ could profitably be used even in Swendsen--Wang simulations
of the Potts model, as an alternative to $\mathcal{F}$.
\end{remark}

\bigskip

\begin{remark}
For any $q>0$ and any $\langle x y \rangle \in E(G)$ we have the identity 
\begin{equation}
q\,\langle n_{x y} \rangle  \;=\; 
p \left[(q-1)\langle\gamma_{x y}\rangle +1 \right].
\label{n test}
\end{equation}
This is most easily seen by first considering the Edwards--Sokal joint
measure when $q$ is an integer $\ge 2$,
and then combining the identities
$\langle n_{xy}\rangle=p \langle\delta_{\sigma_x,\sigma_y}\rangle$ and 
$\langle\gamma_{xy}\rangle=\langle \bsigma_x \cdot \bsigma_y \rangle$
together with (\ref{hypertetrahedral  identity}).
See e.g.\ \cite{SokalLectures}.
This clearly proves (\ref{n test}) for all $q\in\{2,3,\ldots\}$;
and since both sides are rational functions of $q$,
it follows that the equality must hold for all $q\in\mathbb{C}$. 
Summing over all edges we then obtain
\begin{equation}
\langle\N\rangle  \;=\; p\frac{q-1}{q}\langle\E\rangle \,+\, \frac{p}{q} B \;.
\label{N test}
\end{equation}
As a consistency check on the correctness of our simulations 
we numerically tested the identity (\ref{N test}) to high precision 
by measuring the observable 
\begin{equation}
\N \,-\, p\frac{q-1}{q} \E \,-\, \frac{p}{q}B  \;.
\end{equation} 
Clearly this observable should have mean zero,
and in all our simulations this was observed to be the case 
within statistical errors.
\end{remark}

\section{Statistical analysis of the Monte Carlo data}
\label{statistics}

Consider a generic observable $\obs$ with expectation
$\langle\obs\rangle=\mu_{\obs}$.
Suppose that, after equilibration, we have a sequence of $T$
Monte Carlo measurements of $\obs$, denoted $\{\obs_t\}_{t=1}^T$.
Since the Monte Carlo process is assumed equilibrated,
the $\{\obs_t\}_{t=1}^T$ are a sample from a stationary stochastic process
(also called a stationary time series).
In this section we recall the basic principles of statistical
time-series analysis, and describe the standard estimators
for various quantities associated with $\obs$.
For more details, see e.g.\ \cite{SokalLectures,Priestley94}.

The natural estimator for the expectation $\mu_{\obs}$ is the sample mean
\begin{equation}
\overline{\obs} \;=\; \frac{1}{T}\sum_{t=1}^T\obs_t.
\end{equation}
This is an unbiased estimator and has a variance
\begin{subeqnarray}
   \var(\overline{\obs})
   & = &
   \frac{1}{T^2}\sum_{s,t=1}^T C_{\obs\obs}(t-s)
       \\
   & = &
   \frac{1}{T}\sum_{t=-(T-1)}^{T-1}\left(1-\frac{|t|}{T}\right)C_{\obs\obs}(t)
     \\
   & \approx &
   \frac{1}{T}\,2\,\tau_{\text{int},\obs}\,C_{\obs\obs}(0)\quad
   \text{for}\quad T\gg\tau_{\text{int},\obs} \;.
\end{subeqnarray}
This implies that the variance of $\overline{\obs}$ is a factor
$2\,\tau_{\text{int},\obs}$ larger than it would be if the measurements
were uncorrelated. Therefore, in order to obtain accurate error
bars for the estimator of the static quantity $\mu_{\obs}$, we must obtain
an accurate estimate of the dynamic quantity $\tau_{\text{int},\obs}$.

The natural estimator for the autocovariance function is
\begin{equation}
\widehat{C}_{\obs\obs}(t) 
   \;=\;
 \frac{1}{T-|t|}\,\sum_{s=1}^{T-|t|}(\obs_s-\mu_{\obs})(\obs_{s+t}-\mu_{\obs})
\end{equation}
if the expectation $\mu_{\obs}$ is known, and 
\begin{equation}
\widehat{\widehat{C}}_{\obs\obs}(t) 
   \;=\;
 \frac{1}{T-|t|}\,\sum_{s=1}^{T-|t|}(\obs_s-\overline{\obs})(\obs_{s+t}-\overline{\obs})
\end{equation}
if $\mu_{\obs}$ is unknown. The estimator $\widehat{C}_{\obs\obs}(t)$
is unbiased, and the bias of $\widehat{\widehat{C}}_{\obs\obs}(t)$ is
of order $1/T$. 
To leading order for $T\gg \tau_{\text{int},\obs}$ 
the covariance matrices of $\widehat{C}_{\obs\obs}$ and $\widehat{\widehat{C}}_{\obs\obs}$ are equal 
and it can be shown \cite{Anderson71,Priestley94} that
\begin{equation}
\begin{split}
\cov(\widehat{C}_{\obs\obs}(t),\widehat{C}_{\obs\obs}(u))
&=
\frac{1}{T}\,\sum_{s=-\infty}^{\infty}
[C_{\obs\obs}(s)\,C_{\obs\obs}(s+u-t) \,+\, C_{\obs\obs}(s+u)\,C_{\obs\obs}(s-t)
\\
&\qquad\qquad\qquad +\,
\kappa(t,s,s+u)]
\:+\: o(1/T) \;,
\end{split}
\label{autocovariance error}
\end{equation}
where $t,u\ge 0$ and $\kappa$ is the connected 4-point autocorrelation
function
\begin{equation}
\begin{split}
\kappa(r,s,t)
& \;=\; \langle(\obs_i-\mu_{\obs})(\obs_{i+r}-\mu_{\obs})(\obs_{i+s}-\mu_{\obs})(\obs_{i+t}-\mu_{\obs})\rangle
\\
&\quad-C_{\obs\obs}(r)C_{\obs\obs}(t-s)-C_{\obs\obs}(s)C_{\obs\obs}(t-r)-C_{\obs\obs}(t)C_{\obs\obs}(s-r).
\end{split}
\end{equation}

Similarly, the natural estimator for the autocorrelation function is 
\begin{equation}
\widehat{\rho}_{\obs\obs}(t) \;=\;
 \frac{\widehat{C}_{\obs\obs}(t)}{\widehat{C}_{\obs\obs}(0)}
\end{equation}
if $\mu_{\obs}$ is known, and
 \begin{equation}
\widehat{\widehat{\rho}}_{\obs\obs}(t) \;=\;
 \frac{\widehat{\widehat{C}}_{\obs\obs}(t)}{\widehat{\widehat{C}}_{\obs\obs}(0)}
\end{equation}
if $\mu_{\obs}$ is unknown.
Both the estimators $\widehat{\rho}_{\obs\obs}(t)$ and
$\widehat{\widehat{\rho}}_{\obs\obs}(t)$ have bias of order
$1/T$. The covariance matrices of $\widehat{\rho}_{\obs\obs}$ and $\widehat{\widehat{\rho}}_{\obs\obs}$ 
are the same to leading order for large $T$. If the process is
Gaussian, this covariance matrix is given in the large $T$ limit by
\cite{Priestley94}
\begin{equation}
\begin{split}
\cov(\widehat{\rho}_{\obs\obs}(t),\widehat{\rho}_{\obs\obs}(u))
&=\frac{1}{T}
\sum_{s=-\infty}^{\infty}[\rho_{\obs\obs}(s)\rho_{\obs\obs}(s+t-u)+\rho_{\obs\obs}(s+u)\rho_{\obs\obs}(s-t)
\\
&+2\,\rho_{\obs\obs}(t)\rho_{\obs\obs}(u)\rho_{\obs\obs}^2(s)
-2\,\rho_{\obs\obs}(t)\rho_{\obs\obs}(s)\rho_{\obs\obs}(s-u)
\\
&-2\,\rho_{\obs\obs}(u)\rho_{\obs\obs}(s)\rho_{\obs\obs}(s-t)]
  \:+\:  o(1/T)  \;.
\end{split}
\label{autocorrelation error}
\end{equation}
If the process is not Gaussian, then there are additional terms
proportional to the fourth cumulant $\kappa(s,t,t-u)$. The simplest
assumption is to assume the stochastic process to be ``not too far
from Gaussian'' and simply drop all the terms involving $\kappa$. This is
what we will do. If this assumption is not justified, then we are
introducing a bias into the estimate of
$\cov(\widehat{\rho}_{\obs\obs}(t),\widehat{\rho}_{\obs\obs}(u))$.

Finally, we take the estimator for the integrated autocorrelation time
to be \cite{MadrasSokal88}
\begin{equation}
\widehat{\tau}_{\text{int},\obs}  \;=\;
 \frac{1}{2}\,\sum_{t=-M}^{M}\widehat{\rho}_{\obs\obs}(t)
\end{equation}
if $\mu_{\obs}$ is known, 
or the analogous object defined in terms of
$\widehat{\widehat{\rho}}_{\obs\obs}$ is $\mu_{\obs}$ is unknown;
here $M$ ($\le T-1$) is a suitably chosen number,
which we call the {\em window width}\/.
One's first thought might be to use all the data, i.e.\ take $M = T-1$;
but this turns out to be a very bad estimator,
which has a variance of order 1 even as $T\to\infty$.
Loosely speaking (see e.g.\ \cite{MadrasSokal88,Priestley94}),
this is because the terms in $\widehat{\rho}_{\obs\obs}(t)$
with large $t$ (namely, $t \gg \tau_{{\text{exp}},\obs}$)
have variance of order $1/T$ that does not vanish as $t$ grows,
cf.\ (\ref{autocorrelation error}), and there are of order $T$ of them.
These terms thus contribute much ``noise'' but very little ``signal'',
since $\rho_{\obs\obs}(t)$ is tiny when $t \gg \tau_{{\text{exp}},\obs}$.
To obtain a good estimator, we should instead choose the window width $M$
to be large enough so that we do not lose too much ``signal'',
i.e.\ $\rho_{\obs\obs}(t)$ is tiny for all $t > M$,
but not too much larger than this.
In general, a window of width $M$ creates a bias given by
\begin{equation}
\text{bias}
(\widehat{\tau}_{\text{int},\obs})
 \;=\; -\frac{1}{2}\,\sum_{|t|>M}
\rho_{\obs\obs}(t)  \:+\: o(1/T)  \;.
   \label{bias_tauint}
\end{equation}
The variance of the estimator $\widehat{\tau}_{\text{int},\obs}$
can be computed from (\ref{autocorrelation error});
assuming that $\tau_{\text{int},\obs} \ll M \ll T$,
the final result is \cite{MadrasSokal88}
\begin{equation}
\var(\widehat{\tau}_{\text{int},\obs}) \;\approx\;
  \frac{2(2M+1)}{T}\tau_{\text{int},\obs}^2  \;.
   \label{var_tauint}
\end{equation}

To choose the window width $M$ we used the automatic windowing
algorithm introduced in \cite{MadrasSokal88}, in which one sets 
\begin{equation}
M \;=\; \min\{m\in\mathbb{Z}:m\ge c \,\widehat{\tau}_{\text{int},\obs}(m)\}
\end{equation}
where $c$ is a suitably chosen constant. If the normalized
autocorrelation function is approximately a pure exponential,
then a choice in the range $c\approx 6-8$ is reasonable.
Indeed \cite{Salas_Sokal_Potts3},
if we take $\rho_{\cal OO}(t) = e^{-|t|/\tau}$
and minimize the mean-square error
\begin{equation}
\hbox{MSE}(\widehat{\tau}_{{\rm int},{\cal O}}) \;\equiv\;
{\rm bias}(\widehat{\tau}_{{\rm int},{\cal O}})^2 +
{\rm var}(\widehat{\tau}_{{\rm int},{\cal O}})
\end{equation}
using \reff{bias_tauint}/\reff{var_tauint}, we find that the optimal
window width is
\begin{equation}
M_{\rm opt} \;=\; {\tau \over 2} \log \left( {n \over 2 \tau} \right) -1 \;.
\end{equation}
For $n/\tau \approx 10^4,10^5,10^6,10^7,10^8$ with $\tau \gg 1$,
we have $M_{\rm opt}/\tau \approx 4.26,5.41,6.56,7.71,8.86$, respectively.

In this paper we chose $c=6$ for the observable $\mathcal{E}'$,
which has the slowest-decaying autocorrelation function of
all the observables we measured and whose autocorrelation function
is in fact very close to a pure exponential.
We then used the same window width $M$, computed by applying
the automatic windowing algorithm to $\mathcal{E}'$,
for all the other observables.
We think that this procedure is preferable to applying the
automatic windowing algorithm directly to the other observables,
since some of the latter have an autocorrelation decay that is
far from a pure exponential (this is especially so for
 $\mathcal{F}'$, $\mathcal{C}_2$ and $\mathcal{C}_3$).

\bigskip

\begin{remark}
If $\{\obs^{(1)},\dots,\obs^{(m)}\}$ is a family of observables, 
we estimate composite observables of the form 
$f(\langle \obs^{(1)}\rangle,\dots,\langle \obs^{(m)}\rangle)$  
by 
$f(\overline{\obs^{(1)}},\dots, \overline{\obs^{(m)}})$ 
and approximate the corresponding variance
by assuming the fluctuations from the mean are small,
so that 
\begin{equation}
\var\, f(\overline{\obs^{(1)}},\dots, \overline{\obs^{(m)}})
\;\approx\;
g^2\, \var \,\overline{Z},
\end{equation}
where
\begin{eqnarray}
Z_{t} & = & \sum_{i=1}^m\,a_i\,\obs^{(i)}_t
   \\
a_i
  & = &
\frac{1}{g(\langle \obs_1\rangle,\dots,\langle \obs_m\rangle)} \;
\frac{\partial f}{\partial x_i}(\langle \obs_1\rangle,\dots,\langle \obs_m\rangle)
\end{eqnarray}
and 
$g = g(\langle \obs_1\rangle,\dots,\langle \obs_m\rangle)$
is just a convenient common factor of the derivatives of $f$,
which may be simply $1$ in practice.
Obviously in practice we compute $a_i$ using $\overline{\obs}_i$
in place of $\langle \obs_i \rangle$.
We used this method to compute the estimate and standard error of $\xi$.

In principle we can also estimate the specific heat using this procedure,
but in practice it turns out that a better estimator for the specific heat
is provided by
\begin{equation}
\widehat\varN  \;=\; {1 \over B} \, (\N-\langle \N\rangle)^2  \;.
\label{specific heat estimator}
\end{equation}
The mean of (\ref{specific heat estimator}) is clearly $C_H^{(1)}$
[cf. \reff{CH1}], and its variance determines the
error bar on the resulting estimate of $C_H^{(1)}$.
We prefer this estimator since it avoids computing the quantity
$\overline{\N^2}-\overline{\N}^2$ which can be a numerically perilous
object when $\overline{\N^2}$ and $\overline{\N}^2$ are close in value.
\end{remark}

\bigskip

\begin{remark}
For the larger values of $L$ we actually performed a number of independent runs rather than
one long run. The best estimate of each autocorrelation function $\rho_{\obs \obs}(t)$ was 
then computed, for each fixed $t$, by averaging the 
${\widehat{\widehat\rho}}_{\obs\obs}(t)$ from the individual runs, with
weights proportional to the run lengths. The windowing procedure
described above was then applied to this best estimate of $\rho_{\obs
  \obs}(t)$ in order to obtain the final value for the integrated
autocorrelation time.
\end{remark}

\section{Description of the simulations}
\label{simulations}

We implemented the Chayes--Machta (CM) algorithm for the random-cluster model
on an $L \times L$ square lattice with periodic boundary conditions.
We performed all our runs at the exact critical temperature
$p_c=\sqrt{q}/(1+\sqrt{q})$.
We studied lattice sizes $16 \le L \le 1024$ in powers of 2,
and parameters $1.25\le q \le 4$ in steps of $0.25$.
For each $(q,L)$ pair we studied all values of $k$
(the number of active colors) in the range $1 \le k \le \lfloor q \rfloor$.
We thus performed 189 runs in total.

For each triplet $(q,k,L)$
we performed between $2 \times 10^7$ and $10^8$ CM iterations:
more precisely, we peformed $10^8$ iterations for $16 \le L \le 256$,
$5 \times 10^7$ iterations for $L = 512$,
and $2\times 10^7$ iterations for $L = 1024$.
In terms of the autocorrelation time
$\tau_{\rm exp} \approx \tau_{{\rm int},\E}$,
our total data set at each triplet $(q,k,L)$
ranges in length from $\approx 7 \times 10^7 \tau$
on the smallest lattices at small $q$ (i.e.\ $L=16$, $q=1.25$)
to $\approx 7 \times 10^3 \tau$
on the largest lattices at large $q$ (i.e.\ $L=1024$, $q=4$, $k=1$);
in 90\% of cases (171 out of 189 runs)
our run length is at least $10^5 \tau$.
These statistics are high enough to permit a high accuracy in our
estimates of the static (error $\ltapprox$ 0.1\%)
and dynamic (error $\ltapprox$ 1\%) quantities,
except for the largest lattices at large $q$.

Our results for the principal static observables,
obtained by combining the data for all available $k$ values for each pair
$(q,L)$, are reported in
Tables~\ref{static N data}--\ref{static C1 data}.
Our results for the dynamic quantities $\tau_{{\rm int},\obs}$
for the case $k=1$ and the most important observables $\obs$
are reported in Tables~\ref{dynamic EP data}--\ref{dynamic C3 data}.
The complete set of static and dynamic data,
for all observables and all values of $k$,
is contained as a file {\tt cm2data.tar.gz}
in the preprint version of this paper at {\tt arXiv.org}.\footnote{
   The {\tt tar} file unpacks to make a directory {\tt cm2data}
   with subdirectories {\tt static} and {\tt dynamic}.
   Each of these subdirectories is in turn subdivided according to
   the value of $q$ and, in the dynamic case, the value of $k$.
   Individual files have names like {\tt C1\_Q1.25.txt}
   or {\tt tau\_C1\_Q4\_K1.txt}
   and have three fields on each line: $L$, value and error bar.
}

The initial configuration of each run was all-bonds-occupied,
except for the run at $(q,k,L) = (3.75,3,1024)$,
for which it was all-bonds-vacant.
The number of iterations discarded at the beginning of each run
in order to allow the system to reach equilibrium
was $j \times 10^5$ for $j=1,2,3$ or 4,
except for eight runs ($2 \le q \le 2.75$, $L=256$ and $k=1,2$)
for which it was $10^4$.\footnote{
   Please don't ask why the discard interval was chosen in this
   erratic way --- we don't remember anymore!
}
Thus, the the discard interval was in all cases at least
$100 \, \tau_{{\rm int},{\cal E}'}$
and in most cases at least $1000 \, \tau_{{\rm int},{\cal E}'}$.
Since our data suggest that $\tau_{{\rm exp}}$ is very close to
$\tau_{{\rm int},{\cal E}'}$,
it follows that our discard interval was in all cases at least
$100 \, \tau_{{\rm exp}}$, which is more than sufficient
for the systematic error from thermalization to be negligible.\footnote{
   Unless there exists a vastly slower mode of which we are unaware,
   so that in fact $\tau_{{\rm exp}} \gg \tau_{{\rm int},{\cal E}'}$.
}

%
Our program was written in Fortran 77
and run on a 1266 MHz Pentium III Tualatin processor
using the g77 Fortran compiler.
Our program requires approximately $(12+26d) L^d$ bytes memory
for a lattice of linear size $L$ in $d$ dimensions.
%
%
%
%
%
The CPU time required by our program was approximately
0.35~$\mu$s/iteration/lattice site;
this value is weakly dependent on the lattice size and on $k$ and $q$.
In particular, the CPU time per site rises sharply on very small lattices
due to the ``fixed costs'' of the algorithm
(i.e.\ those that do not scale with the volume);
it rises gradually on lattices $L \gtapprox 128$
due to ``cache misses'' coming from
the nonlocal nature of the Chayes--Machta algorithm,
when the lattice no longer fits in the computer's 512 KB cache.
The complete set of runs reported in this paper
used approximately 14 yr CPU time.

One delicate issue concerns the choice of the pseudo-random-number generator.
We learned by bitter experience \cite{Ossola-Sokal_systematic}
that linear congruential pseudo-random-number generators can cause
systematic errors in Monte Carlo simulations using the Swendsen--Wang
(or Chayes--Machta) algorithm,
if the lattice size is a multiple of a very large power of~2
and one random number is used per bond in a periodic manner.
These systematic errors arise from correlations within
a single bond-update half-sweep:
see \cite{Ossola-Sokal_systematic} for details.
These errors can be eliminated (or at least radically reduced)
by updating the bonds in a random order or in an aperiodic manner,
and by using a generator of large modulus (e.g.\ 60 or more bits). 
In the present project, therefore, we used a linear congruential generator
\begin{equation}
   x_{n+1}  \;=\;  ax_n + c  \, \pmod{m}
 \label{def_LCRG}
\end{equation}
with modulus $m=2^{64}$, increment $c=1$,
and multiplier $a=3202034522624059733$.
This multiplier gives good results on the spectral test in low dimensions
\cite{Lecuyer_99}.
Moreover --- and perhaps even more importantly ---
we used an ``aperiodic'' updating,
in which the random-number subroutine is called
{\em only}\/ if the two spins are equal and of an ``active color'',
i.e.\ for an edge $e = \langle xy \rangle$

\begin{tabbing}
\samepage
\qquad \= \qquad \= \qquad \= \qquad \= \+\+ \kill
   {\bf if} $\sigma_x = \sigma_y \le k$ {\bf then}  \\
\>   {\bf if} $\hbox{{\rm ran}(\,)} \le p$ {\bf then}  \\
\> \>   $n_e \leftarrow 1$ \\
\>   {\bf else} \\
\> \>   $n_e \leftarrow 0$ \\
\>   {\bf endif} \\
   {\bf endif}
\end{tabbing}

\noindent
To the best of our knowledge \cite{Ossola-Sokal_systematic}
these choices suffice to make the systematic errors negligible.

\section{Data analysis: Static quantities}
\label{sec.data_static}

Our general methodology for analyzing the Monte Carlo data
(both static and dynamic) is as follows:
For each quantity $Q$ of interest we impose an ansatz of the form
\begin{equation}
   Q(L)  \;=\;  A_1 L^{p_1} + A_2 L^{p_2} + \ldots + A_k L^{p_k}
\label{ansatz}
\end{equation}
(or a logarithmic variant thereof),
where some of the parameters may be fixed and others free.
The precise ansatz will be motivated in each case
by finite-size-scaling theory,
possibly together with some exactly known exponents.
Typically our ans\"atze will have 2--4 free parameters.
We then perform the nonlinear fits corresponding to the chosen ansatz
by using {\sc Mathematica}'s function \texttt{NonlinearModelFit}.
As a precaution against correction-to-scaling terms that we have failed
to include in our chosen ansatz,
we impose a lower cutoff $L\ge \lmin$ on the data points admitted in the fit,
and we systematically study the the effect on the $\chi^2$ value of
increasing $\lmin$.  In general, our preferred fit for any given ansatz
corresponds to the smallest $\lmin$ for which the goodness of fit is
reasonable and for which subsequent increases in $\lmin$ do not cause
the $\chi^2$ value to drop by vastly more than one unit per degree of freedom.
In practice, by ``reasonable'' we mean that the confidence level is
$\gtapprox$~10--20\%.\footnote{
   ``Confidence level'' is the probability that $\chi^2$ would exceed
   the the observed value, assuming that the underlying statistical model
   is correct.  An unusually low confidence level (e.g. less than 5\%)
   thus suggests that the underlying statistical model is {\em incorrect}\/;
   in our context it suggests that the terms we have chosen to retain
   in the ansatz (\ref{ansatz}) are insufficient to explain our data,
   i.e.\ there are unincluded corrections to scaling whose contribution
   is comparable to or larger than the statistical errors in our data.
}
We do not allow fits with zero degrees of freedom,
since there would then be no way of testing the goodness of fit.
As a last step, we consider the effect of including different terms
in the ansatz, by comparing the fits obtained with different ans\"atze.

The exact values of the static critical exponents are of course
already known [cf.\ \reff{eq.nu.exact}--\reff{eq.gammaovernu.exact}].
Our analysis of the static Monte Carlo data therefore has three major goals:
\begin{enumerate}
   \item To test how accurately we would be able to estimate the
       static critical exponents if we did {\em not}\/ know them exactly.
       Here our knowledge of the exact values allows us to evaluate
       systematic errors (including those due to unknown causes)
       as well as statistical ones.
       This study of the static critical exponents serves as
       a comparison case for our subsequent study
       of the dynamic critical exponents
       (Section~\ref{sec.data_dynamic} below).
       In particular, the behavior of
       the autocorrelation times is closely
       related to that of the specific heat,
       as is evident from the Li--Sokal bound \reff{SW_Li--Sokal}
       and the empirical fact of its near-sharpness.
   \item To estimate universal amplitude ratios such as
       $x^\star = \lim\limits_{L\to\infty} \xi(L)/L$.
   \item To study the structure of the corrections to scaling.
       In particular, are we able to see the leading non-analytic
       correction-to-scaling term $L^{-\Delta_1}$
       with exponent given by \reff{eq.Delta1.exact}?
       And what additional corrections are present?
       Depending on the observable,
       one might expect to see analytic corrections such as $L^{-1}$
       or regular background contributions.
\end{enumerate}

We begin by summarizing the finite-size-scaling (FSS) ans\"atze
that motivate our fits (Section~\ref{subsec.FSS}).
We then study successively
the correlation length $\xi$ (Section~\ref{subsec.xi}),
the susceptibility $\chi$ (Section~\ref{subsec.chi}),
the largest cluster $C_1$ (Section~\ref{subsec.C1}),
and the specific heat $C_H$ (Section~\ref{subsec.CH}).


\subsection{Corrections to finite-size-scaling}   \label{subsec.FSS}

Let us work exactly at the critical point,
and consider a quantity $Q(L)$ whose asymptotic behavior as $L \to\infty$ is
\begin{equation}
   Q(L)  \;=\;  A L^\psi \, ( 1 \,+\, \hbox{corrections} )
   \;,
 \label{eq.FSS.leading}
\end{equation}
where $\psi > 0$ is the leading critical exponent
and $A$ is the leading critical amplitude.\footnote{
   For the specific heat we have $\psi = \alpha/\nu < 0$ when $q < 2$.
   The same formulae will apply in this case,
   but the meanings of ``dominant'' and ``subdominant'' will be interchanged.
}
Then finite-size-scaling (FSS) theory
predicts the following types of contributions to the ``corrections''
in \reff{eq.FSS.leading}:\footnote{
   See e.g.\ \cite[Section~3]{Salas_Sokal_Ising_v1} for a review
   and citations to the original literature.
   We are also indebted to Youjin Deng for extremely helpful discussions
   of these points.
}
\begin{enumerate}
   \item ``Non-analytic'' corrections to scaling $L^{y_i}$
      coming from irrelevant scaling fields, i.e.\ those with $y_i < 0$.
      The leading such contribution is given by $y_i = y_{T2} = -\Delta_1$
      [cf.\ \reff{eq.Delta1.exact}].
      There are also higher-order ``non-analytic'' terms
      of the general form $L^{y_i + y_j + \ldots}$,
      but these are so far down as to be undetectable in our data.
   \item ``Analytic'' corrections to scaling coming from nonlinear mixing
      between the thermal and magnetic scaling fields.
      The leading such contribution behaves as
      $L^{y_{T1} - 2y_{H1}} = L^{(\alpha-1-\gamma)/\nu}$.
   \item A regular background term $L^{-\psi}$
      [i.e.\ a contribution of order $L^0$ to $Q(L)$].
   \item Corrections with negative integer exponents,
      i.e.\ $L^{-1}$, $L^{-2}$, etc.
      Some theoretical frameworks \cite{Guo_87} suggest that
      these contributions are absent.\footnote{
   See \cite[Section~3.2]{Salas_Sokal_Ising_v1} for further discussion.
   Interestingly, a recent study \cite{Feng_08}
   of percolation on two-dimensional lattices,
   using both transfer-matrix and Monte Carlo methods,
   found no evidence of any $L^{-1}$ correction
   but found indirect evidence of an $L^{-2}$ correction
   (namely, an apparent term $L^{-2} \log L$ that could result from
    mixing between the $L^{-2}$ and $L^{y_{T2}} = L^{-2}$ terms when $q=1$).
}
\end{enumerate}

Looking at the known exact values
\reff{eq.nu.exact}--\reff{eq.gammaovernu.exact}
of these exponents (see Table~\ref{table_exact_exponents}),
we can predict the pattern of corrections to scaling for different observables:
\begin{itemize}
   \item For strongly divergent quantities such as the susceptibility
     ($\psi = \gamma/\nu \approx 1.75$)
     and the mean cluster size
     ($\psi = d-\beta/\nu \approx 1.88$),
     the strongest correction to scaling at large $L$
     is expected to come from the leading irrelevant term $L^{-\Delta_1}$
     whenever $q > (3+\sqrt{5})/2 \approx 2.618$;
     indeed, this behavior will hold for all $q \gtapprox 1.294$
     (resp.\ $q \gtapprox 1.136$)
     if the $L^{-1}$ correction is absent.
     However, for $q \ltapprox 1.5$
     the irrelevant, regular-background and $L^{-2}$ contributions
     are all of roughly the same order
     and will be extremely difficult to disentangle.
   \item The same pattern is expected to hold for the correlation length,
     since it is built out of the more fundamental quantities
     $\chi$ and $F$ [cf.\ \reff{correlation length definition}]
     and therefore inherits their corrections to scaling
     (we expect $F$ to behave like $\chi$ in this context).
   \item For weakly divergent quantities such as the specific heat
     for $q \ge 2$, the strongest correction to scaling at large $L$
     is expected to come from the regular background term when
     $2 \le q < 2 + \sqrt{2} \approx 3.414$
     and from the leading irrelevant term when $q > 2 + \sqrt{2}$.
   \item For non-divergent quantities such as the specific heat
     for $q < 2$, the dominant contribution is the regular background,
     and the leading singular term $L^{\alpha/\nu}$ becomes the
     principal subdominant contribution.
\end{itemize}

One very interesting question concerns the presence or absence
of the correction-to-scaling term $L^{-\Delta_1} = L^{-4/3}$
in the case of the Ising model ($q=2$).
Plausible physical arguments \cite{Blote_88}
suggest that such a correction should be present in at least some models
in the Ising universality class, even though special symmetries might cause it
to be absent in simple exactly-soluble models such as the nearest-neighbor
square-lattice model.
A transfer-matrix study of the square-lattice random-cluster model
for $q \approx 2$ suggested \cite{Blote_88}
that such a correction is indeed present for all $q \neq 2$
but has an amplitude that vanishes linearly with $q-2$ when $q \to 2$.
If this is the case, then this correction will be observable
in derivatives $d/dq$ of standard Potts-model observables evaluated at $q=2$.
Note also that such derivatives can be expressed as bond observables in the
random-cluster model;  therefore, in this scenario, at least some
bond observables in the $q=2$ random-cluster model will exhibit
the correction-to-scaling term $L^{-4/3}$.

\subsection{The correlation length $\xi$}   \label{subsec.xi}

Finite-size-scaling (FSS) theory predicts that $\xi(L)/L$
for the random-cluster model on a torus
tends to a universal value $x^\star(q)$ as $L \to\infty$.\footnote{
   In general this value depends on the aspect ratio of the torus.
   Here we are considering only the case of aspect ratio 1.
}
More precisely, we expect the behavior
\begin{equation}
   \xi/L  \;=\;  x^\star  \,+\,  B L^{-p} \,+\, \ldots
 \label{def_FSS_xi}
\end{equation}
where $p > 0$ is one of the correction exponents
discussed in the preceding subsection.
For $q=2$, the exact value of $x^\star$ is known from conformal-invariance
theory \cite{DiFrancesco_87,DiFrancesco_88}
together with numerical integration \cite{Salas_Sokal_Ising_v2} to be
\begin{equation}
   x^\star(q=2) \;=\; 0.90504 88292 \pm 0.00000 00004   \;.
 \label{eq.xstar.exact.q=2}
\end{equation}
For $q \neq 2$ the exact value of $x^\star$ is unknown.
In this section we will use our Monte Carlo data to estimate
the universal values $x^\star(q)$ and to test the agreement of the
corrections to scaling with the FSS ansatz \reff{def_FSS_xi}.

\begin{figure}
\vspace*{-5mm}\hspace*{-1.5cm}
\begin{tabular}{c@{\qquad}c}
\includegraphics[width=240pt]{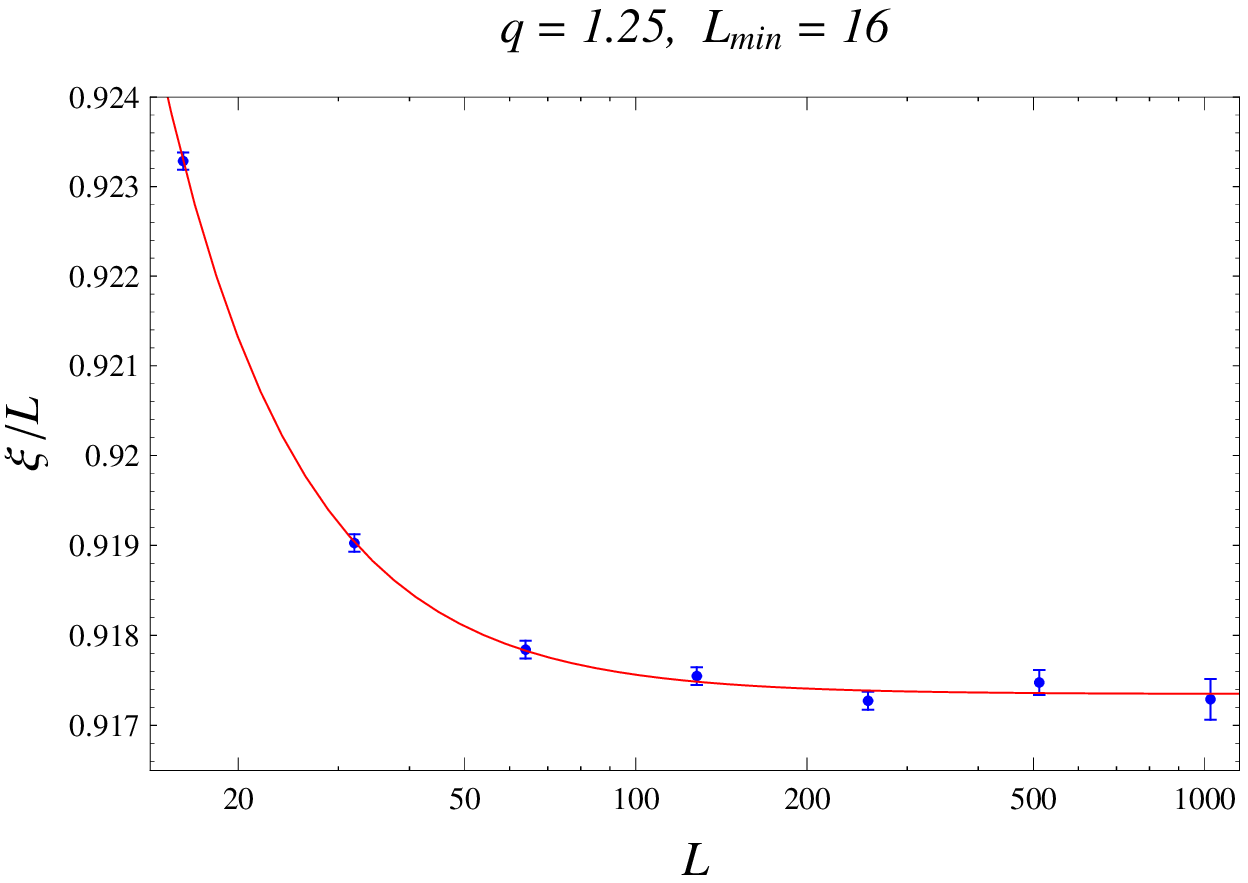} &
\includegraphics[width=240pt]{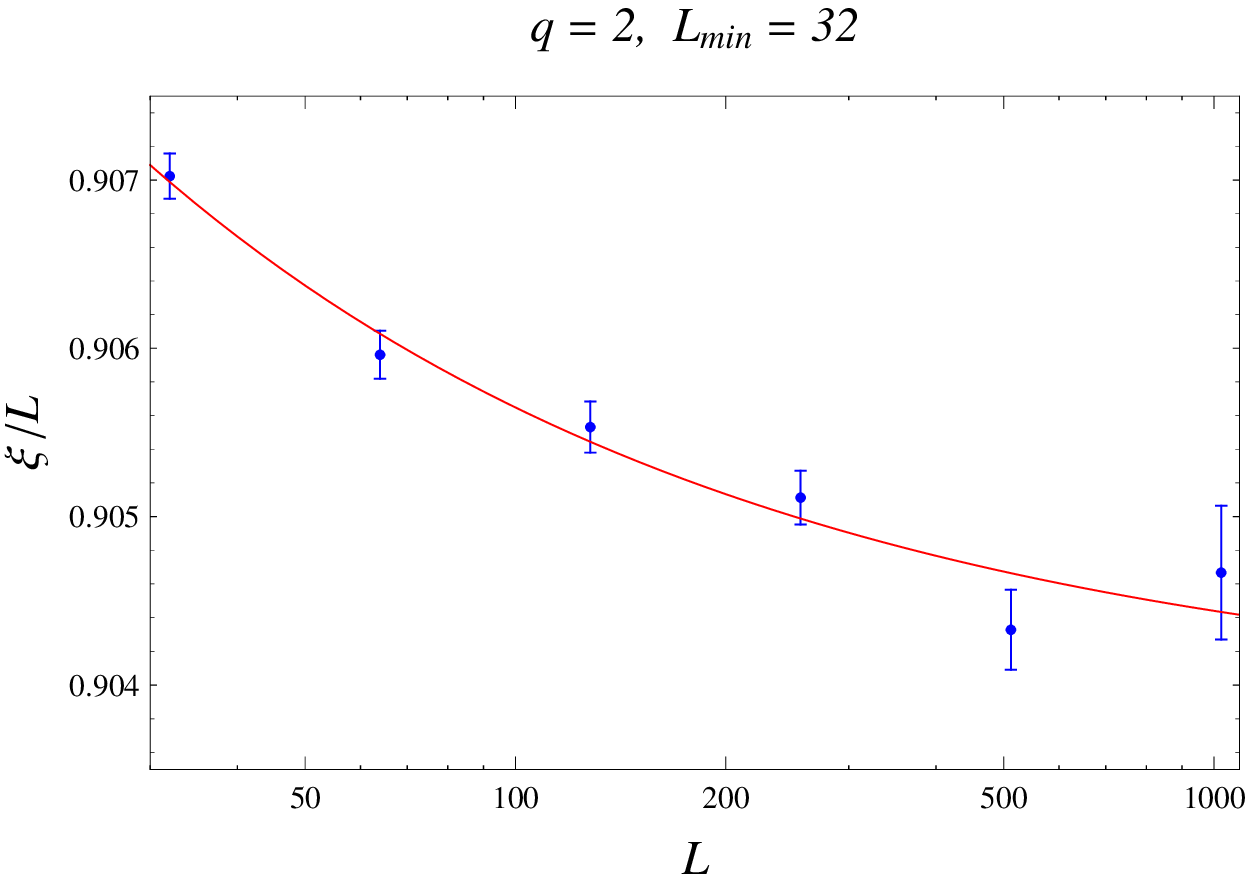} \\
\phantom{(((a)}(a) & \phantom{(((a)}(b) \\[8mm]
\includegraphics[width=240pt]{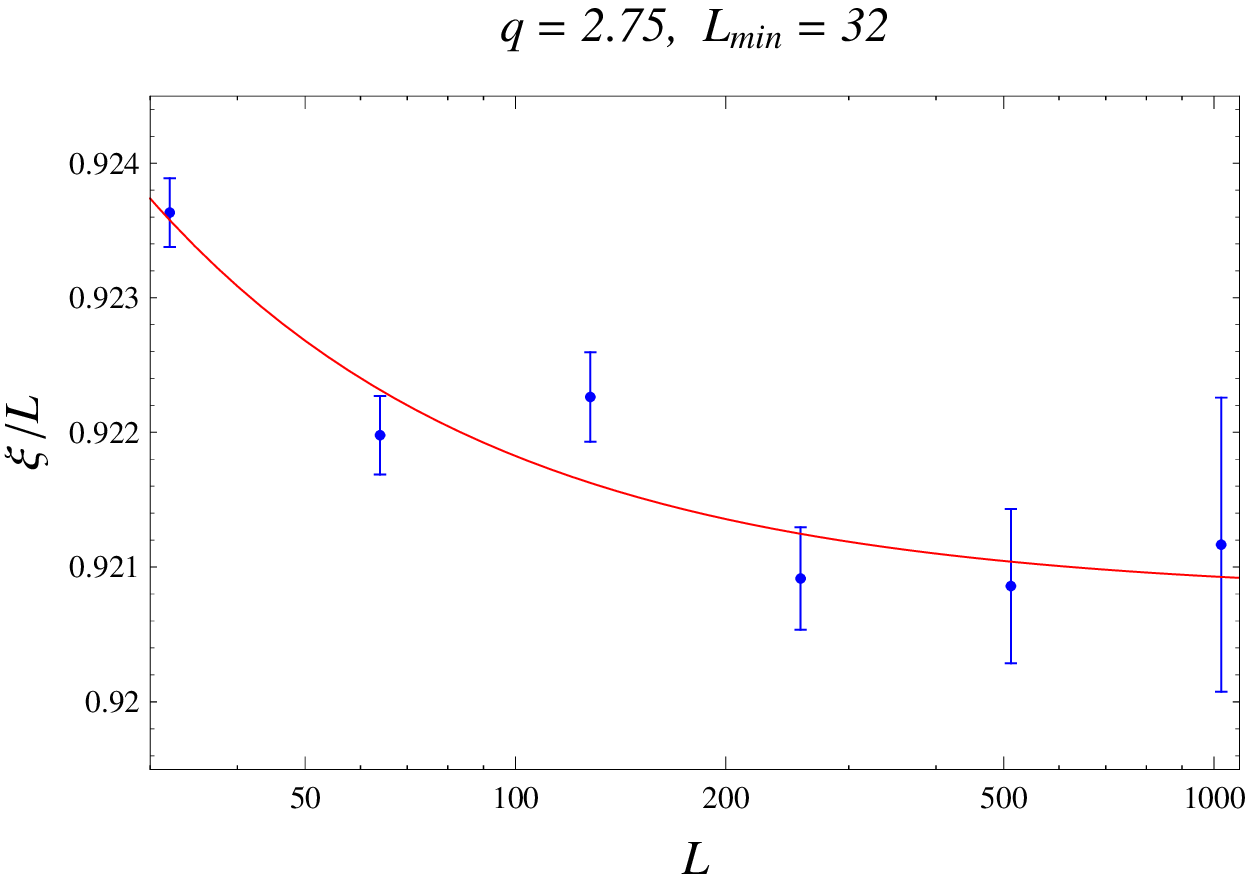} &
\includegraphics[width=240pt]{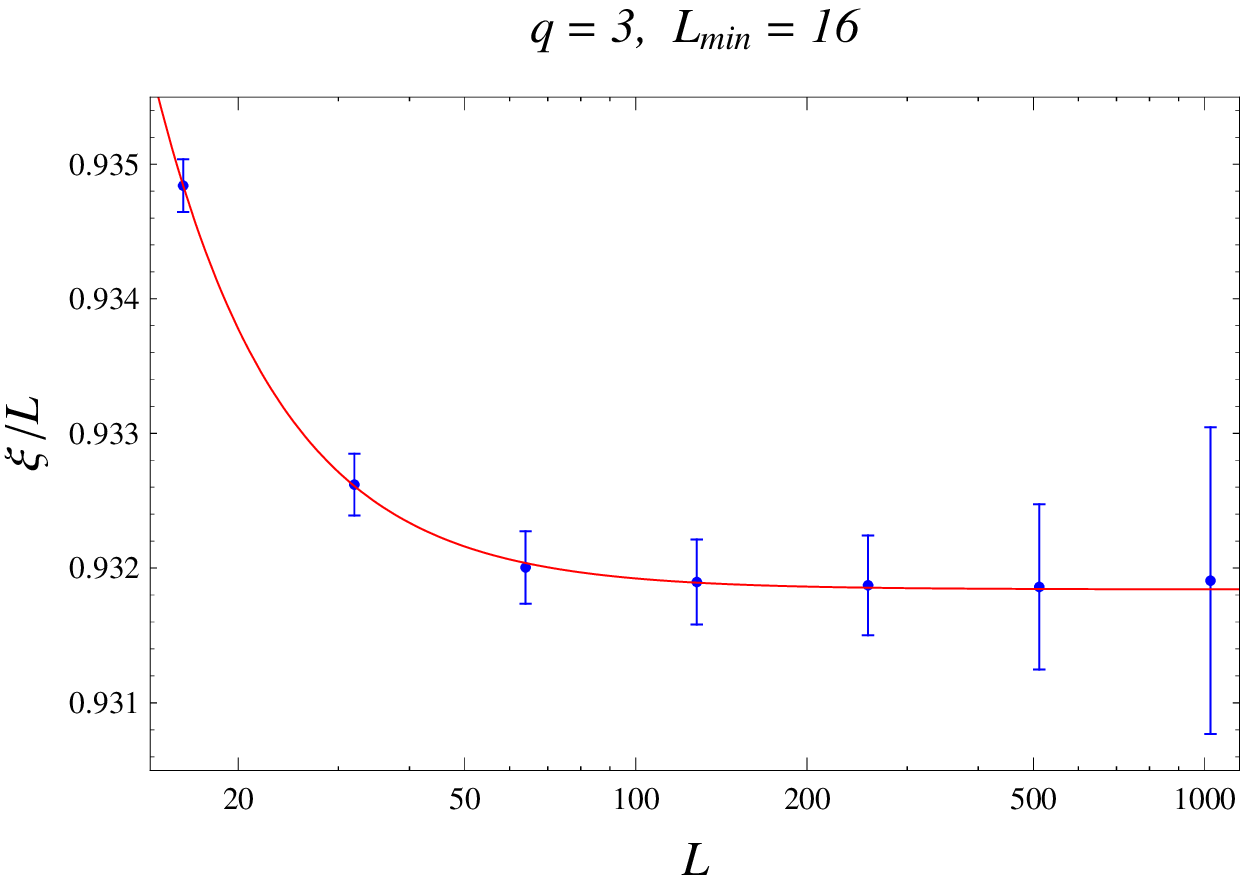} \\
\phantom{(((a)}(c) & \phantom{(((a)}(d) \\[8mm]
\includegraphics[width=240pt]{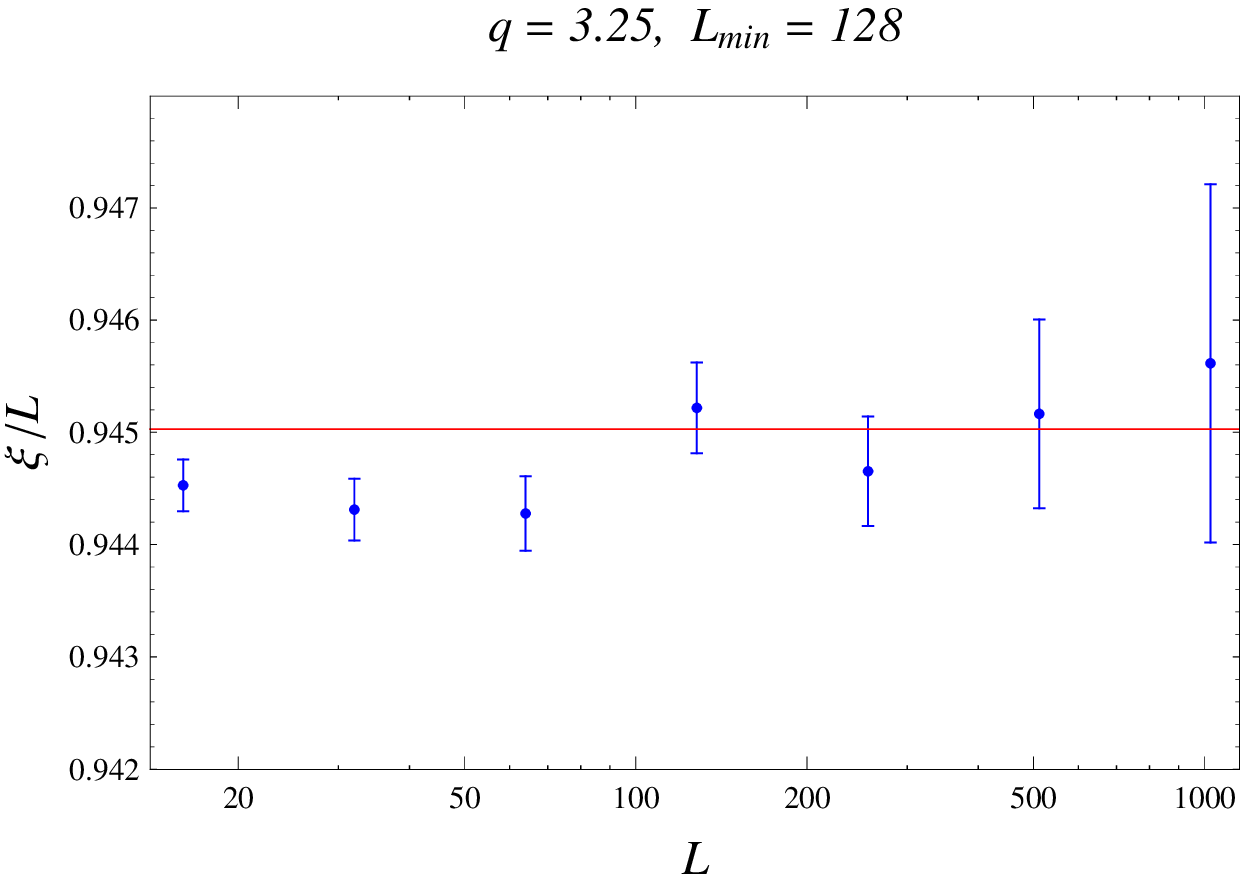} &
\includegraphics[width=240pt]{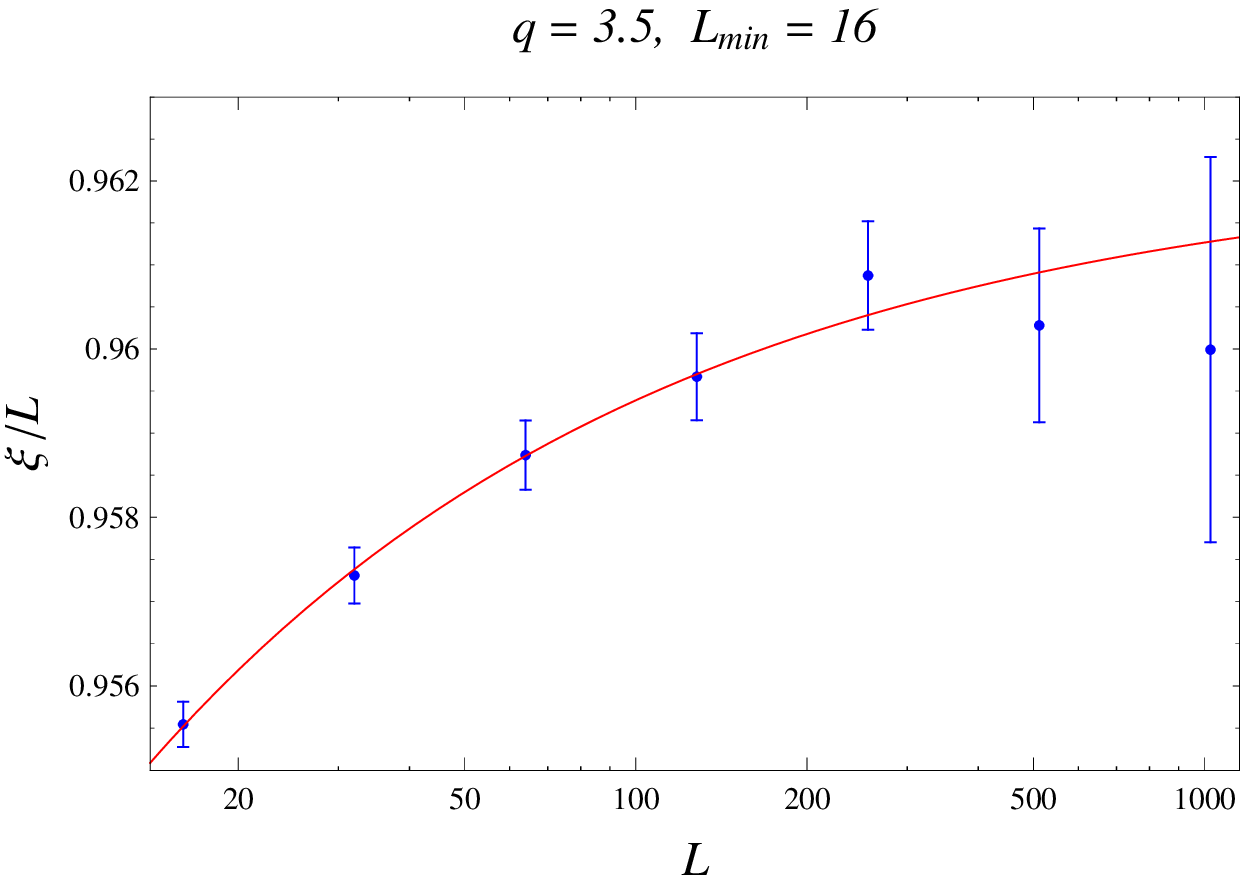} \\
\phantom{(((a)}(e) & \phantom{(((a)}(f) \\[3mm]
\end{tabular}
\caption{
   Fits of $\xi/L$ to $x^\star + B L^{-p}$.
   (a) $q=1.25$, (b) $q=2$, (c) $q=2.75$, (d) $q=3$, (e) $q=3.25$, (f) $q=3.5$.
   For $q=3.25$ the fit is actually to a constant $x^\star$.
}
\label{fig_xioverL_plots}
\end{figure}

In Table~\ref{table_xioverL_analysis}
we report fits of $\xi/L$
to the ans\"atze $x^\star$ and $x^\star + B L^{-p}$.
For $q=1.25$ and 1.50, the fits to both ans\"atze are excellent,
and the estimated values of $p$ agree closely
with the known exact values of $\Delta_1$;
the correction-to-scaling amplitude $B$ is positive and
more than three standard deviations away from zero.
In particular, there is no evidence of any $L^{-1}$ correction.
For $q=1.75$, the behavior is similar
but the quality of the fit is poorer
(possibly due to a statistical fluctuation that made the $L=1024$ point
 $2\sigma$ too high).
For $2 \le q \le 3.25$, the fits to a constant are excellent
provided that $\lmin$ is chosen large enough,
but the fits to the ansatz $x^\star + B L^{-p}$ are mostly badly behaved:
either they fail completely to converge (as for $q=3.25$),
or they exhibit estimates for $p$ that seem much too low ($q=2$)
or much too high ($q=2.5,3$) or somewhat too high ($q=2.25$);
only $q=2.75$ behaves well in these respects.
Also, the estimated correction amplitudes $B$
are mostly within two standard deviations of zero.
Examination of plots of these fits (see Figure~\ref{fig_xioverL_plots})
suggests that the problem arises because $\xi/L$
is a very flat and possibly nonomonotonic
function of $L$ at large $L$ in this range of $q$.\footnote{
   Or, perhaps, the statistical errors are just too large,
   compared to the small ``signal'', for us to estimate the
   corrections to scaling accurately.
}
This suggests that the ansatz \reff{def_FSS_xi} might be too simple,
and that we are seeing the combined effect of two (or more)
correction-to-scaling terms of opposite sign.
However, our statistical errors are too large for us
to resolve this question clearly.
For $3.5 \le q \le 4$, by contrast,
the fits to both ans\"atze are again excellent,
and the estimated values of $p$ agree well
with the known exact values of $\Delta_1$
(except at $q=4$, where the corrections to scaling
 $\log\log L/\log L$ and $1/\log L$ \cite{Salas_Sokal_Potts4}
 are mimicked by a small inverse power of $L$).
Moreover, the correction-to-scaling amplitude $B$ is now clearly negative.
This suggests that $B$ passed through zero somewhere in the range
$2 \ltapprox q \ltapprox 3.4$ and was small throughout this range,
thereby explaining why our fits to $x^\star + B L^{-p}$ were so unstable.

It is curious that our estimate for $x^\star(q=2)$ agrees
slightly {\em less well}\/ in absolute terms
with the exact value \reff{eq.xstar.exact.q=2}
than the estimate a decade ago by
Salas and Sokal \cite{Salas_Sokal_Ising_v2},
despite the fact that our raw-data error bars are a factor of 3--7 smaller
than theirs.
Most likely we were the unlucky victim of a statistical fluctuation
on our $L=512$ and $L=1024$ data points that placed them
roughly 2--3$\sigma$ and 1$\sigma$ too low, respectively,
thereby causing us to choose $\lmin=512$ rather than 256
in the fit to a constant and hence to privilege these discrepant values
[see Figure~\ref{fig_xioverL_plots}(b)].

\subsection{The susceptibility $\chi$}   \label{subsec.chi}


In Table~\ref{table_chi_analysis}
we report fits of the susceptibility $\chi$
to the pure-power-law ansatz $AL^{\gamma/\nu}$,
just to see how accurately we would be able to estimate the
exponent $\gamma/\nu$ if we did not know its exact value.
The last two columns show the deviation of the estimated $\gamma/\nu$
from the known exact value \reff{eq.gammaovernu.exact},
in absolute terms and in units of its standard deviation.

For small $q$ we are able to estimate $\gamma/\nu$ with extraordinary
accuracy: both statistical and systematic errors are of order 0.0001 or less.
However, as $q$ grows, both the systematic and the statistical errors grow:
the statistical errors grow because the critical slowing-down
is becoming gradually worse (see Section~\ref{sec.data_dynamic} below);
and the systematic errors grow (even faster than the statistical errors)
because the corrections to scaling are becoming much stronger
as the correction-to-scaling exponent $\Delta_1$ decreases
(see Table~\ref{table_exact_exponents}).
As a consequence, at $q \approx 3$
the statistical and systematic errors are of order 0.0003 and 0.001,
respectively;
at $q=3.75$ the errors are roughly twice this;
and at $q=4$ the statistical and systematic errors are of order 0.0015 and 0.01,
respectively.
(The large systematic errors at $q=4$ are of course not surprising
 in view of the multiplicative logarithmic correction $(\log L)^{-1/8}$
 and the additive logarithmic corrections 
 $\log\log L/\log L$, $1/\log L$, etc.\ \cite{Salas_Sokal_Potts4}.
 Indeed, what is somewhat surprising is that the systematic errors
 are not larger!
 See Section~\ref{subsec.CH} below for the case of the specific heat.)

\begin{figure}
\hspace*{-1.5cm}
\begin{tabular}{c@{\qquad}c}
\includegraphics[width=240pt]{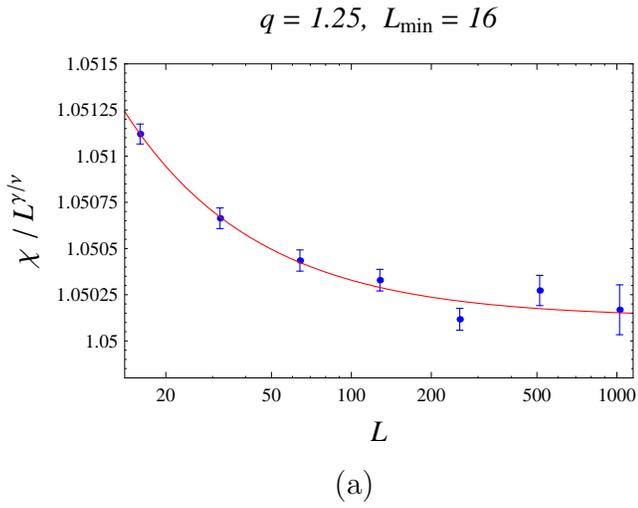} &
\includegraphics[width=240pt]{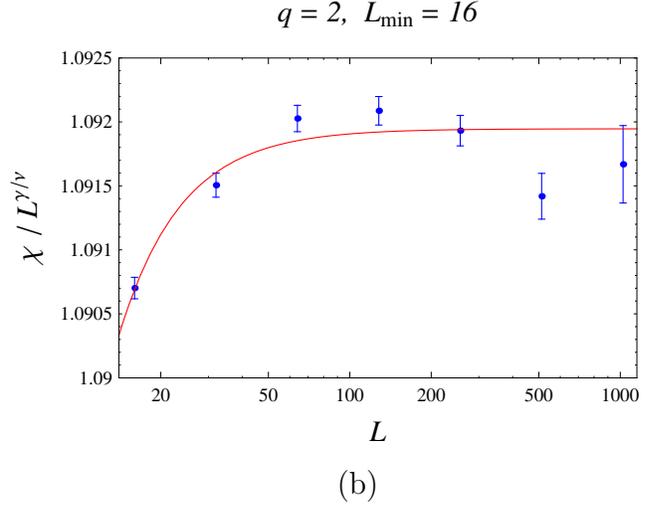} \\[1mm]
\phantom{(((a)}(a) & \phantom{(((a)}(b) \\[12mm]
\includegraphics[width=240pt]{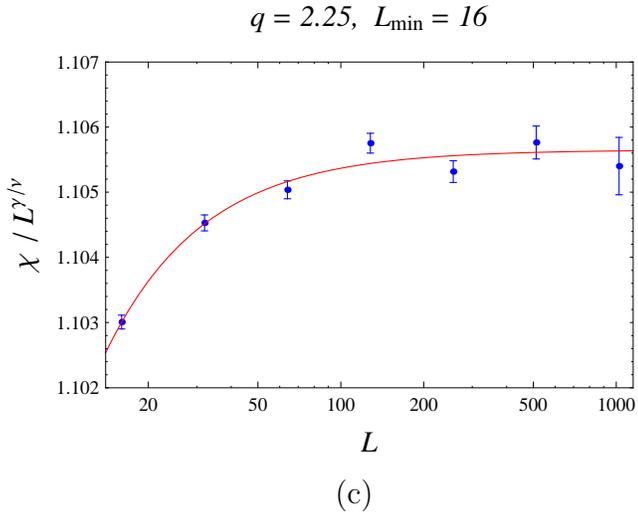} &
\includegraphics[width=240pt]{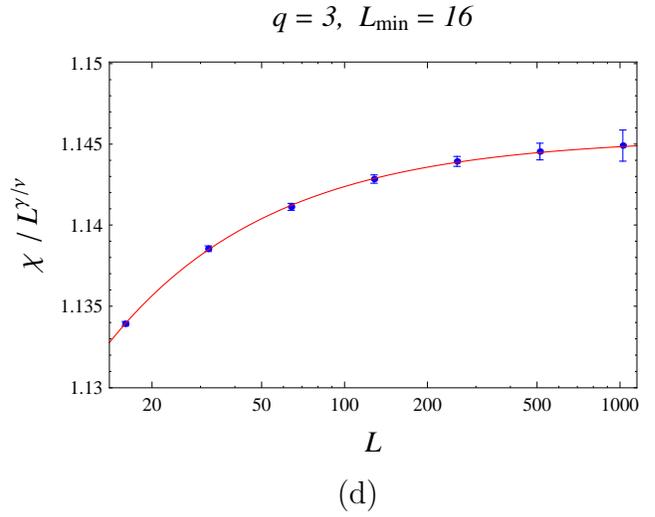} \\[1mm]
\phantom{(((a)}(c) & \phantom{(((a)}(d) \\[8mm]
\end{tabular}
\caption{
   Fits of $\chi/L^{\gamma/\nu}$ to $A + B L^{-p}$.
   (a) $q=1.25$, (b) $q=2$, (c) $q=2.25$, (d) $q=3$.
}
\label{fig_chi_plots}
\end{figure}

We next imposed the known exact value \reff{eq.gammaovernu.exact}
of $\gamma/\nu$ and attempted to extract the corrections to scaling.
%
Table~\ref{table_chi_analysis_corrections} shows
fits of $\chi/L^{\gamma/\nu}$
to the ans\"atze $A$ and $A + B L^{-p}$.
Plots for some selected values of $q$ are shown in Figure~\ref{fig_chi_plots}.
For $q=1.25$ and 1.5,
the correction to scaling at small $L$ is clearly positive in sign,
though very small in absolute magnitude (of order 0.001 at $L=16$):
see Figure~\ref{fig_chi_plots}(a).
This suggests that the correction-to-scaling exponent is fairly large,
in agreement with the theoretical prediction.
However, the fits produce unusually small estimates for the correction
exponent $p$;
this is a possible behavior of the ``effective exponent''
when there are two correction-to-scaling terms of opposite sign
(e.g.\ $L^{-1}$ and $L^{-\Delta_1}$).
For $q=1.75$ and 2,
the corrections to scaling are small and erratic
and it was not possible to fit them to a single inverse power;
indeed, for $q=2$ the behavior may be nonmonotonic in $L$
[see Figure~\ref{fig_chi_plots}(b)],
suggesting again the presence of
two correction-to-scaling terms of opposite sign.
For $q \ge 2.25$
the correction to scaling is clearly negative in sign,
and it gets larger in absolute magnitude as $q$ grows
(evaluated at $L=16$,
 the magnitude of the correction grows from
 $\approx 0.003$ at $q=2.25$ to $\approx 0.018$ at $q=3.75$);
furthermore, the fits produce estimates for the correction exponent $p$
that are in decent agreement with the exact value of $\Delta_1$.
At $q=4$ the correction to scaling is extremely strong (as expected)
and it is not possible to fit it to an inverse power.

In summary, for $q \ge 2.25$ we obtain modest evidence
that a correction $L^{-\Delta_1}$ is present, with negative amplitude.
For $q \le 2$ we are unable to say much about the corrections to scaling
except that the amplitude is probably positive for $q \le 1.5$
(we say ``probably'' because it is not clear that the positive contribution
 at small $L$ comes from the {\em same}\/ correction-to-scaling term
 that is dominant at large $L$).
But our data are at least compatible with the scenario \cite{Blote_88}
that the correction-to-scaling amplitude passes through zero at $q=2$.

\subsection{The largest cluster $C_1$}   \label{subsec.C1}

In Table~\ref{table_C1_analysis}
we report fits of the mean size of the largest cluster,
$C_1 = \langle \mathcal{C}_1 \rangle$,
to the pure-power-law ansatz $AL^{d-\beta/\nu}$,
just to see how accurately we would be able to estimate the
exponent $\beta/\nu$ if we did not know its exact value.
The last two columns show the deviation of the estimated $\beta/\nu$
from the known exact value \reff{eq.betaovernu.exact},
in absolute terms and in units of its standard deviation.

The results are extremely similar to those obtained for the susceptibility.
For small $q$,
both the statistical and the systematic errors are of order 0.0001 or less.
As $q$ grows, the statistical errors grow and the systematic errors
grow even more.
But the results are remarkably good except at $q=4$.

\subsection{The specific heat $C_H$}   \label{subsec.CH}

In Table~\ref{table_C_H_analysis}
we report fits of the specific heat $C_H$
to the ans\"atze $AL^{\alpha/\nu}$ and $AL^{\alpha/\nu} + B$.
The exact value \reff{eq.alphaovernu.exact} of $\alpha/\nu$
is shown for comparison in the next-to-last column.
The last two columns show the deviation of the estimated $\alpha/\nu$
(taken from the fit $AL^{\alpha/\nu} + B$) from the known exact value,
in absolute terms and in units of its standard deviation.

For $q \ltapprox 2.75$ the fits to a pure power law are horrible.
This is not surprising: after all,
for $q < 2$ the specific heat increases to a finite value as $L \to\infty$,
so the fit will indicate a positive value of $\alpha/\nu$
(but with a ridiculously poor goodness of fit)
while the true value of $\alpha/\nu$ is actually negative.
And for $2 < q \ltapprox 2.75$ the additive constant is very important,
so a pure power law has poor goodness of fit.

Good fits to $AL^{\alpha/\nu} + B$ can be obtained for all values of $q$
other than $q=2$ by appropriate choice of $\lmin$.
For $q=2$ the fit fails to converge:
initial guesses with $A > 0$, $B < 0$ and $\alpha/\nu$ slightly positive
always get driven to $\alpha/\nu \downarrow 0$
with $A \approx -B \uparrow +\infty$.
The fit is therefore suggesting the correct behavior
$C_H \approx A \log L + B$.
Indeed, a fit to this ansatz is good already for $\lmin = 64$ and yields
\begin{equation}
   A \;=\; 0.6371(9), \; B \;=\; 0.1732(42)
   \qquad
   (\chi^2 = 1.49, \hbox{3 DF}, \hbox{CL = 68\%})
   \;,
\end{equation}
in good agreement with the known exact values
$A = 2/\pi \approx 0.6366$ and $B \approx 0.1778$
\cite{Ferdinand_69,Salas_01}.\footnote{
   These values of $A$ and $B$ can be extracted from
   \cite{Ferdinand_69} or \cite{Salas_01}
   after translating their conventions to ours.
   We use the definition $C_H^{(2)}$ of specific heat [cf.\ \reff{CH2}].
   Salas \cite{Salas_01} uses the definition $C_H^{(3)}$ [cf.\ \reff{CH3}],
   which is 4 times ours when $q=2$.
   Ferdinand and Fisher use $C_H^{(3)}$ multiplied by $K^2$,
   where $K$ is the nearest-neighbor coupling in the Ising normalization;
   the critical point is given by $K_c = {1 \over 2} \log(1 + \sqrt{2})$.
}
If we fit instead to $A \log^2 L + B \log L + C$, we get a good fit
for $\lmin = 32$,
\begin{equation}
   A \;=\; -0.0012(6), \;
   B \;=\; 0.6493(55), \; C \;=\; 0.1415(125)
   \qquad
   (\chi^2 = 2.36, \hbox{3 DF}, \hbox{CL = 50\%})
   \;,
\end{equation}
in which the extremely small value of $A$
correctly suggests that the $\log^2 L$ term is absent.

Although the fits to $AL^{\alpha/\nu} + B$ are good for all $q \neq 2$,
some of the estimated values of $\alpha/\nu$ deviate significantly
from the known exact value
(see the last two columns of Table~\ref{table_C_H_analysis}).
For $q < 2$ the estimates of $\alpha/\nu$ are surprisingly good,
given that we are estimating a subleading singular contribution
underneath a nonsingular background.
For $2.25 \le q \le 2.75$ the estimates of $\alpha/\nu$ are also excellent.
However, for $q \ge 3$ the estimates of $\alpha/\nu$ 
deviate by more than three standard deviations from the exact value.
This may be due to the effect of corrections to scaling:
in particular, those governed by the exponent $\Delta_1$,
whose known exact value is given by \reff{eq.Delta1.exact}.
Indeed, for $q > 2 + \sqrt{2} \approx 3.414$ ($g > 7/2$)
we have $\Delta_1 < \alpha/\nu$,
so that the correction-to-scaling contribution $L^{\alpha/\nu - \Delta_1}$
is larger than the nonsingular background.
For $3 \le q \le 3.75$ we therefore tried fits to the ans\"atze
$AL^{\alpha/\nu} + BL^{\alpha/\nu - \Delta_1}$ and
$AL^{\alpha/\nu} + BL^{\alpha/\nu - \Delta_1} + C$,
in which $\Delta_1$ is fixed at its known exact value
and $\alpha/\nu$ is free:

(i)
For $q=3$ it is silly not to include the constant background,
since $\alpha/\nu - \Delta_1 = -0.4 \ll 0$;
nevertheless, we obtain a good fit to
$AL^{\alpha/\nu} + BL^{\alpha/\nu - \Delta_1}$ when $\lmin = 64$:
\begin{equation}
   \alpha/\nu \;=\; 0.4152(15), \;
   A \;=\; 1.397(14), \; B \;=\; -2.346(153)
   \qquad
   (\chi^2 = 0.75, \hbox{2 DF}, \hbox{CL = 69\%})
   \;.
\end{equation}
However, the estimate for $\alpha/\nu$ is 10 standard deviations away from
the correct value!
If we include the constant background
we get a good fit already when $\lmin = 16$,
\begin{eqnarray}
   & &
   \alpha/\nu \;=\; 0.4036(34) \;
   A \;=\; 1.555(44),  \; B \;=\; -0.787(188),  \; C \;=\; -0.795(152)
   \qquad  \nonumber \\
   & & \hspace*{7cm}
   (\chi^2 = 1.07, \hbox{3 DF}, \hbox{CL = 78\%})
   \;,
\end{eqnarray}
and the estimate for $\alpha/\nu$ is in excellent agreement with
the correct answer.

(ii)
For $q=3.25$ the correction-to-scaling contribution is only slightly smaller
than the nonsingular background, and it will probably be difficult for the fit
to separate the two.
The fit to $AL^{\alpha/\nu} + BL^{\alpha/\nu - \Delta_1}$
is good already when $\lmin = 16$,
\begin{equation}
   \alpha/\nu \;=\; 0.4972(7) \;
   A \;=\; 1.201(5) \; B \;=\; -0.945(19)
   \qquad
   (\chi^2 = 2.58, \hbox{4 DF}, \hbox{CL = 63\%})
   \;,
\end{equation}
but the estimate for $\alpha/\nu$ is almost six standard deviations away from
the correct value $\approx 0.5013$.
If we include the constant background
we also get a good fit when $\lmin = 16$,
\begin{eqnarray}
   & &
   \alpha/\nu \;=\; 0.4988(44), \;
   A \;=\; 1.186(42),  \; B \;=\; -1.089(396),  \; C \;=\; 0.133(365)
   \qquad  \nonumber \\
   & & \hspace*{7cm}
   (\chi^2 = 2.45, \hbox{3 DF}, \hbox{CL = 49\%})
   \;,
\end{eqnarray}
but the estimate for the constant $C$ is consistent with zero.
The estimate for $\alpha/\nu$ is now less than one standard deviation
away from the correct value, but this is principally because
the standard deviation has become much larger, not because the estimated
value has actually moved much closer to the true value!

(iii)
For $q=3.5$ the correction-to-scaling contribution is now slightly larger
than the nonsingular background, and it will again likely be difficult
for the fit to separate the two.
The fit to $AL^{\alpha/\nu} + BL^{\alpha/\nu - \Delta_1}$
is good when $\lmin = 64$,
\begin{equation}
   \alpha/\nu \;=\; 0.5945(39), \;
   A \;=\; 0.926(25), \; B \;=\; -0.017(91)
   \qquad
   (\chi^2 = 0.91, \hbox{2 DF}, \hbox{CL = 64\%})
   \;,
\end{equation}
but the estimate for the correction-to-scaling amplitude $B$
is consistent with zero.
Indeed, the result from this fit is virtually identical
to what was obtained from the ansatz $AL^{\alpha/\nu} + B$,
and the estimate for $\alpha/\nu$ is again about four standard deviations
away from the correct value $\approx 0.6101$.
By contrast, the fit to $AL^{\alpha/\nu} + BL^{\alpha/\nu - \Delta_1} + C$
is good already when $\lmin = 16$,
\begin{eqnarray}
   & &
   \alpha/\nu \;=\; 0.6095(67), \;
   A \;=\; 0.805(43),  \; B \;=\; 2.961(368),  \; C \;=\; -3.509(390)
   \qquad  \nonumber \\
   & & \hspace*{7cm}
   (\chi^2 = 2.32, \hbox{3 DF}, \hbox{CL = 51\%})
   \;,
\end{eqnarray}
and the estimate for $\alpha/\nu$ is now in excellent agreement with
the correct answer.
Interestingly, the estimates for the amplitudes $B$ and $C$
are {\em not}\/ consistent with zero;
rather, they are strongly nonzero but of opposite signs.
Clearly, what happened is that when we performed a fit with a single
correction term (whether $B$ or $BL^{\alpha/\nu - \Delta_1}$,
which are anyway nearly the same) this {\em pair}\/
of correction terms combined to make an ``effective'' correction term
(in the given range of $L$) with a nearly zero amplitude;
but this gave a {\em biased}\/ estimate of the leading exponent $\alpha/\nu$.
What is slightly surprising is that our fit was able to separate
the correction-to-scaling contribution $B L^{\approx 0.090}$
from the nonsingular background $C$.
Perhaps the stunning agreement of the estimated value of $\alpha/\nu$
with the exact answer is a fluke and ought not be taken too seriously.

(iv)
For $q=3.75$ the correction-to-scaling contribution is significantly larger
than the nonsingular background.
The fit to $AL^{\alpha/\nu} + BL^{\alpha/\nu - \Delta_1}$
is good when $\lmin = 64$,
\begin{equation}
   \alpha/\nu \;=\; 0.7181(90), \;
   A \;=\; 0.568(42), \; B \;=\; 0.413(76)
   \qquad
   (\chi^2 = 1.26, \hbox{2 DF}, \hbox{CL = 53\%})
   \;,
\end{equation}
and the estimate for $\alpha/\nu$ is only about two standard deviations
away from the true value $\approx 0.7376$.
By contrast, the fit to $AL^{\alpha/\nu} + BL^{\alpha/\nu - \Delta_1} + C$
is good already when $\lmin = 16$:
\begin{eqnarray}
   & &
   \alpha/\nu \;=\; 0.7856(237), \;
   A \;=\; 0.288(68),  \; B \;=\; 1.085(64),  \; C \;=\; -1.046(37)
   \qquad  \nonumber \\
   & & \hspace*{7cm}
   (\chi^2 = 0.30, \hbox{3 DF}, \hbox{CL = 96\%})
   \;.
\end{eqnarray}
Here the estimated amplitudes $B$ and $C$
have opposite signs and are apparently nonzero;
but the estimate for $\alpha/\nu$ has now far overshot the correct value
(it is again two standard deviations away, but with a much larger
standard deviation).
The poor performance of this two-correction-term fit
--- in a case where the two terms $B L^{\approx 0.388}$ and $C$ 
should have been much {\em easier}\/ to separate than they were for $q=3.5$ ---
suggests that the good result obtained for $q=3.5$ was indeed a fluke
and that the mediocre result obtained for $q=3.75$ is what should
ordinarily be expected.

Finally, for $q = 4$ the true leading behavior is known
\cite{Salas_Sokal_Potts4} to be $L (\log L)^{-3/2}$,
but with corrections to scaling down by
$\log\log L/\log L$, $1/\log L$, etc.
It is clearly hopeless to try to fit to such an ansatz
unless one has data for colossally large values of $L$.
It is of course not surprising that fits to
$AL^{\alpha/\nu}$ or $AL^{\alpha/\nu} + B$
gave estimates of $\alpha/\nu$ near 0.8,
far off from the correct value 1;
the factor $(\log L)^{3/2}$ is imitating a power $L^{\approx 0.2}$
in our range of $L$.

\section{Data analysis: Dynamic quantities}
\label{sec.data_dynamic}

In this section we analyze the dynamic data by the same general methods
as were used in the preceding section to analyze the static data.
Our main goal is to estimate the dynamic critical exponents
$z_{\text{int},\mathcal{O}}$ associated to the
integrated autocorrelation times $\tau_{\text{int},\mathcal{O}}$ for
various observables $\mathcal{O}$.

We proceed as follows:
First we discuss the dependence of the autocorrelation times
on the number $k$ of active colors (Section~\ref{subsec.dependence_on_k}),
and we give an overview of the qualitative behavior
of the autocorrelation times for different observables
(Section~\ref{subsec.qualitative}).
Then we present a detailed analysis of the dynamic critical exponent
$z_{\text{int},\E}$ (Section~\ref{subsec.zintE});
in particular we discuss the sharpness of the Li--Sokal bound
(Section~\ref{subsec.li-sokal}) and the correctness of the
Ossola--Sokal conjecture (Section~\ref{subsec.ossola-sokal}).
Finally, we analyze briefly the dynamic critical exponent
$z_{\text{int},\mathcal{O}}$ for other observables $\mathcal{O}$
(Section~\ref{subsec.otherO}).

\subsection{Dependence on $k$}  \label{subsec.dependence_on_k}

We began by analyzing the dependence of $\tau_{{\rm int},\mathcal{O}}$
on the number $k$ of active colors.
Of course we expect that all values of $k$ lie in the same
dynamic universality class:
that is, we expect that the ratios of $\tau_{{\rm int},\mathcal{O}}$
for different $k$ tend to nonzero finite constants as $L \to\infty$.
Moreover, it is intuitively reasonable to think that
an update with $k$ active colors
does roughly ``$k$ times as much work''
as an update with one active color;
therefore, we expect that $\tau_{{\rm int},\mathcal{O}}$
should be roughly proportional to $1/k$.

We tested these expectations by analyzing the ratios
$\tau_{{\rm int},\mathcal{O}}(1) / \tau_{{\rm int},\mathcal{O}}(k)$
as a function of $L$
for each $(q,\mathcal{O})$ and each allowable $k$.
In all cases the ratios are fairly close to $k$, as expected;
but in general they are not exactly equal to from $k$.
The ratios also show some dependence on $L$,
but tend as $L \to\infty$ to a limiting value, again as expected.
Roughly speaking, for the smaller values of $q$
the $L$-dependence is fairly strong,
and the ratios are comparatively far from $k$
(which is perhaps not surprising because the values of
 $\tau_{{\rm int},\mathcal{O}}$ are themselves quite small);
the limiting values also appear to be different from $k$,
though this conclusion is only tentative because of the
strong corrections to scaling.
For the larger values of $q$
the $L$-dependence is weaker,
and the ratios are closer to $k$;
in particular, the limiting values are compatible with $k$
within our statistical errors.
In Table~\ref{table_kratios_examples}
we show typical examples of these two behaviors,
namely $\mathcal{O} = \mathcal{E}'$ with $q=2$ and $q=3.25$.
In Table~\ref{table_kratios_fits}
we show our best estimates for the limiting ratios
$\tau_{{\rm int},\mathcal{E}'}(1) / \tau_{{\rm int},\mathcal{E}'}(k)$,
obtained by fitting the ratio to a constant
and increasing $L_{\rm min}$ until a decent fit is obtained.
The behavior for the other observables is qualitatively similar.

Having confirmed that all values of $k$ lie in the same dynamic
universality class, we henceforth analyze the data for each value of $k$
separately and then compute a weighted average of the resulting
exponent estimates.

\subsection{Summary of qualitative behavior}  \label{subsec.qualitative}

Let us begin by summarizing the qualitative behavior of
$\tau_{\text{int},\mathcal{O}}$ for different observables $\mathcal{O}$:

1) For nearly every triplet $(q,k,L)$, we find that $\E$ is the
observable (of those we have measured)
that has the largest $\tau_{\text{int}}$.
The only exceptions are $q=4$, $L=16$, $1 \le k \le 4$
(for which $\tau_{\text{int},\mathcal{S}_2}$ is slightly larger than
$\tau_{\text{int},\E}$)
and $q=4$, $L=1024$, $k=1$
(for which $\tau_{\text{int},\mathcal{C}_1}$ is slightly larger than
$\tau_{\text{int},\E}$).
But these differences are extremely small
and may well represent statistical fluctuations.

2) For every triplet $(q,k,L)$, we find that $\mathcal{C}_2$ is the
observable (of those we have measured)
that has the smallest $\tau_{\text{int}}$.

3) For every triplet $(q,k,L)$ we find that
\begin{equation}
   \tau_{\text{int},\E}  \;>\;  \tau_{\text{int},\N}
   \;.
\end{equation}
When $q$ is an integer and $k=q$,
this inequality is easily proved rigorously for the SW algorithm
\cite{Salas_Sokal_Potts3},
so it is not surprising that it holds here for CM.
However, we do not yet have a rigorous proof
(not even in the case when $q$ is an integer and $k < q$).

4) In general, the observables $\obs$ that we have measured
fall into four groups according to their integrated autocorrelation times
$\tau_{\text{int},\mathcal{O}}$:
\begin{enumerate}
  \item $\E$, $\N$, $\mathcal{S}_2$ and $\mathcal{C}_1$
have the largest values of $\tau_{\text{int}}$,
and they are all fairly close to each other
(all are at least $\approx 0.75$ times that of $\E$,
 and usually much closer);
  \item $\mathcal{F}'$ has an intermediate value of $\tau_{\text{int}}$,
of order 0.5--0.9 times that of $\E$;
  \item $\mathcal{C}_3$ has a slightly lower value of $\tau_{\text{int}}$,
of order 0.4--0.8 times that of $\E$;
  \item $\mathcal{C}_2$ has the smallest $\tau_{\text{int}}$,
of order 0.35--0.8 times that of $\E$.
\end{enumerate}
Indeed, for all triplets $(q,k,L)$ we have
$\tau_{\text{int},\mathcal{O}_1} > \tau_{\text{int},\mathcal{O}_2}
   > \tau_{\text{int},\mathcal{O}_3} > \tau_{\text{int},\mathcal{O}_4}$
whenever $\mathcal{O}_1, \mathcal{O}_2, \mathcal{O}_3, \mathcal{O}_4$
belong to groups 1,2,3,4,
with the exception that for a few triplets at $q=4$, $L \ge 256$
we have $\tau_{\text{int},\mathcal{F}'} < \tau_{\text{int},\mathcal{C}_3}$.


\bigskip

\begin{figure}
\centering
\includegraphics[width=0.9\textwidth]{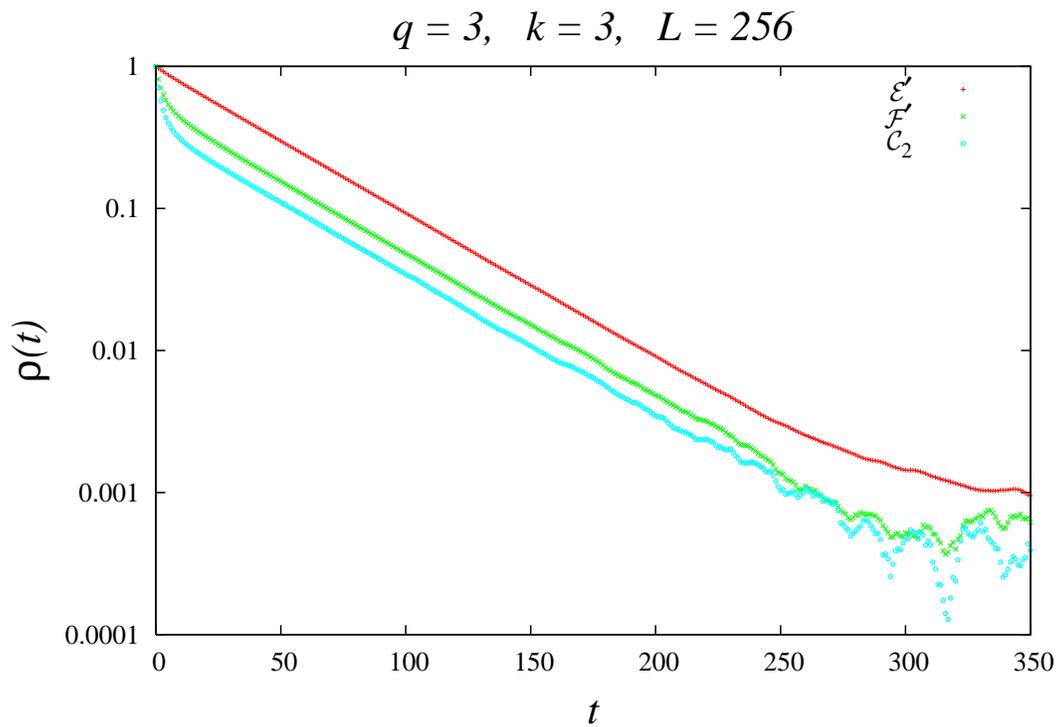}
\caption{
   Normalized autocorrelation function $\rho_{\obs \obs}(t)$
   for the observables $\obs = \E$, $\F$ and $\mathcal{C}_2$,
   in the case $q=3$, $k=3$, $L=256$.
}
   \label{fig_rho_plot}
\end{figure}

These behaviors can be better understood by looking at the
normalized autocorrelation functions $\rho_{\obs \obs}(t)$.
A typical example is shown in Figure~\ref{fig_rho_plot}.
We see that $\rho_{\E \E}(t)$ is nearly a pure exponential,
so that $\tau_{\text{int},\E} \approx \tau_{\text{exp}}$.
By contrast, the autocorrelation functions
for the observables $\obs$ in groups 2, 3 and 4
exhibit an initial fast decay, followed by a decay at the same
exponential rate $\tau_{\rm exp}$
but with an amplitude $A_{\obs}$ that is significantly less than 1.
What we do not know is whether $A_{\obs}$ tends to a nonzero value
as $L \to\infty$
(in which case we will have $z_{\text{int},\obs} = z_{\text{exp}}$)
or tends to zero as an inverse power of $L$
(in which case we will have $z_{\text{int},\obs} < z_{\text{exp}}$).
See Section~\ref{subsec.otherO} for further analysis of this question;
and see \cite[Section~5.2]{OssolaSokal04} for a more detailed analysis
in the case of the Swendsen--Wang dynamics
for the three-dimensional Ising model.

\bigskip

In the following subsections
we shall fit $\tau_{\text{int},\mathcal{O}}$,
for each observable $\mathcal{O}$, to a variety of ans\"atze, notably:
\begin{itemize}
   \item {\em Fits for $z=0$:}  $\tau_{\text{int},\mathcal{O}} = A$
       or $\tau_{\text{int},\mathcal{O}} = A + B L^{-p}$.
   \item {\em Fits for $z=0$ with a multiplicative logarithm:}
       $\tau_{\text{int},\mathcal{O}} = A \log L + B$ or
       $\tau_{\text{int},\mathcal{O}} = A \log^2 L + B \log L + C$.
   \item {\em Fits for $z>0$:}
       $\tau_{\text{int},\mathcal{O}} = AL^z$ or
       $\tau_{\text{int},\mathcal{O}} = AL^z + B$.
\end{itemize}
Note that the fits to $A + B L^{-p}$ and $AL^z + B$
are in fact the {\em same}\/ fit in different notation!

We shall begin (Sections~\ref{subsec.zintE}--\ref{subsec.ossola-sokal})
by focusing on the observable $\mathcal{E}'$,
which has the largest autocorrelation time
of all the observables we measured.
Then (Section~\ref{subsec.otherO})
we shall discuss, more briefly, the other observables,
with emphasis on $\mathcal{C}_2$.

\subsection{Dynamic critical exponent $z_{\text{int},\mathcal{E}'}$}
   \label{subsec.zintE}

In this subsection we fit the integrated autocorrelation time
$\tau_{\text{int},\E}$ to a variety of ans\"atze
in an effort to estimate the dynamic critical exponent $z_{\text{int},\E}$.
We begin by presenting our fits, in order of increasing $q$,
without much comment.
Then we go back and try to interpret what these fits
might be telling us about the dynamic critical behavior
of the Chayes--Machta algorithm as a function of $q$.

For $q=1.25$ the behavior is fairly clear:
$\tau_{\text{int},\mathcal{E}'}$ tends to a finite constant as $L \to\infty$.
The fits to the ansatz $\tau_{\text{int},\mathcal{E}'} = A$ are horrible
(except of course the fit $\lmin = 1024$ that has zero degrees of freedom);
but if we fit to $\tau_{\text{int},\mathcal{E}'} = A + B L^{-p}$
we get a decent fit for $\lmin = 128$:
\begin{equation}
   A \;=\; 2.19(8), \; B \;=\; -1.61(5), \; p \;=\; 0.213(35)
   \qquad
   (\chi^2 = 0.49, \hbox{1 DF}, \hbox{CL = 48\%})
   \;.
 \label{eq.fit_q=1.25}
\end{equation}
Finally, the fit to $AL^z$ yields an estimated exponent $z \approx 0.05$
but has poor goodness of fit even when $\lmin = 256$:
$z = 0.053(1)$, $A = 1.27(1)$ with $\chi^2 = 3.76$, 1 DF, CL = 5\%.

For $q=1.5$ the fits to a constant $A$ are again horrible;
but if we fit to $A + B L^{-p}$ we get a good fit already for $\lmin = 32$:
\begin{equation}
   A \;=\; 17.09(2.96), \; B \;=\; -15.86(2.92), \; p \;=\; 0.034(7)
   \qquad
   (\chi^2 = 0.82, \hbox{3 DF}, \hbox{CL = 85\%})
   \;.
 \label{eq.fit_q=1.5}
\end{equation}
We also tried a fit to $A \log L + B$:
a decent fit is obtained for $\lmin = 128$, namely
\begin{equation}
   A \;=\; 0.440(4), \; B \;=\; 1.486(21)
   \qquad
   (\chi^2 = 0.83, \hbox{2 DF}, \hbox{CL = 66\%})
   \;.
\end{equation}
A good fit to $A \log^2 L + B \log L + C$ is obtained already for $\lmin = 32$,
but with a slightly {\em negative}\/ value of $A$
(which is of course impossible for the actual asymptotics):
$A = -0.008(2)$, $B = 0.53(2)$, $C = 1.24(4)$
with $\chi^2 = 0.80$, 3 DF, CL = 85\%.
Finally, the fit to $AL^z$ yields an estimated exponent $z \approx 0.10$
but again has poor goodness of fit even when $\lmin = 256$:
$z = 0.104(2)$, $A = 2.20(2)$ with $\chi^2 = 3.84$, 1 DF, CL = 5\%.

For $q=1.75$ (and all larger $q$) the fit to a constant $A$ is again horrible.
The fit to $A \log L + B$ is mediocre even when $\lmin = 256$:
$A = 1.36(2)$, $B=0.32(12)$ with $\chi^2 = 2.73$, 1 DF, CL = 10\%.
A decent fit to $A \log^2 L + B \log L + C$
is obtained already for $\lmin = 16$:
$A = 0.037(3)$, $B = 0.915(23)$, $C = 1.689(45)$
with $\chi^2 = 4.14$, 4 DF, CL = 39\%.
The fit to $AL^z$ is good when $\lmin = 256$:
$z = 0.158(2)$, $A = 3.28(5)$ with $\chi^2 = 0.21$, 1 DF, CL = 65\%.
The fit to $AL^z + B$ is decent already for $\lmin = 16$
($\chi^2 = 4.63$, 4 DF, CL = 33\%),
but the $\chi^2$ drops notably
from $\lmin = 64$ ($\chi^2 = 2.86$) to $\lmin = 128$ ($\chi^2 = 0.33$),
so that our preferred fit is $\lmin = 128$:
\begin{equation}
   z \;=\; 0.085(26), \; A \;=\; 9.66(4.42), \; B \;=\; -7.56(4.82)
   \qquad
   (\chi^2 = 0.33, \hbox{1 DF}, \hbox{CL = 57\%})
   \;.
\end{equation}

For $q \ge 2$ we have data from more than one value of $k$.
For simplicity we discuss here in words only the case $k=1$;
the other cases are qualitatively similar and are reported in
Tables~\ref{table_fits_ALz} and \ref{table_fits_ALz+B}.

For $q=2$ the fit to $A \log L + B$ is poor (CL $< 2\%$)
even when $\lmin = 256$.
The fit to $A \log^2 L + B \log L + C$ is good already for $\lmin = 16$:
$A = 0.206(6)$, $B = 1.046(53)$, $C = 2.572(103)$
with $\chi^2 = 1.40$, 4 DF, CL = 84\%.
The fit to $AL^z$ is good for $\lmin = 256$:
$z = 0.215(3)$, $A=4.47(9)$ with $\chi^2 = 0.06$, 1 DF, CL = 81\%.
Finally, the fit to $AL^z + B$ is good already for $\lmin = 16$:
\begin{equation}
   z \;=\; 0.145(4), \; A \;=\; 10.4(5), \; B \;=\; -8.4(6)
   \qquad
   (\chi^2 = 2.32, \hbox{4 DF}, \hbox{CL = 68\%})
   \;.
\end{equation}

For $q=2.25$ the fit to $A \log L + B$ is again poor (CL $< 5\%$)
even when $\lmin = 256$.
The fit to $A \log^2 L + B \log L + C$ is decent already for $\lmin = 32$:
$A = 0.77(3)$, $B=-0.84(24)$, $C = 7.07(53)$
with $\chi^2 = 3.15$, 3 DF, CL = 37\%.
The fit to $AL^z$ yields an estimated exponent $z \approx 0.29$
but has mediocre goodness of fit even when $\lmin = 256$:
$z = 0.286(4)$, $A = 5.35(14)$ with $\chi^2 = 2.16$, 1 DF, CL = 14\%.
Finally, the fit to $AL^z + B$ is good already for $\lmin = 32$:
\begin{equation}
   z \;=\; 0.235(8), \; A \;=\; 8.91(61), \; B \;=\; -6.68(86)
   \qquad
   (\chi^2 = 1.62, \hbox{3 DF}, \hbox{CL = 66\%})
   \;.
\end{equation}

For $q=2.5$ (and all larger $q$) the fit to $A \log L + B$ is poor
(CL $< 3\%$) even when $\lmin = 256$.
The fit to $A \log^2 L + B \log L + C$ is decent for $\lmin = 128$:
$A = 2.7(3)$, $B=-13.7(3.1)$, $C=37.4(8.6)$
with $\chi^2 = 0.66$, 1 DF, CL = 42\%.
However, the error bars on $B$ and $C$ are very large
in absolute magnitude (and in particular large compared to the value of $A$),
which renders the fit somewhat dubious.
The fit to $AL^z$ is good for $\lmin = 256$:
$z = 0.353(6)$, $A=6.35(23)$ with $\chi^2 = 0.02$, 1 DF, CL = 90\%.
Finally, the fit to $AL^z + B$ is excellent already for $\lmin = 32$:
\begin{equation}
   z \;=\; 0.315(8), \; A \;=\; 9.01(59), \; B \;=\; -6.75(1.00)
   \qquad
   (\chi^2 = 0.21, \hbox{3 DF}, \hbox{CL = 98\%})
   \;.
\end{equation}

For $q=2.75$ the fit to $A \log^2 L + B \log L + C$ is decent for $\lmin = 64$:
$A = 5.9(3)$, $B=-31.9(2.7)$, $C=71.0(6.5)$
with $\chi^2 = 2.56$, 2 DF, CL = 28\%.
But the error bars on $B$ and $C$ are again quite large.
The fit to $AL^z$ is good for $\lmin = 256$:
$z = 0.424(8)$, $A=7.19(34)$ with $\chi^2 = 0.07$, 1 DF, CL = 80\%.
Finally, the fit to $AL^z + B$ is excellent already for $\lmin = 16$:
\begin{equation}
   z \;=\; 0.411(5), \; A \;=\; 8.19(30), \; B \;=\; -4.76(57)
   \qquad
   (\chi^2 = 1.06, \hbox{4 DF}, \hbox{CL = 90\%})
   \;.
\end{equation}

For $q=3$
the fit to $A \log^2 L + B \log L + C$ is decent for $\lmin = 128$,
but with huge error bars:
$A = 16.0(1.5)$, $B=-112.6(16.5)$, $C=257.5(44.5)$
with $\chi^2 = 0.77$, 1 DF, CL = 38\%.
(The same behavior persists for all larger values of $q$,
 with even larger values of the coefficients $A,B,C$ and their error bars;
 we refrain from reporting the gory results.)
The fit to $AL^z$ is good for $\lmin = 128$:
$z = 0.505(6)$, $A=7.52(23)$ with $\chi^2 = 1.18$, 2 DF, CL = 56\%.
Finally, the fit to $AL^z + B$ is excellent already for $\lmin = 32$:
\begin{equation}
   z \;=\; 0.481(10), \; A \;=\; 9.08(63), \; B \;=\; -6.79(1.68)
   \qquad
   (\chi^2 = 0.84, \hbox{3 DF}, \hbox{CL = 84\%})
   \;.
\end{equation}

For $q=3.25$
the fit to $AL^z$ is decent for $\lmin = 128$:
$z = 0.590(7)$, $A=7.66(30)$ with $\chi^2 = 3.14$, 2 DF, CL = 21\%.
Finally, the fit to $AL^z + B$ is decent for $\lmin = 64$:
\begin{equation}
   z \;=\; 0.558(21), \; A \;=\; 9.8(1.4), \; B \;=\; -11.9(6.0)
   \qquad
   (\chi^2 = 2.93, \hbox{2 DF}, \hbox{CL = 23\%})
   \;.
\end{equation}

For $q=3.5$
the fit to $AL^z$ is decent for $\lmin = 128$:
$z = 0.676(10)$, $A=7.77(39)$ with $\chi^2 = 1.07$, 2 DF, CL = 58\%.
Finally, the fit to $AL^z + B$ is decent for $\lmin = 64$:
\begin{equation}
   z \;=\; 0.648(25), \; A \;=\; 9.5(1.6), \; B \;=\; -14.4(8.9)
   \qquad
   (\chi^2 = 1.15, \hbox{2 DF}, \hbox{CL = 56\%})
   \;.
\end{equation}

For $q=3.75$
the fit to $AL^z$ is good already for $\lmin = 32$:
$z = 0.779(4)$, $A=7.14(13)$ with $\chi^2 = 0.94$, 4 DF, CL = 92\%.
Finally, the fit to $AL^z + B$ is excellent already for $\lmin = 16$:
\begin{equation}
   z \;=\; 0.790(9), \; A \;=\; 6.7(3), \; B \;=\; 3.0(1.5)
   \qquad
   (\chi^2 = 0.33, \hbox{4 DF}, \hbox{CL = 99\%})
   \;.
\end{equation}

For $q=4$
the fit to $AL^z$ is good for $\lmin = 128$:
$z = 0.916(16)$, $A=5.44(46)$ with $\chi^2 = 0.15$, 2 DF, CL = 93\%.
Finally, the fit to $AL^z + B$ is excellent already for $\lmin = 32$:
\begin{equation}
   z \;=\; 0.935(18), \; A \;=\; 4.8(5), \; B \;=\; 20.5(5.1)
   \qquad
   (\chi^2 = 0.66, \hbox{3 DF}, \hbox{CL = 88\%})
   \;.
\end{equation}

Let us now comment on what we think these fits show.

For $q=1.25$ it seems fairly clear that
$\tau_{\text{int},\mathcal{E}'}$ converges as $L \to\infty$
to a finite value, i.e.\ $z_{\text{int},\mathcal{E}'} = 0$.
For $q=1.5$ the behavior is unclear:
perhaps $\tau_{\text{int},\mathcal{E}'}$ converges to a finite value,
but with extremely strong corrections to scaling
(e.g.\ $A + BL^{-p}$ with $p > 0$ extremely small);
or perhaps it grows logarithmically, like $\log L$ or $\log^2 L$;
or perhaps it grows with a very small positive power ($z \ltapprox 0.1$)
together with very strong corrections to scaling.
The evidence points weakly towards the first scenario,
but a divergence like $\log L$ or an extremely small positive power of $L$
is also a possibility.
For $1.75 \le q \ltapprox 2.25$ a reasonable fit is obtained with the ansatz
$A \log^2 L + B \log L + C$; but it seems to us implausible on theoretical
grounds that we would have such a logarithmic growth
for an entire {\em interval}\/ of $q$.
Much more likely is that there exists {\em one}\/ value $q_\star$
such that $z=0$ for $q < q_\star$ and $z>0$ for $q > q_\star$,
in which case $\tau_{\text{int},\mathcal{E}'}$ might grow logarithmically
{\em at}\/ $q = q_\star$ (but only there).
Our data suggest that $q_\star$ lies between 1.25 and 1.75;
our best guess would be $\approx 1.6$,
based on linearly interpolating the exponent estimates
produced by our fits for $q=1.5$ and $q=1.75$.

\begin{figure}
\centering
\includegraphics[width=0.9\textwidth]{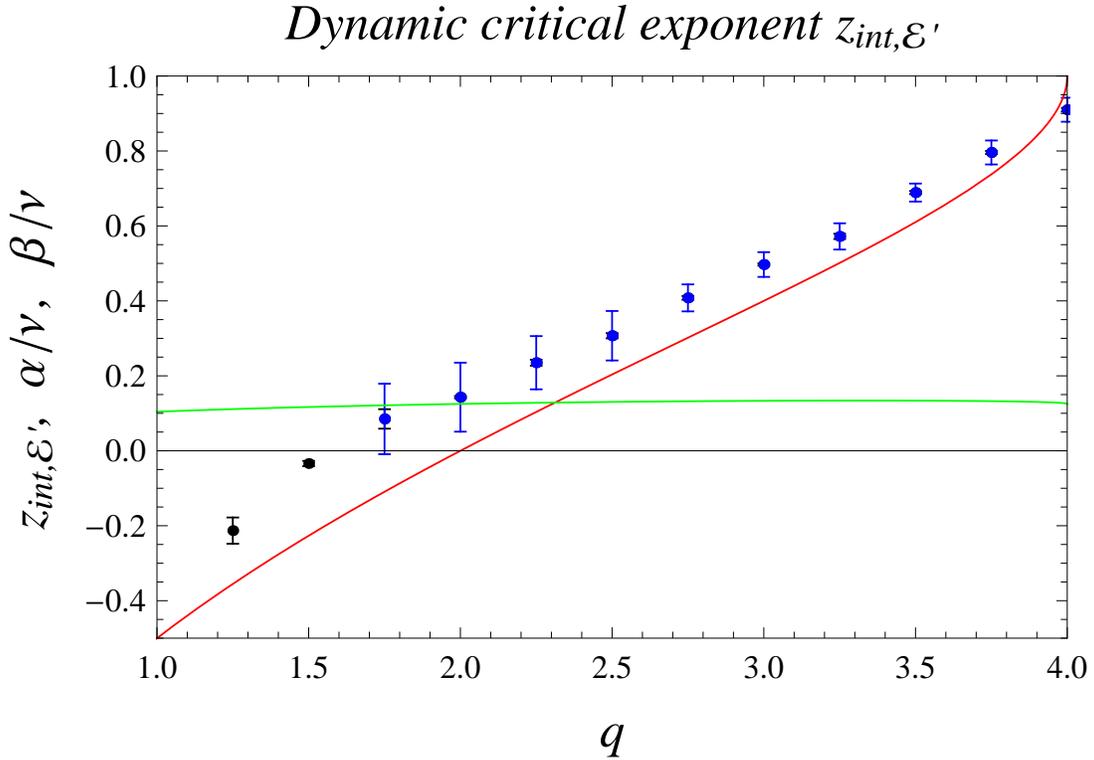}
\caption{
   Our best estimates of the dynamic critical exponent
   $z_{\text{int},\mathcal{E}'}$.
   The purely statistical error bar is indicated in black
   and is usually about the same size as the symbol;
   the combined statistical and systematic error bar is indicated in blue.
   (For $q=1.25,1.50$ we are unable to estimate the systematic errors.)
   The static exponents $\alpha/\nu$ and $\beta/\nu$ are shown for comparison
   (red and green curves, respectively).
}
   \label{fig_z_estimates}
\end{figure}

For all $q \ge 1.75$ we are able to obtain decent fits
to the ans\"atze $AL^z$ and $AL^z + B$ with an exponent $z>0$;
these fits are reported in
Tables~\ref{table_fits_ALz} and \ref{table_fits_ALz+B}, respectively,
and the results obtained by averaging over $k$
are reported in Table~\ref{table_fits_allk}.
For the smaller values of $q$ in this table,
the discrepancy between the fits with and without a constant term $B$
is fairly large:
this is not surprising because $z$ is fairly small
and hence the effect of the constant term is very strong.
As $q$ grows, the discrepancy between the two fits decreases:
from $\approx 0.1$ at $q=1.75$ and $\approx 0.07$ at $q=2$
to approximately zero at $q = 3.5$;
for $q > 3.5$ the discrepancy has the opposite sign but remains small.
Correspondingly, the estimated value of $B$
appears to go through zero (and change sign) at $q \approx 3.5$.
We have a slight preference for the fits to $AL^z + B$,
for the simple reason that such a constant term must surely be present,
if only because the definition of $\tau_{\text{int},\mathcal{O}}$
is somewhat arbitrary
(should one include the contribution ${1 \over 2}$ from $t=0$ or not?).
But our data are insufficient to resolve clearly the discrepancy
between the two fits.
We therefore choose to report our results for $q \ge 1.75$ in the form
\begin{equation}
   \hbox{best estimate $\,\pm\,$  statistical error $\,\pm\,$ systematic error}
   \;,
\end{equation}
where ``statistical error'' denotes the one-standard-deviation error bar
from the fit to ${AL^z + B}$
(after averaging over $k$ the fits from the chosen values of $\lmin$);
and ``systematic error'' is a 68\% subjective confidence interval
defined as the absolute value of the discrepancy between the
fits $AL^z$ and $AL^z + B$
(after averaging over $k$ the fits from the chosen values of $\lmin$)
plus 0.02.
The final results are therefore:
\begin{subeqnarray}
   q=1.75 \colon
     &  &  z_{\text{int},\mathcal{E}'} \;=\; 0.085 \pm 0.026 \pm 0.094  \\
   q=2.00 \colon
     &  &  z_{\text{int},\mathcal{E}'} \;=\; 0.143 \pm 0.003 \pm 0.092  \\
   q=2.25 \colon
     &  &  z_{\text{int},\mathcal{E}'} \;=\; 0.231 \pm 0.008 \pm 0.071  \\
   q=2.50 \colon
     &  &  z_{\text{int},\mathcal{E}'} \;=\; 0.307 \pm 0.007 \pm 0.066  \\
   q=2.75 \colon
     &  &  z_{\text{int},\mathcal{E}'} \;=\; 0.408 \pm 0.005 \pm 0.036  \\
   q=3.00 \colon
     &  &  z_{\text{int},\mathcal{E}'} \;=\; 0.497 \pm 0.003 \pm 0.033  \\
   q=3.25 \colon
     &  &  z_{\text{int},\mathcal{E}'} \;=\; 0.572 \pm 0.007 \pm 0.035  \\
   q=3.50 \colon
     &  &  z_{\text{int},\mathcal{E}'} \;=\; 0.689 \pm 0.004 \pm 0.024  \\
   q=3.75 \colon
     &  &  z_{\text{int},\mathcal{E}'} \;=\; 0.796 \pm 0.004 \pm 0.032  \\
   q=4.00 \colon
     &  &  z_{\text{int},\mathcal{E}'} \;=\; 0.910 \pm 0.005 \pm 0.032
 \label{eq.zint.final}
\end{subeqnarray}
In Figure~\ref{fig_z_estimates} we plot these estimates versus $q$,
and compare them with the static exponents $\alpha/\nu$ and $\beta/\nu$.
The Ossola--Sokal conjecture $z \ge \beta/\nu$ appears to be violated
for $1 \le q \ltapprox 1.95$
(but see Section~\ref{subsec.ossola-sokal} for an alternative fit
 that is compatible with the conjecture).
The Li--Sokal bound $z \ge \alpha/\nu$ is obeyed for $q \neq 4$
and appears to be non-sharp
(see Section~\ref{subsec.li-sokal} for a more detailed analysis).
The apparent violation of the Li--Sokal bound at $q=4$ is manifestly
due to the multiplicative logarithmic corrections:
it is known \cite{Salas_Sokal_Potts4} that $C_H \sim L (\log L)^{-3/2}$
but fits to a power law (with or without a constant background)
yield an effective exponent $\approx 0.8$
(see the last line of Table~\ref{table_C_H_analysis});
it is therefore not surprising that $\tau_{\text{int},\mathcal{E}'}$
shows a similar behavior.
If one looks directly at the ratio $\tau_{\text{int},\mathcal{E}'}/C_H$
one finds that the Li--Sokal bound (which is after all a rigorous theorem!)\ 
is obeyed (see Section~\ref{subsec.li-sokal}).

For $q=1.25$ and 1.50 our best estimates suggest that
$\tau_{\text{int},\mathcal{E}'}$ is bounded as $L \to\infty$,
i.e.\ that $z_{\text{int},\mathcal{E}'} = 0$.
Our fits to $A + B L^{-p}$ suggest values for the correction exponent $p$
[cf.\ \reff{eq.fit_q=1.25}/\reff{eq.fit_q=1.5}],
but we do not know how reliable these estimates are.
     
Since the autocorrelation function $\rho_{\E \E}(t)$
is nearly a pure exponential,
we expect that the dynamic critical exponent $z_{\text{exp}}$
is either exactly equal or almost exactly equal
to $z_{\text{int},\mathcal{E}'}$.

\subsection{Sharpness of Li--Sokal bound}  \label{subsec.li-sokal}


The estimates of $z_{\text{int},\mathcal{E}'}$ summarized in
Figure~\ref{fig_z_estimates} suggest that the Li--Sokal bound
$z_{\text{int},\mathcal{E}'} \ge \alpha/\nu$
holds {\em as a strict inequality}\/ over the entire range $1 \le q < 4$,
i.e.\ that it is {\em non-sharp by a power}\/.
But this conclusion is weakened by the fact that our fits of
the specific heat $C_H$ give estimates of $\alpha/\nu$
that deviate significantly from the known exact values
(Table~\ref{table_C_H_analysis}).
It is therefore of interest to
study directly the ratio $\tau_{\text{int},\mathcal{E}'}/C_H$,
in an effort to fit its behavior as $L\to\infty$ to one of the
following ans\"atze:
\begin{itemize}
   \item[1)]
Asymptotically constant with additive corrections to scaling
$A + B L^{-p}$.
   \item[2)]
A logarithmic growth, either as $A \log L + B$ or as $A \log^p L$.
   \item[3)]
A power-law growth $A L^p$ or $A L^p + B$.
\end{itemize}

Unfortunately our time-series analysis does not produce statistically valid
error bars for composite static-dynamic quantities such as
$\tau_{\text{int},\mathcal{E}'}/C_H$.
We therefore conducted the analysis in this subsection
under the crude assumption that the statistical fluctuations
on our estimators of $\tau_{\text{int},\mathcal{E}'}$ and $C_H$ are
uncorrelated.  In fact it is likely that these fluctuations are
{\em positively}\/ correlated, so that the true error bars on the
ratio $\tau_{\text{int},\mathcal{E}'}/C_H$ are {\em smaller}\/ than
we have supposed.  If so, this means that the true confidence level
of our fits is {\em smaller}\/ than what we report.\footnote{
   This failure to provide statistically valid error bars is embarrassing.
   We should have used the batch-means method
   \cite[Section~4.2]{Salas_Sokal_Potts3}
   to obtain such error bars.
   Unfortunately, the raw data from our simulations are no longer accessible,
   so we are unable to conduct such a reanalysis.
}

The fits to $A + B L^{-p}$ with $p > 0$ were always bad:
either they had a horrible confidence level,
or they converged to a value $p < 0$
(indicating that the true leading behavior is a power-law {\em growth}\/).
We therefore focussed on comparing
the logarithmic-growth and power-law-growth scenarios.
We chose $A \log L + B$ and $A L^p$ as the ans\"atze
in order to compare fits with an equal number of free parameters.
The results are shown in Table~\ref{table_tauint_CH_ratio}.
We see that both fits are in general good
(though the confidence levels may be overestimated as noted above).
However, the power-law fits are in general better.
In particular, there are no cases in which the power-law fit
has a confidence level less than 25\%;
but there are quite a few cases in which the logarithmic fit
exhibits such a low confidence level (sometimes much lower).
We therefore conclude that the fits to $\tau_{\text{int},\mathcal{E}'}/C_H$
also provide weak evidence that the Li--Sokal bound is non-sharp by a power.

\subsection{Test of Ossola--Sokal conjecture}  \label{subsec.ossola-sokal}


In the Introduction we argued that the
Ossola--Sokal conjecture $z_{\text{CM}} \ge \beta/\nu$
is probably false, on the grounds that it fails for $q=1$
(where $z_{\text{CM}}= 0$ and $\beta/\nu > 0$)
and that $z_{\text{CM}}$ is presumably a continuous function of $q$.
But this latter assumption is far from certain:
it is possible, for instance, that $z_{\text{CM}} = \beta/\nu$ exactly
for all $q$ near 1, but with an amplitude that vanishes as $q \downarrow 1$.
In this subsection we would like to test this scenario against our data
for $1.25 \le q \le 2$.

We first tried fits to $\tau_{\text{int},\mathcal{E}'} = A L^{\beta/\nu} + B$.
We then tried fits to
$\tau_{\text{int},\mathcal{E}'} = A L^{\beta/\nu} + B + CL^{-p}$
where $p>0$ is fixed and $A,B,C$ are free.
Let us report our results from these fits in decreasing order of~$q$:

\bigskip

1)  For $q=2$ ($k=1$) the fit to $A L^{\beta/\nu} + B$ is decent already
for $\lmin = 32$ ($\chi^2 = 3.90$, 4~DF, CL = 42\%)
and better for $\lmin = 64$:
\begin{equation}
   A \;=\; 13.27(7), \; B \;=\; -11.83(13)
   \qquad
   (\chi^2 = 1.51, \hbox{3 DF}, \hbox{CL = 68\%})
   \;.
\end{equation}
This behavior is not surprising, because our fit to $A L^z + B$
with $z$ free yielded $z = 0.145(4)$ with $\lmin = 16$,
which is not very far from $\beta/\nu = 0.125$.

\begin{figure}
\hspace*{-1.2cm}
\begin{tabular}{c@{\qquad\qquad}c}
\includegraphics[width=220pt]{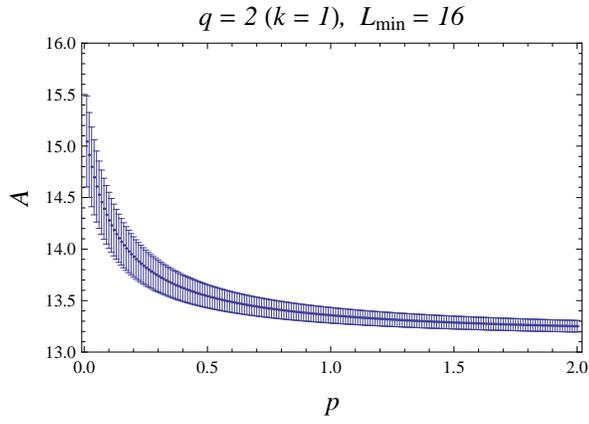} &
\includegraphics[width=220pt]{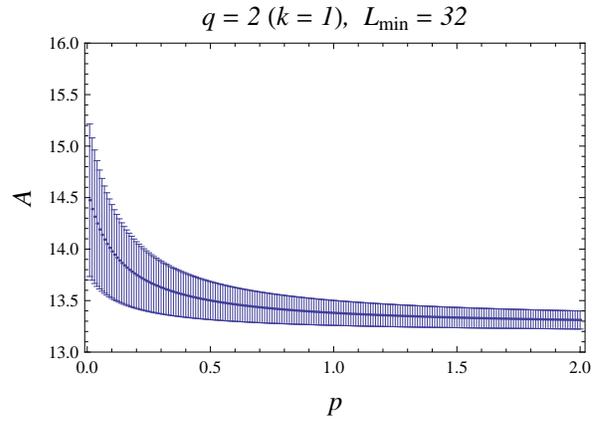} \\[3mm]
\includegraphics[width=220pt]{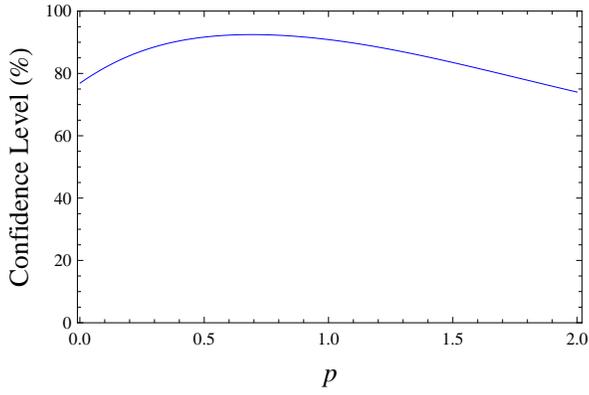} &
\includegraphics[width=220pt]{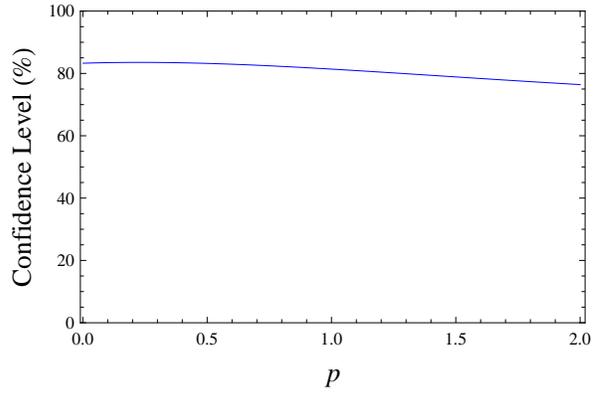} \\[3mm]
\phantom{(((a)}(a) & \phantom{(((a)}(b) \\[8mm]
\end{tabular}
\caption{
   Fits to $\tau_{\text{int},\mathcal{E}'} = A L^{\beta/\nu} + B + CL^{-p}$
   for $q=2$ ($k=1$).
   (a) $\lmin = 16$, (b) $\lmin=32$.
}
\label{fig_ossola_q=2}
\end{figure}

In Figure~\ref{fig_ossola_q=2} we plot the results of our fits to
$A L^{\beta/\nu} + B + CL^{-p}$ as a function of $p$,
for $\lmin=16$ and 32:
\begin{itemize}
   \item
For $\lmin=16$ an excellent confidence level is obtained over
the whole range $0 < p < 2$ (and indeed beyond),
with an optimum at $p \approx 0.692$ (CL = 92\%).
The estimated amplitudes $A$ are positive and very far from zero;
at the optimum we have $A = 13.44(9)$.
As $p \to\infty$ the amplitude $A$ tends to a value $\approx 13$,
which is close to that obtained from the fit to $A L^{\beta/\nu} + B$.
   \item
For $\lmin=32$ an excellent confidence level is again obtained over
the whole range $0 < p < 2$ (and indeed beyond),
with an optimum at $p \approx 0.228$ (CL = 84\%)
but with a very broad peak.
The estimated amplitudes $A$ are again positive and very far from zero;
at the optimum we have $A = 13.71(30)$.
As $p \to\infty$ the amplitude $A$ again tends to $\approx 13$.
\end{itemize}

2)  For $q=1.75$ the fit to $A L^{\beta/\nu} + B$ is decent already
for $\lmin = 128$ ($\chi^2 = 2.31$, 2~DF, CL = 32\%)
and excellent for $\lmin = 256$:
\begin{equation}
   A \;=\; 5.36(8), \; B \;=\; -2.62(17)
   \qquad
   (\chi^2 = 0.002, \hbox{1 DF}, \hbox{CL = 97\%})
   \;.
\end{equation}
Once again this behavior is not surprising, because our fit to $A L^z + B$
with $z$ free yielded $z = 0.085(26)$ with $\lmin = 128$,
which is not very far from $\beta/\nu \approx 0.121$.

\begin{figure}
\hspace*{-1.2cm}
\begin{tabular}{c@{\qquad\qquad}c}
\includegraphics[width=220pt]{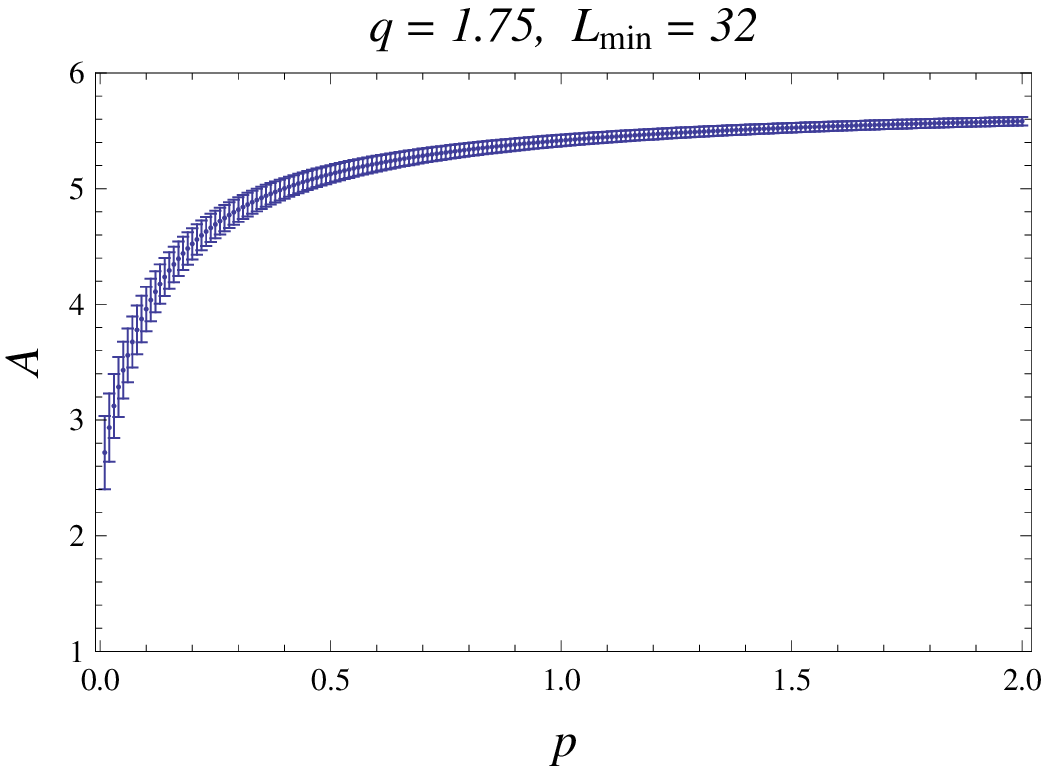} &
\includegraphics[width=220pt]{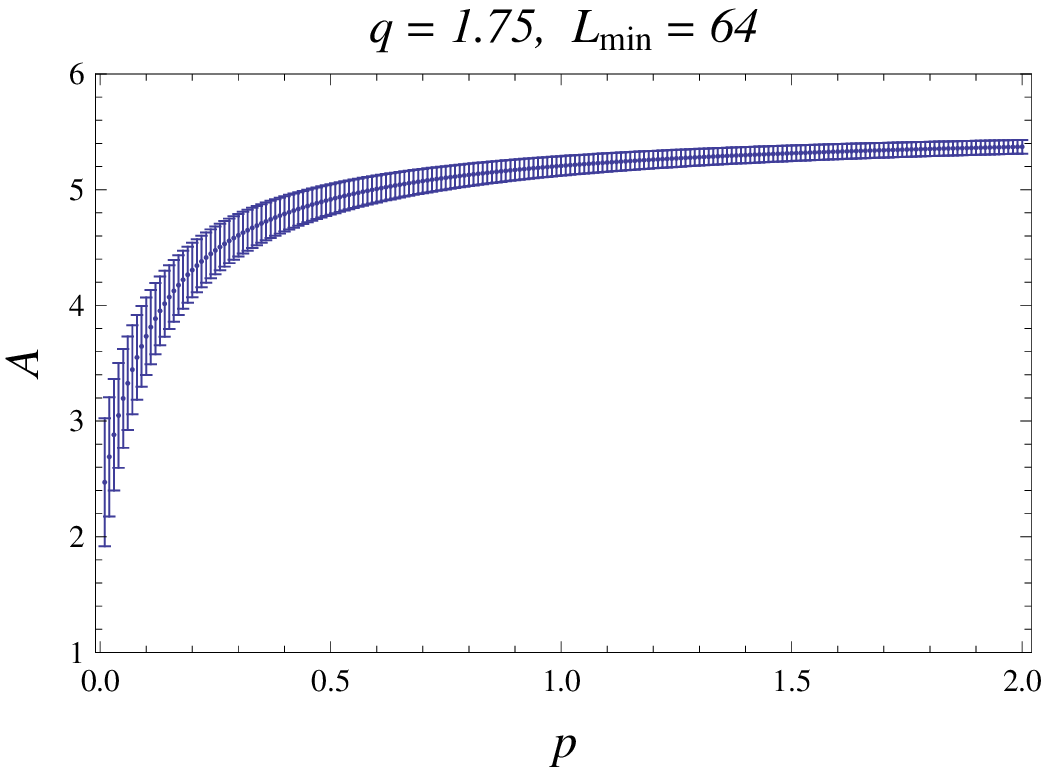} \\[3mm]
\includegraphics[width=220pt]{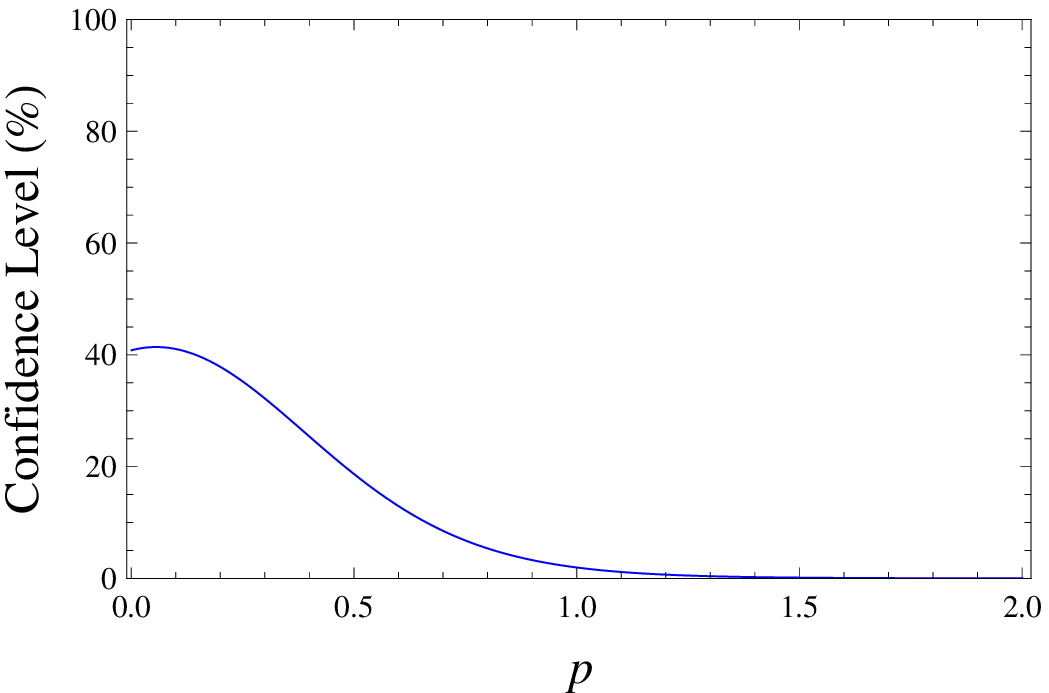} &
\includegraphics[width=220pt]{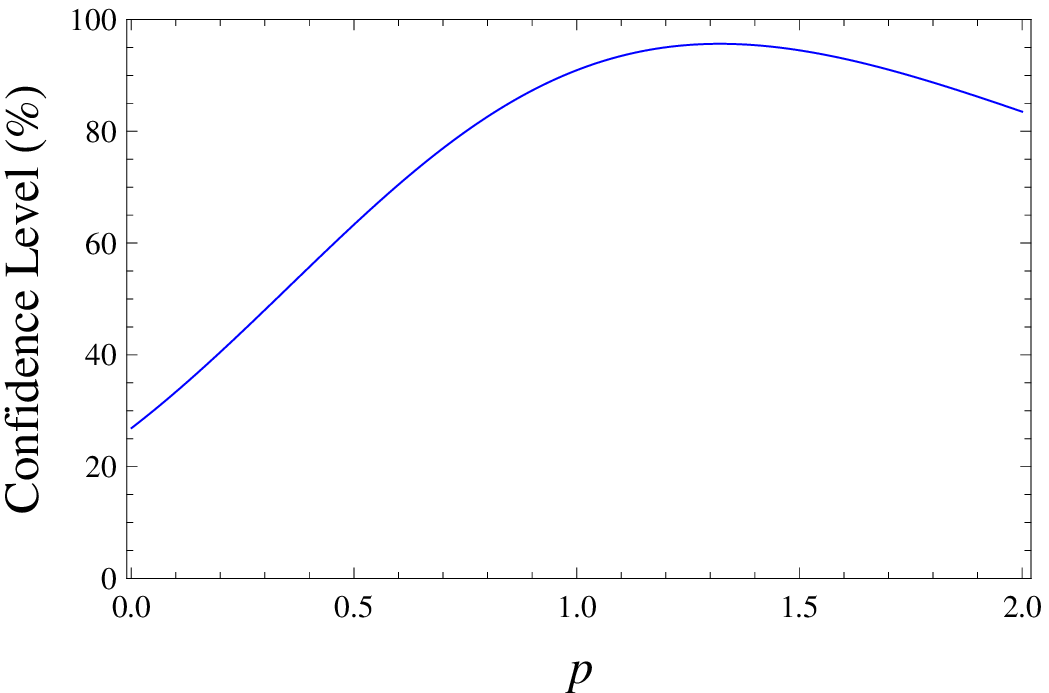} \\[3mm]
\phantom{(((a)}(a) & \phantom{(((a)}(b) \\[8mm]
\end{tabular}
\caption{
   Fits to $\tau_{\text{int},\mathcal{E}'} = A L^{\beta/\nu} + B + CL^{-p}$
   for $q=1.75$.
   (a) $\lmin = 32$, (b) $\lmin=64$.
}
\label{fig_ossola_q=1.75}
\end{figure}

In Figure~\ref{fig_ossola_q=1.75} we plot the results of our fits to
$A L^{\beta/\nu} + B + CL^{-p}$ as a function of $p$,
for $\lmin=32$ and 64:
\begin{itemize}
   \item
For $\lmin=32$ a decent confidence level is obtained for $p \ltapprox 0.6$,
with an optimum at $p \approx 0.056$ (CL = 41\%).
The estimated amplitudes $A$ are positive and far from zero;
at the optimum we have $A = 3.51(24)$.
As $p \to\infty$ the amplitude $A$ tends to a value $\approx 5.8$,
which is close to that obtained from the fit to $A L^{\beta/\nu} + B$.
   \item
For $\lmin=64$ a decent confidence level is obtained over the entire range
$0 < p \le 2$ (or even beyond),
with an optimum at $p \approx 1.321$ (CL = 96\%).
The estimated amplitudes $A$ are again positive and far from zero,
and in fact quite close to those obtained from $\lmin=32$.
At the optimum, we have $A = 5.28(7)$.
\end{itemize}

\begin{figure}
\hspace*{-1.2cm}
\begin{tabular}{c@{\qquad\qquad}c}
\includegraphics[width=220pt]{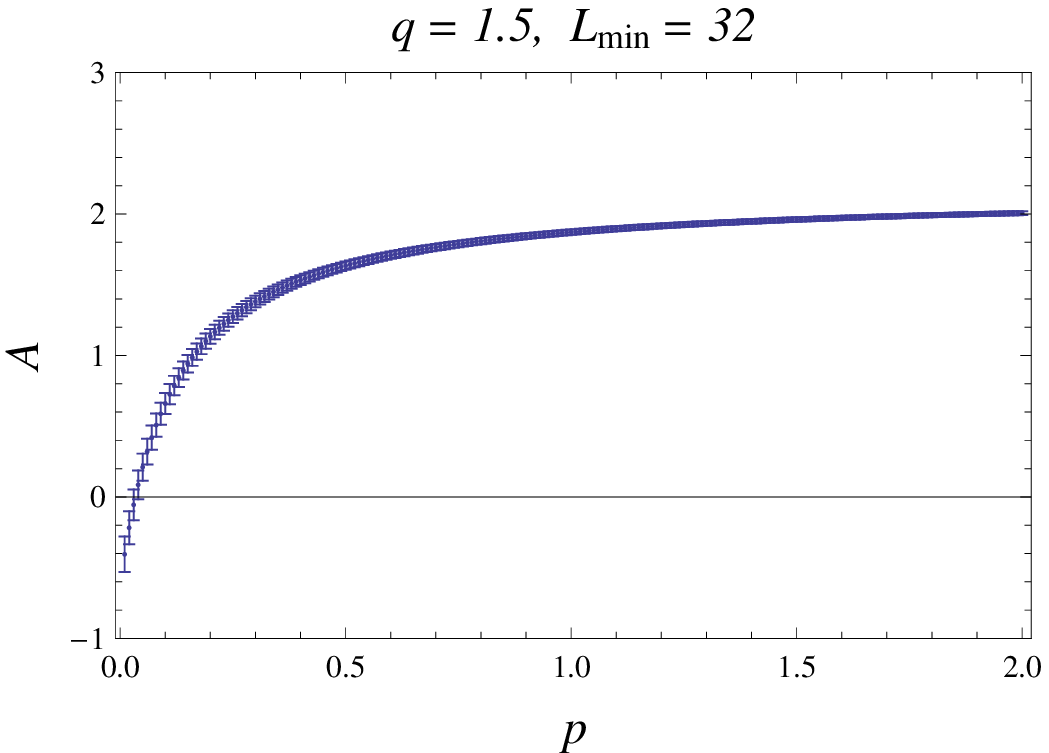} &
\includegraphics[width=220pt]{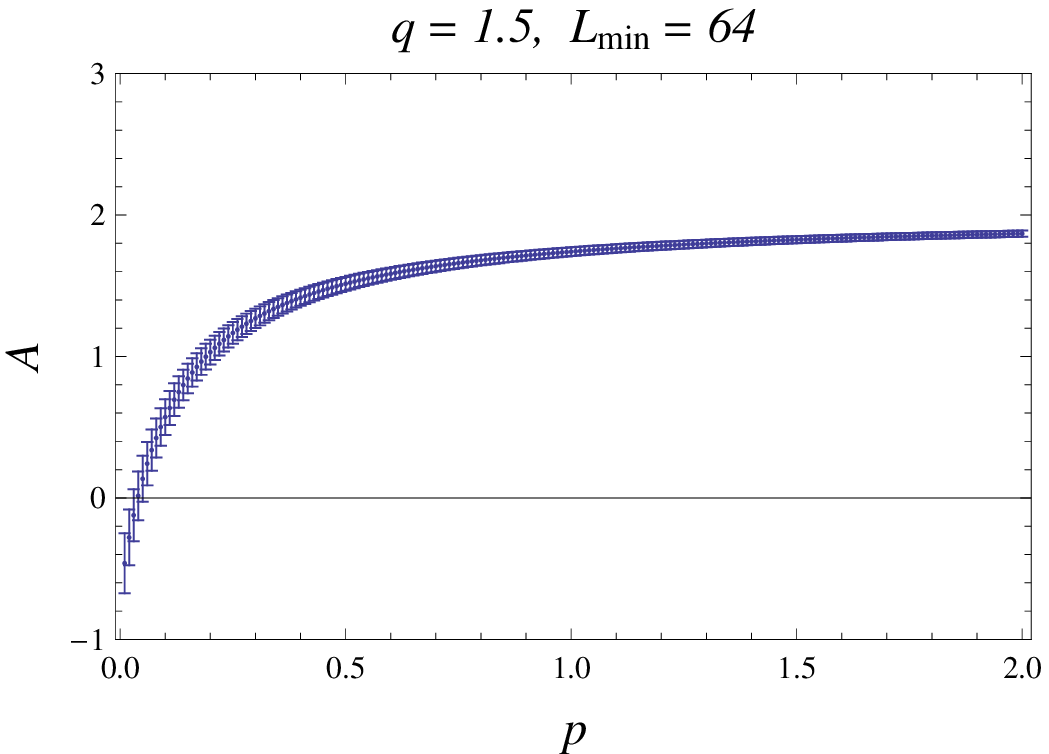} \\[3mm]
\includegraphics[width=220pt]{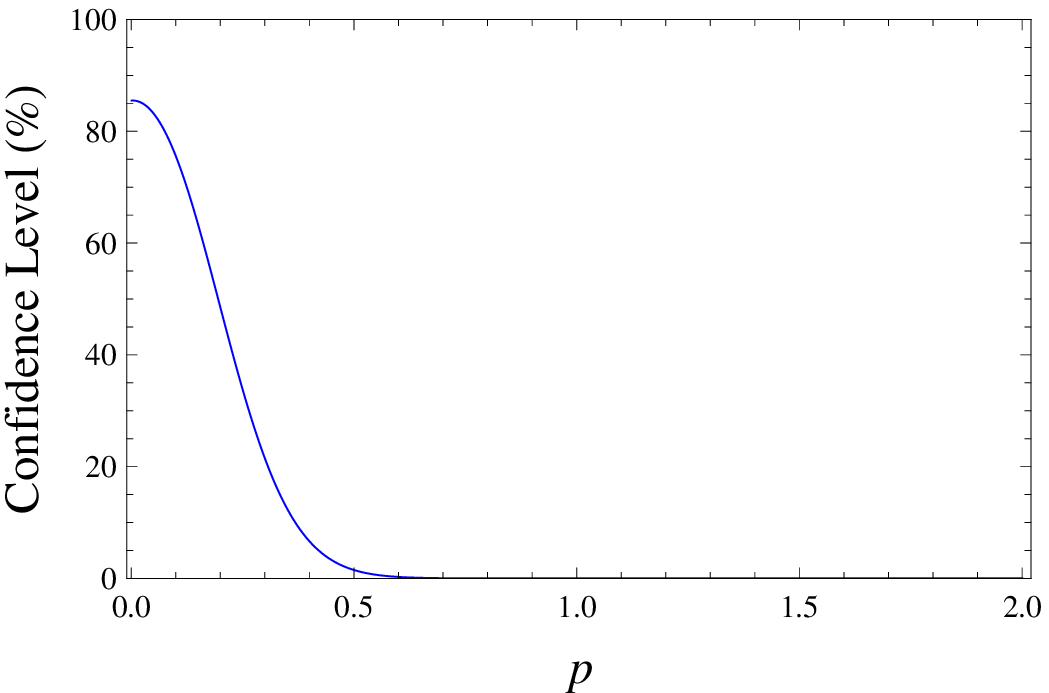} &
\includegraphics[width=220pt]{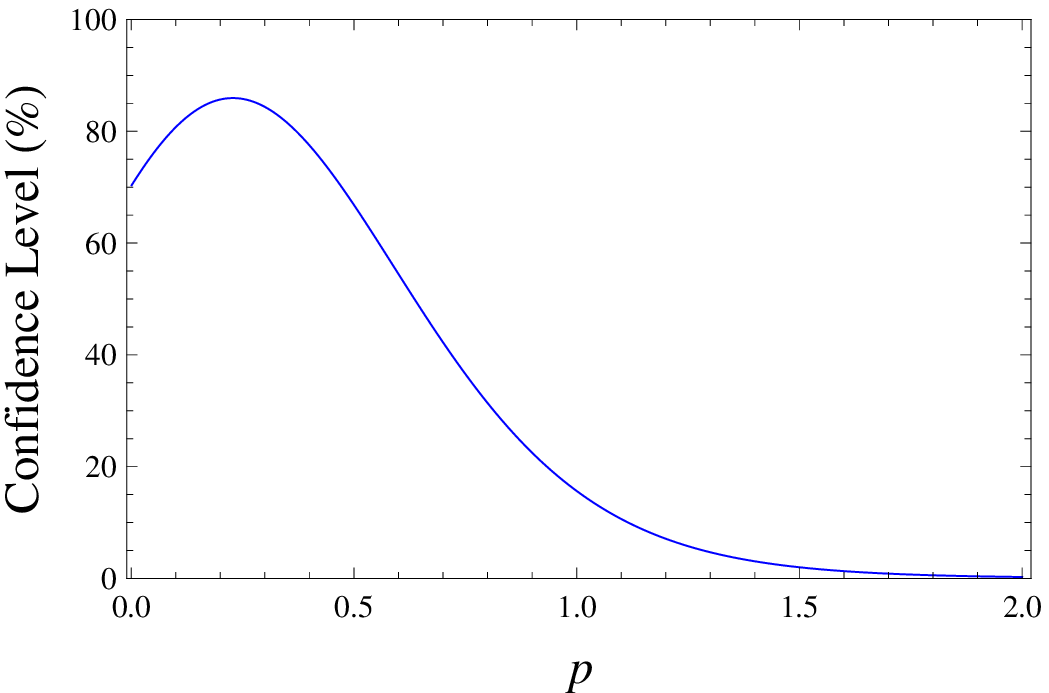} \\[3mm]
\phantom{(((a)}(a) & \phantom{(((a)}(b) \\[8mm]
\end{tabular}
\caption{
   Fits to $\tau_{\text{int},\mathcal{E}'} = A L^{\beta/\nu} + B + CL^{-p}$
   for $q=1.5$.
   (a) $\lmin = 32$, (b) $\lmin=64$.
}
\label{fig_ossola_q=1.5}
\end{figure}

3) For $q=1.5$ the fit to $A L^{\beta/\nu} + B$ is poor even for
$\lmin = 256$ (CL = 3\%).
In Figure~\ref{fig_ossola_q=1.5} we plot the results of our fits to
$A L^{\beta/\nu} + B + CL^{-p}$ as a function of $p$,
for $\lmin=32$ and 64:
\begin{itemize}
   \item
For $\lmin=32$ a decent confidence level is obtained for $p \ltapprox 0.3$,
but with the optimum attained at the ridiculously small value
$p \approx 0.004$ (CL = 86\%).
Also, the estimated amplitude $A$ is negative for $p \ltapprox 0.033$,
which is obviously impossible.
However, for $0.08 \ltapprox p \ltapprox 0.3$ a decent fit is obtained
with an amplitude $A$ that is positive and far from zero.
   \item
For $\lmin=64$ a decent confidence level is obtained for
$0 < p \ltapprox 0.9$,
with an optimum at $p \approx 0.228$ (CL = 86\%).
The estimated amplitudes $A$ are again negative for $p \ltapprox 0.038$
but positive and far from zero for $p \gtapprox 0.1$.
At the optimum, we have $A = 1.11(8)$.
\end{itemize}

\begin{figure}
\centering
\includegraphics[width=220pt]{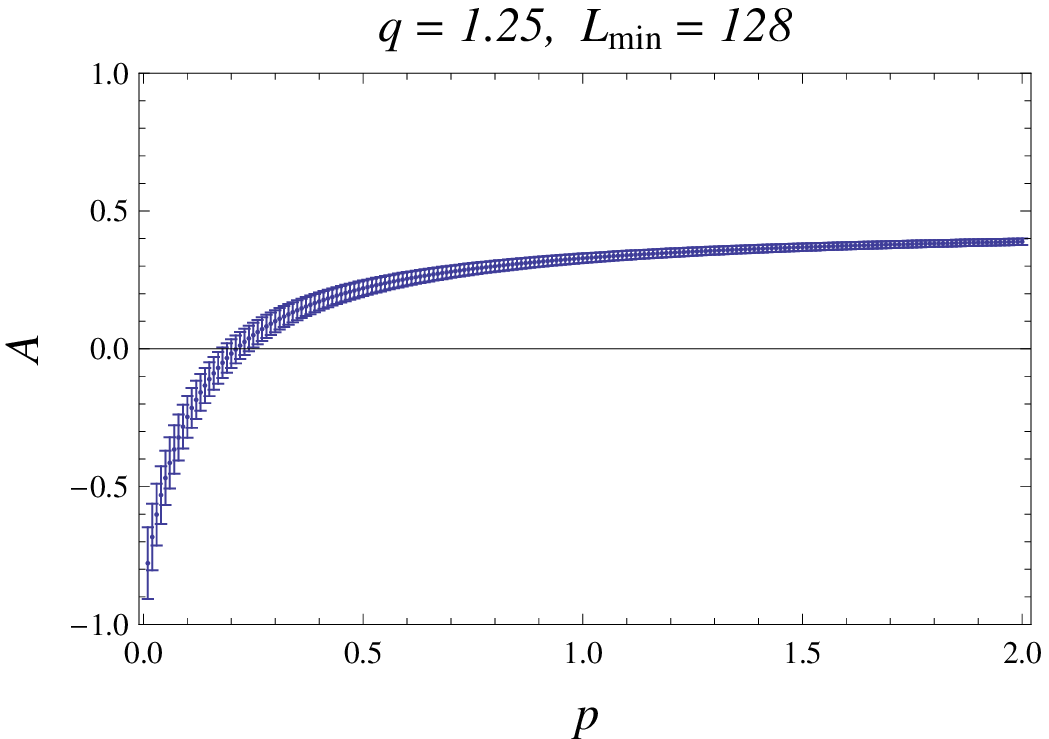} \\[3mm]
\includegraphics[width=220pt]{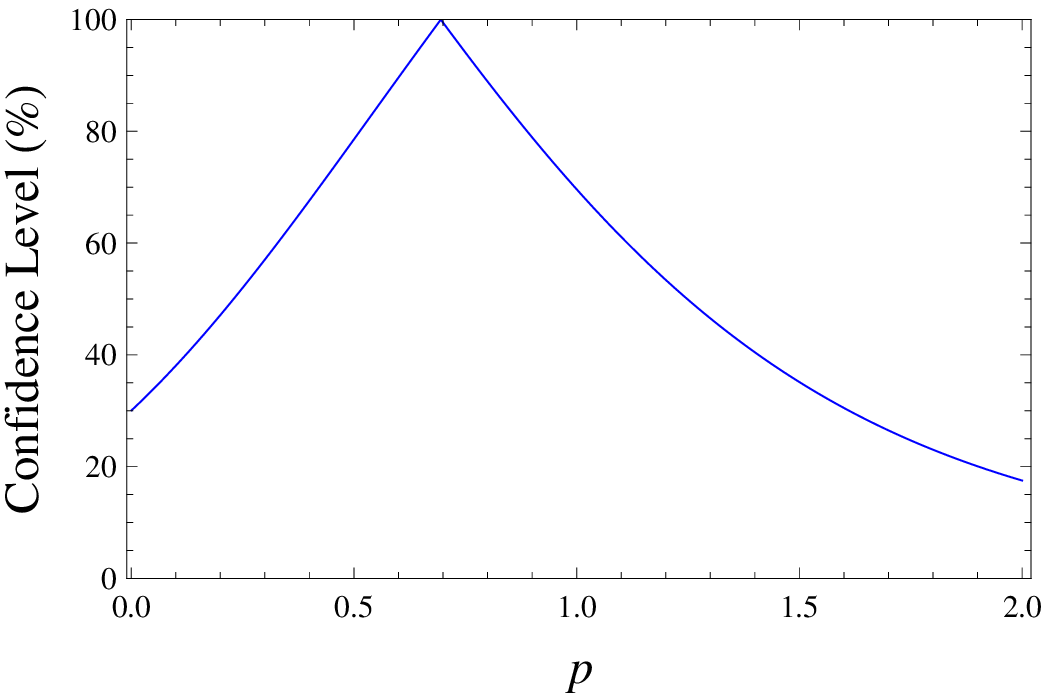} \\[8mm]
\caption{
   Fits to $\tau_{\text{int},\mathcal{E}'} = A L^{\beta/\nu} + B + CL^{-p}$
   for $q=1.25$, with $\lmin = 128$.
}
\label{fig_ossola_q=1.25}
\end{figure}

4) For $q=1.25$ the fit to $A L^{\beta/\nu} + B$ is poor even for
$\lmin = 256$ (CL = 1\%).
The fits to $A L^{\beta/\nu} + B + CL^{-p}$ are poor
(CL $< 6\%$) for $\lmin=32$ and 64 for all $p > 0$.
For $\lmin = 128$, however, we are able to obtain decent fits
over the whole range $0 < p \ltapprox 2$,
as shown in Figure~\ref{fig_ossola_q=1.25}.
The optimum lies at $p \approx 0.695$ (CL = 99.9997\%).
The estimated amplitudes $A$ are negative for $p \ltapprox 0.211$
but positive and far from zero for $p \gtapprox 0.4$.
At the optimum, we have $A = 0.28(2)$.

\bigskip

The foregoing fits show that the
Ossola--Sokal conjecture $z_{\text{CM}} \ge \beta/\nu$
is {\em not}\/ ruled out by our data at $1.25 \le q \le 2$.
Indeed, our data are consistent with the possibility that
$z_{\text{CM}} = \beta/\nu$ exactly for $1 < q \ltapprox 2$,
but with an amplitude that vanishes as $q \downarrow 1$,
perhaps proportional to $q-1$
(the fits for $q=1.25$ and 1.50 are consistent with this latter behavior).

\subsection{Dynamic critical exponents $z_{\text{int},\mathcal{O}}$ for
   other $\mathcal{O}$}   \label{subsec.otherO}

Let us now look briefly at the dynamic critical exponents
$z_{\text{int},\mathcal{O}}$
for observables $\mathcal{O}$ other than $\E$.
For $\obs = \N, \mathcal{S}_2, \mathcal{C}_1$,
the values of $\tau_{\text{int},\mathcal{O}}$ are very close to
those of $\tau_{\text{int},\E}$, so the estimates of
$z_{\text{int},\mathcal{O}}$ will be nearly the same;
little would be gained by going through these fits in detail.
Instead, it seems sensible to look at the observable that has the
{\em smallest}\/ autocorrelation time, namely $\mathcal{C}_2$ ---
the idea being that if any differences in $z_{\text{int},\mathcal{O}}$
between different observables are to be detected, they will be detected here.

{}From Figure~\ref{fig_rho_plot} we see that the autocorrelation function
of $\E$ is nearly a pure exponential (and this is so for all $q,k,L$),
so that $\tau_{\text{int},\E} \approx \tau_{\text{exp}}$
and hence $z_{\text{int},\E} = z_{\text{exp}}$ or nearly so.
The autocorrelation function of $\mathcal{C}_2$, by contrast,
exhibits an initial fast decay, followed by a decay at the {\em same}\/
exponential rate $\tau_{\rm exp}$ as for the other observables
but with an amplitude $A_{\mathcal{C}_2}$ that is significantly less than 1
(e.g.\ around 0.4 in the plot shown).
The key question is whether this amplitude tends to a nonzero value
when $L \to\infty$
(in which case we will have
 $z_{\text{int},\mathcal{C}_2} = z_{\text{exp}}$)
or tends to zero as an inverse power of $L$
(in which case we will have
 $z_{\text{int},\mathcal{C}_2} < z_{\text{exp}}$).\footnote{
   See \cite[Section~5.2]{OssolaSokal04}
   for a more detailed analysis of this kind,
   for the Swendsen--Wang dynamics for the three-dimensional Ising model.
}

In Tables~\ref{table_fits_C2_ALz} and \ref{table_fits_C2_ALz+B}
we show the fits for $\tau_{\text{int},\mathcal{C}_2}$
to the ans\"atze $AL^z$ and $AL^z + B$.
In Table~\ref{table_fits_C2_allk} we show the results
for $z_{\text{int},\mathcal{C}_2}$ obtained by averaging over $k$.

The estimates for $z_{\text{int},\mathcal{C}_2}$
are indeed less than those for $z_{\text{int},\E}$,
by an amount that is $\approx 0.1$ for $q \approx 2$
and tends to zero as $q \uparrow 4$.
The question is:  Are these differences real,
or are they artifacts of corrections to scaling at small $L$?
The fact that the differences are smaller for $q \gtapprox 3$,
where the autocorrelation times are larger,
suggests that perhaps the differences will disappear as $L \to\infty$
but that at small $q$ we have to go to larger $L$ to see this.
But this is far from clear;
we will only know the truth by doing simulations at significantly larger
values of $L$.


\section{Discussion}
\label{discussion}

In this paper we have studied the dynamic critical behavior of
the Chayes--Machta algorithm as a function of $q$
over the whole range $1.25\le q \le 4$.
We have obtained estimates of the dynamic critical exponent
$z_{\text{int},\mathcal{E}'}$ as a function of $q$:
see \reff{eq.zint.final}, Table~\ref{table_fits_allk}
and Figure~\ref{fig_z_estimates}.
Since the autocorrelation function $\rho_{\E \E}(t)$
is nearly a pure exponential,
we also expect that the dynamic critical exponent $z_{\text{exp}}$
is either exactly equal or almost exactly equal
to $z_{\text{int},\mathcal{E}'}$.

By simultaneously studying the whole range of values of $q$,
we were able to gain some insights that would not have been available
had we studied only a single value of $q$ (such as the Ising value $q=2$)
or even all integer values of $q$
as in past studies of the Swendsen--Wang algorithm.
For instance:

1) The autocorrelation time $\tau_{\text{int},\mathcal{E}'}$ at $q=2$
can be plausibly fit with the ansatz $A \log^2 L + B \log L + C$,
suggesting that the Li--Sokal bound might be sharp modulo a logarithm;
this agrees with the conclusions of the paper \cite{Salas_Sokal_Ising_v1},
where the data were in fact found to slightly favor the
non-sharp-by-a-logarithm ansatz over the non-sharp-by-a-power ansatz.
But we now find that the good fit to $A \log^2 L + B \log L + C$
persists over the whole range $1.75 \le q \ltapprox 2.25$.
And it seems to us implausible on theoretical
grounds that we would have such a logarithmic growth
for an entire {\em interval}\/ of $q$;
rather, we expect that there exists {\em one}\/ value $q_\star$
such that $z=0$ for $q < q_\star$ and $z>0$ for $q > q_\star$,
with a possible (poly)logarithmic growth {\em at}\/ $q = q_\star$.
Our data suggest that $q_\star$ lies between 1.25 and 1.75,
with our best guess being around 1.6.
If this scenario is correct, it follows that the Li--Sokal bound 
is non-sharp by a power for $q_\star < q \le 2$.
This then suggests (but of course does not prove)
that it might be non-sharp by a power over the whole interval
$q_\star < q < 4$
(possibly reverting to non-sharp by a logarithm {\em at}\/ $q = 4$).

2) By considering the Ossola--Sokal conjecture $z_{\text{CM}} \ge \beta/\nu$
simultaneously for all $q$,
we can see immediately that it fails at $q=1$
(where $z_{\text{CM}} = 0$ but $\beta/\nu > 0$)
and hence fails also for $q$ near 1
{\em if}\/ the dynamic critical exponent $z_{\text{CM}}$
is a continuous function of~$q$.
Indeed, our pure power-law fits suggest that
the conjecture fails for $1 \le q \ltapprox 1.95$.
(In particular, if $q_\star > 1$ as just suggested,
 then the conjecture fails spectacularly in the interval
 $1 \le q < q_\star$, where $z_{\text{CM}} = 0$ but $\beta/\nu > 0$.)
However --- and perhaps surprisingly ---
our data are also compatible with an alternative scenario in which
$z_{\text{CM}} = \beta/\nu$ exactly for all $q$ near 1,
but with an amplitude that vanishes as $q \downarrow 1$
(Section~\ref{subsec.ossola-sokal}).

However, the behavior of the Chayes--Machta autocorrelation time
for $1 < q < 2$ is still unclear:
though our data strongly suggest
that $\tau_{\text{int},\mathcal{E}'}$
diverges as $L \to\infty$ for $q$ slightly below 2,
and is nondivergent as $L \to\infty$ for $q$ slightly above 1,
we cannot rule out the possibility that the data at $L \le 1024$
are misleading and that the true asymptotic behavior is different
from what we conjecture.
Future work at larger values of $L$ would of course be desirable.

Likewise, though our data suggest that the Li--Sokal bound
is non-sharp by a power for $1.6 \ltapprox q < 4$,
it is also true that the exponent estimates have been dropping over time
as data becomes available at larger values of $L$
and as we try ans\"atze other than a pure power law
(see Table~\ref{table_fits_allk} for the effect of the ansatz).
Our data suggest most clearly the non-sharpness of the Li--Sokal bound
(when the non-sharpness is measured in units of the standard deviation
of our estimate) at $q=2.75$ and $q=3$.
At these values of $q$, the critical slowing-down is strong enough
that interference from the regular background term is less important
than it is at smaller $q$
(i.e.\ the $AL^z$ and $AL^z + B$ fits show less discrepancy)
but modest enough that we can have reasonably good data on large lattices
(contrary to the situation at larger $q$);
furthermore, the correction-to-scaling exponent $\Delta_1$
is still fairly large.
It would therefore be of great interest to perform
high-precision simulations at these values of $q$,
going to very high values of $L$.

\section*{Acknowledgments}

The authors would like to thank Youjin Deng and Jonathan Machta
for many helpful discussions.
We would also like to thank Mulin Ding for helping us to recover
some of the data from this project.

This research was supported in part
by U.S.\ National Science Foundation grants PHY--0116590 and PHY--0424082.


\clearpage

\newpage
\def\kk{\phantom{1}}
\addcontentsline{toc}{section}{Tables}


%
%
\begin{table}
\begin{center}
\begin{tabular}{|c|cccc|}
\hline
$q$  &     $\alpha/\nu$ &   $\beta/\nu$ & $\gamma/\nu$ &   $\Delta_1$ \\
\hline
1.25     & $-$0.35527    &        0.11118        &        1.77764        &
  1.80702       \\
1.50     & $-$0.22663    &        0.11678        &        1.76644        &
  1.63551       \\
1.75     & $-$0.10929    &        0.12131        &        1.75738        &
  1.47905       \\
2.00     &        0.00000        &        0.12500        &        1.75000
 &        1.33333       \\
2.25     &        0.10363        &        0.12798        &        1.74404
 &        1.19517       \\
2.50     &        0.20357        &        0.13034        &        1.73932
 &        1.06191       \\
2.75     &        0.30168        &        0.13212        &        1.73576
 &        0.93110       \\
3.00     &        0.40000        &        0.13333        &        1.73333
 &        0.80000       \\
3.25     &        0.50126        &        0.13393        &        1.73214
 &        0.66499       \\
3.50     &        0.61007        &        0.13377        &        1.73246
 &        0.51991       \\
3.75     &        0.73760        &        0.13242        &        1.73517
 &        0.34986       \\
4.00     &        1.00000        &        0.12500        &        1.75000
 &        0.00000       \\
\hline
\end{tabular}
\caption{
   Exact values of the static critical exponents,
   rounded to five decimal places.
}
   \label{table_exact_exponents}
\end{center}
\end{table}

\clearpage

\begin{table}
\centering
      {\small
        \begin{tabular}{rr@{}lr@{}lr@{}lr@{}l}
          \hline
          $L$ & \multicolumn{2}{c}{$q=1.25$} &
\multicolumn{2}{c}{$q=1.50$} & \multicolumn{2}{c}{$q=1.75$}&
\multicolumn{2}{c}{$q=2.00$} \\
          \hline
          16& $ 0.5011732$ &(38)& $ 0.5025161$ &(59)& $ 0.5040267$
&(83)& $ 0.5056788$ &(68)\\
          32& $ 0.5005202$ &(20)& $ 0.5011648$ &(32)& $ 0.5019472$
&(47)& $ 0.5028427$ &(41)\\
          64& $ 0.5002301$ &(10)& $ 0.5005395$ &(17)& $ 0.5009352$
&(26)& $ 0.5014281$ &(24)\\
          128& $ 0.5001022$ &(5)& $ 0.5002494$ &(9)& $ 0.5004485$
&(14)& $ 0.5007133$ &(14)\\
          256& $ 0.5000444$ &(3)& $ 0.5001146$ &(5)& $ 0.5002159$
&(8)& $ 0.5003559$ &(8)\\
          512& $ 0.5000199$ &(2)& $ 0.5000530$ &(3)& $ 0.5001046$
&(6)& $ 0.5001763$ &(6)\\
          1024& $ 0.5000089$ &(2)& $ 0.5000243$ &(3)& $ 0.5000510$
&(5)& $ 0.5000884$ &(6)\\
\hline
        \end{tabular}
      }
  \caption{\label{static N data} Static data for
$\langle\mathcal{N}\rangle$, from the Monte Carlo simulations at the
critical point of the 2-dimensional random cluster model, as a
function of $L$ and $q$,
    obtained by combining the data for all available $k$ values for
each $q$. The quoted error bar corresponds to one standard deviation.
  }
  \medskip
\end{table}

\begin{table}
\centering
      {\small
	\begin{tabular}{rr@{}lr@{}lr@{}lr@{}l}
	  \hline
	  $L$ & \multicolumn{2}{c}{$q=1.25$} & \multicolumn{2}{c}{$q=1.50$} & \multicolumn{2}{c}{$q=1.75$}& \multicolumn{2}{c}{$q=2.00$} \\
	  \hline
	  16& $   0.5533\,\pm\,$ & $ 0.0004$& $   1.0380\,\pm\,$ & $ 0.0006$& $   1.4902\,\pm\,$ & $ 0.0008$& $   1.9285\,\pm\,$ & $ 0.0006$\\
	  32& $   0.6180\,\pm\,$ & $ 0.0004$& $   1.1940\,\pm\,$ & $ 0.0006$& $   1.7753\,\pm\,$ & $ 0.0009$& $   2.3779\,\pm\,$ & $ 0.0008$\\
	  64& $   0.6672\,\pm\,$ & $ 0.0004$& $   1.3284\,\pm\,$ & $ 0.0007$& $   2.0348\,\pm\,$ & $ 0.0010$& $   2.8226\,\pm\,$ & $ 0.0010$\\
	  128& $   0.7058\,\pm\,$ & $ 0.0004$& $   1.4420\,\pm\,$ & $ 0.0007$& $   2.2767\,\pm\,$ & $ 0.0012$& $   3.2651\,\pm\,$ & $ 0.0012$\\
	  256& $   0.7360\,\pm\,$ & $ 0.0004$& $   1.5366\,\pm\,$ & $ 0.0007$& $   2.4957\,\pm\,$ & $ 0.0013$& $   3.7062\,\pm\,$ & $ 0.0014$\\
	  512& $   0.7594\,\pm\,$ & $ 0.0006$& $   1.6200\,\pm\,$ & $ 0.0011$& $   2.7016\,\pm\,$ & $ 0.0020$& $   4.1456\,\pm\,$ & $ 0.0023$\\
	  1024& $   0.7768\,\pm\,$ & $ 0.0010$& $   1.6911\,\pm\,$ & $ 0.0018$& $   2.8922\,\pm\,$ & $ 0.0035$& $   4.5918\,\pm\,$ & $ 0.0043$\\
\hline
	\end{tabular}
      }
  \caption{\label{specific heat data} Static data for $C_H$, from the Monte Carlo simulations at the critical point of the 2-dimensional random cluster model, as a function of $L$ and $q$,
    obtained by combining the data for all available $k$ values for each pair $(q,L)$.
 The quoted error bar corresponds to one standard deviation.
  }
  \medskip
\end{table}

\begin{table}
\hspace*{-7mm}
      {\footnotesize
	\begin{tabular}{rr@{}lr@{}lr@{}lr@{}l}
	  \hline
	  $L$ & \multicolumn{2}{c}{$q=1.25$} & \multicolumn{2}{c}{$q=1.50$} & \multicolumn{2}{c}{$q=1.75$}& \multicolumn{2}{c}{$q=2.00$} \\
	  \hline
	  16& $     145.2585\,\pm\,$ & $    0.0075$& $     142.6486\,\pm\,$ & $    0.0106$& $     140.8411\,\pm\,$ & $    0.0140$& $     139.6098\,\pm\,$ & $    0.0107$\\
	  32& $     497.8233\,\pm\,$ & $    0.0265$& $     485.2393\,\pm\,$ & $    0.0382$& $     476.2855\,\pm\,$ & $    0.0515$& $     469.9356\,\pm\,$ & $    0.0405$\\
	  64& $    1706.4867\,\pm\,$ & $    0.0932$& $    1650.6023\,\pm\,$ & $    0.1357$& $    1610.1083\,\pm\,$ & $    0.1860$& $    1581.4235\,\pm\,$ & $    0.1495$\\
	  128& $    5850.3427\,\pm\,$ & $    0.3249$& $    5614.3344\,\pm\,$ & $    0.4780$& $    5442.1190\,\pm\,$ & $    0.6643$& $    5319.5479\,\pm\,$ & $    0.5429$\\
	  256& $   20054.6950\,\pm\,$ & $    1.1280$& $   19099.3800\,\pm\,$ & $    1.6687$& $   18399.0050\,\pm\,$ & $    2.3542$& $   17890.2105\,\pm\,$ & $    1.9519$\\
	  512& $   68770.5920\,\pm\,$ & $    5.3616$& $   64981.2580\,\pm\,$ & $    7.9347$& $   62211.4880\,\pm\,$ & $   11.2476$& $   60146.9825\,\pm\,$ & $    9.8673$\\
	  1024& $  235766.3800\,\pm\,$ & $   30.2818$& $  221047.8000\,\pm\,$ & $   45.1926$& $  210455.2400\,\pm\,$ & $   64.7251$& $  202355.8500\,\pm\,$ & $   56.1135$\\
\hline
	\end{tabular}
      }
  \caption{\label{static S2 data} Static data for $\chi$, from the Monte Carlo simulations at the critical point of the 2-dimensional random cluster model, as a function of $L$ and $q$,
    obtained by combining the data for all available $k$ values for each pair $(q,L)$.
 The quoted error bar corresponds to one standard deviation.
  }
  \medskip
\end{table}

\begin{table}
\centering
      {\small
        \begin{tabular}{rr@{}lr@{}lr@{}lr@{}l}
          \hline
          $L$ & \multicolumn{2}{c}{$q=1.25$} &
\multicolumn{2}{c}{$q=1.50$} & \multicolumn{2}{c}{$q=1.75$}&
\multicolumn{2}{c}{$q=2.00$} \\
          \hline
          16& $  0.92329$ &(10)& $  0.91489$ &(13)& $  0.91142$ &(17)&
$  0.91137$ &(13)\\
          32& $  0.91903$ &(10)& $  0.91064$ &(13)& $  0.90705$ &(17)&
$  0.90702$ &(13)\\
          64& $  0.91784$ &(10)& $  0.90927$ &(14)& $  0.90559$ &(18)&
$  0.90596$ &(14)\\
          128& $  0.91755$ &(10)& $  0.90869$ &(14)& $  0.90492$
&(19)& $  0.90553$ &(15)\\
          256& $  0.91727$ &(10)& $  0.90868$ &(14)& $  0.90482$
&(20)& $  0.90511$ &(16)\\
          512& $  0.91748$ &(14)& $  0.90867$ &(20)& $  0.90500$
&(27)& $  0.90433$ &(24)\\
          1024& $  0.91729$ &(23)& $  0.90855$ &(33)& $  0.90595$
&(46)& $  0.90467$ &(40)\\
\hline
        \end{tabular}
      }
  \caption{\label{static XI over L data} Static data for $\xi/L$, from
the Monte Carlo simulations at the critical point of the 2-dimensional
random cluster model,
    as a function of $L$ and $q$,
    obtained by combining the data for all available $k$ values for
each $q$. The quoted error bar corresponds to one standard deviation.
  }
  \medskip
\end{table}

\begin{table}
\centering
      {\footnotesize
        \begin{tabular}{rr@{}lr@{}lr@{}lr@{}l}
          \hline
          $L$ & \multicolumn{2}{c}{$q=1.25$} &
\multicolumn{2}{c}{$q=1.50$} & \multicolumn{2}{c}{$q=1.75$}&
\multicolumn{2}{c}{$q=2.00$} \\
          \hline
          16& $   186.2505\,\pm\,$ & $    0.0063$& $
184.4023\,\pm\,$ & $    0.0090$& $   183.1015\,\pm\,$ & $    0.0119$&
$   182.1899\,\pm\,$ & $    0.0092$\\
          32& $   689.6841\,\pm\,$ & $    0.0241$& $
680.2851\,\pm\,$ & $    0.0349$& $   673.4366\,\pm\,$ & $    0.0472$&
$   668.4742\,\pm\,$ & $    0.0373$\\
          64& $  2554.0565\,\pm\,$ & $    0.0912$& $
2509.5209\,\pm\,$ & $    0.1337$& $  2476.5244\,\pm\,$ & $    0.1844$&
$  2452.6604\,\pm\,$ & $    0.1491$\\
          128& $  9458.3576\,\pm\,$ & $    0.3426$& $
9256.6841\,\pm\,$ & $    0.5088$& $  9106.1186\,\pm\,$ & $    0.7136$&
$  8996.7788\,\pm\,$ & $    0.5878$\\
          256& $ 35023.7950\,\pm\,$ & $    1.2827$& $
34147.6420\,\pm\,$ & $    1.9207$& $ 33487.5700\,\pm\,$ & $
2.7425$& $ 32997.3910\,\pm\,$ & $    2.2970$\\
          512& $ 129717.7400\,\pm\,$ & $    6.5774$& $
125972.7400\,\pm\,$ & $    9.8879$& $ 123159.5900\,\pm\,$ & $
14.2165$& $ 121001.7950\,\pm\,$ & $   12.6325$\\
          1024& $ 480361.5500\,\pm\,$ & $   40.1000$& $
464676.1500\,\pm\,$ & $   60.9749$& $ 453086.4700\,\pm\,$ & $
88.8330$& $ 443892.0650\,\pm\,$ & $   78.1385$\\
\hline
        \end{tabular}
      }
  \caption{\label{static C1 data} Static data for
$C_1$, from the Monte Carlo simulations at the
critical point of the 2-dimensional random cluster model, as a
function of $L$ and $q$,
    obtained by combining the data for all available $k$ values for
each $q$. The quoted error bar corresponds to one standard deviation.
  }
  \medskip
\end{table}

\begin{table}
\centering
      {\footnotesize
	\begin{tabular}{rr@{}lr@{}lr@{}lr@{}l}
	  \hline
	  $L$ & \multicolumn{2}{c}{$q=1.25$} & \multicolumn{2}{c}{$q=1.50$} & \multicolumn{2}{c}{$q=1.75$}& \multicolumn{2}{c}{$q=2.00$} \\
	  \hline
	  16& $     1.3601\,\pm\,$ & $    0.0008$& $     2.6552\,\pm\,$ & $    0.0022$& $     4.5086\,\pm\,$ & $    0.0048$& $     7.0555\,\pm\,$ & $    0.0093$\\
	  32& $     1.4516\,\pm\,$ & $    0.0009$& $     2.9749\,\pm\,$ & $    0.0026$& $     5.2959\,\pm\,$ & $    0.0060$& $     8.6748\,\pm\,$ & $    0.0127$\\
	  64& $     1.5433\,\pm\,$ & $    0.0010$& $     3.2988\,\pm\,$ & $    0.0030$& $     6.1255\,\pm\,$ & $    0.0075$& $    10.4931\,\pm\,$ & $    0.0167$\\
	  128& $     1.6225\,\pm\,$ & $    0.0011$& $     3.6209\,\pm\,$ & $    0.0034$& $     7.0060\,\pm\,$ & $    0.0093$& $    12.4925\,\pm\,$ & $    0.0217$\\
	  256& $     1.7014\,\pm\,$ & $    0.0012$& $     3.9284\,\pm\,$ & $    0.0039$& $     7.8884\,\pm\,$ & $    0.0110$& $    14.7173\,\pm\,$ & $    0.0278$\\
	  512& $     1.7679\,\pm\,$ & $    0.0016$& $     4.2341\,\pm\,$ & $    0.0060$& $     8.8097\,\pm\,$ & $    0.0176$& $    17.0928\,\pm\,$ & $    0.0493$\\
	  1024& $     1.8281\,\pm\,$ & $    0.0028$& $     4.5295\,\pm\,$ & $    0.0109$& $     9.8139\,\pm\,$ & $    0.0340$& $    19.8144\,\pm\,$ & $    0.0982$\\
\hline
	\end{tabular}
      }
  \caption{\label{dynamic EP data} 
   Estimates of $\tau_{\text{int},\E}$ from the Monte Carlo simulations at the critical point of the 2-dimensional random cluster model, as a function of $L$ and $q$,
   for $k=1$.
     The quoted error bar corresponds to one standard deviation.
  }
  \medskip
\end{table}

\begin{table}
\centering
      {\footnotesize
	\begin{tabular}{rr@{}lr@{}lr@{}lr@{}l}
	  \hline
	  $L$ & \multicolumn{2}{c}{$q=1.25$} & \multicolumn{2}{c}{$q=1.50$} & \multicolumn{2}{c}{$q=1.75$}& \multicolumn{2}{c}{$q=2.00$} \\
	  \hline
	  16& $     1.2040\,\pm\,$ & $    0.0007$& $     2.2847\,\pm\,$ & $    0.0019$& $     3.8600\,\pm\,$ & $    0.0041$& $     6.0689\,\pm\,$ & $    0.0080$\\
	  32& $     1.2656\,\pm\,$ & $    0.0008$& $     2.5269\,\pm\,$ & $    0.0022$& $     4.5125\,\pm\,$ & $    0.0052$& $     7.4842\,\pm\,$ & $    0.0110$\\
	  64& $     1.3202\,\pm\,$ & $    0.0009$& $     2.7598\,\pm\,$ & $    0.0025$& $     5.1783\,\pm\,$ & $    0.0063$& $     9.0496\,\pm\,$ & $    0.0144$\\
	  128& $     1.3614\,\pm\,$ & $    0.0009$& $     2.9810\,\pm\,$ & $    0.0028$& $     5.8732\,\pm\,$ & $    0.0078$& $    10.7689\,\pm\,$ & $    0.0187$\\
	  256& $     1.4015\,\pm\,$ & $    0.0010$& $     3.1840\,\pm\,$ & $    0.0032$& $     6.5651\,\pm\,$ & $    0.0092$& $    12.6943\,\pm\,$ & $    0.0240$\\
	  512& $     1.4325\,\pm\,$ & $    0.0013$& $     3.3804\,\pm\,$ & $    0.0048$& $     7.2791\,\pm\,$ & $    0.0146$& $    14.7495\,\pm\,$ & $    0.0425$\\
	  1024& $     1.4560\,\pm\,$ & $    0.0022$& $     3.5685\,\pm\,$ & $    0.0086$& $     8.0614\,\pm\,$ & $    0.0279$& $    17.1123\,\pm\,$ & $    0.0848$\\
\hline
	\end{tabular}
      }
  \caption{\label{dynamic N data} 
      Estimates of $\tau_{\text{int},\N}$ from the Monte Carlo simulations at the critical point of the 2-dimensional random cluster model, as a function of $L$ and $q$,
    for $k=1$. The quoted error bar corresponds to one standard deviation.
  }
  \medskip
\end{table}

\begin{table}
\centering
      {\footnotesize
	\begin{tabular}{rr@{}lr@{}lr@{}lr@{}l}
	  \hline
	  $L$ & \multicolumn{2}{c}{$q=1.25$} & \multicolumn{2}{c}{$q=1.50$} & \multicolumn{2}{c}{$q=1.75$}& \multicolumn{2}{c}{$q=2.00$} \\
	  \hline
	  16& $     1.3386\,\pm\,$ & $    0.0008$& $     2.6034\,\pm\,$ & $    0.0021$& $     4.4205\,\pm\,$ & $    0.0047$& $     6.9286\,\pm\,$ & $    0.0091$\\
	  32& $     1.4078\,\pm\,$ & $    0.0009$& $     2.8602\,\pm\,$ & $    0.0025$& $     5.0735\,\pm\,$ & $    0.0058$& $     8.3286\,\pm\,$ & $    0.0122$\\
	  64& $     1.4676\,\pm\,$ & $    0.0010$& $     3.0892\,\pm\,$ & $    0.0028$& $     5.7122\,\pm\,$ & $    0.0070$& $     9.8003\,\pm\,$ & $    0.0156$\\
	  128& $     1.5118\,\pm\,$ & $    0.0010$& $     3.2941\,\pm\,$ & $    0.0031$& $     6.3331\,\pm\,$ & $    0.0084$& $    11.2921\,\pm\,$ & $    0.0196$\\
	  256& $     1.5495\,\pm\,$ & $    0.0011$& $     3.4664\,\pm\,$ & $    0.0034$& $     6.8843\,\pm\,$ & $    0.0096$& $    12.8477\,\pm\,$ & $    0.0243$\\
	  512& $     1.5770\,\pm\,$ & $    0.0015$& $     3.6264\,\pm\,$ & $    0.0051$& $     7.4316\,\pm\,$ & $    0.0149$& $    14.3819\,\pm\,$ & $    0.0415$\\
	  1024& $     1.6009\,\pm\,$ & $    0.0024$& $     3.7573\,\pm\,$ & $    0.0090$& $     7.9652\,\pm\,$ & $    0.0276$& $    16.0737\,\pm\,$ & $    0.0796$\\
\hline
	\end{tabular}
      }
  \caption{\label{dynamic S2 data} 
      Estimates of $\tau_{\text{int},\mathcal{S}_2}$ from the Monte Carlo simulations at the critical point of the 2-dimensional random cluster model, as a function of $L$ and $q$,
    for $k=1$. The quoted error bar corresponds to one standard deviation.
  }
  \medskip
\end{table}

\begin{table}
\centering
      {\footnotesize
	\begin{tabular}{rr@{}lr@{}lr@{}lr@{}l}
	  \hline
	  $L$ & \multicolumn{2}{c}{$q=1.25$} & \multicolumn{2}{c}{$q=1.50$} & \multicolumn{2}{c}{$q=1.75$}& \multicolumn{2}{c}{$q=2.00$} \\
	  \hline
	  16& $     1.1821\,\pm\,$ & $    0.0007$& $     2.1678\,\pm\,$ & $    0.0018$& $     3.5450\,\pm\,$ & $    0.0038$& $     5.4257\,\pm\,$ & $    0.0072$\\
	  32& $     1.2066\,\pm\,$ & $    0.0007$& $     2.2716\,\pm\,$ & $    0.0020$& $     3.8255\,\pm\,$ & $    0.0044$& $     6.0790\,\pm\,$ & $    0.0089$\\
	  64& $     1.2285\,\pm\,$ & $    0.0008$& $     2.3601\,\pm\,$ & $    0.0021$& $     4.1069\,\pm\,$ & $    0.0050$& $     6.7601\,\pm\,$ & $    0.0108$\\
	  128& $     1.2433\,\pm\,$ & $    0.0008$& $     2.4422\,\pm\,$ & $    0.0023$& $     4.3807\,\pm\,$ & $    0.0058$& $     7.4396\,\pm\,$ & $    0.0129$\\
	  256& $     1.2563\,\pm\,$ & $    0.0009$& $     2.5088\,\pm\,$ & $    0.0025$& $     4.6063\,\pm\,$ & $    0.0064$& $     8.1526\,\pm\,$ & $    0.0154$\\
	  512& $     1.2672\,\pm\,$ & $    0.0012$& $     2.5784\,\pm\,$ & $    0.0036$& $     4.8445\,\pm\,$ & $    0.0097$& $     8.8373\,\pm\,$ & $    0.0255$\\
	  1024& $     1.2756\,\pm\,$ & $    0.0019$& $     2.6249\,\pm\,$ & $    0.0063$& $     5.0706\,\pm\,$ & $    0.0176$& $     9.5914\,\pm\,$ & $    0.0475$\\
\hline
	\end{tabular}
      }
  \caption{\label{dynamic FP data} 
      Estimates of $\tau_{\text{int},\mathcal{F}'}$ from the Monte Carlo simulations at the critical point of the 2-dimensional random cluster model, as a function of $L$ and $q$,
    for $k=1$. The quoted error bar corresponds to one standard deviation.
  }
  \medskip
\end{table}

\begin{table}
\centering
      {\footnotesize
	\begin{tabular}{rr@{}lr@{}lr@{}lr@{}l}
	  \hline
	  $L$ & \multicolumn{2}{c}{$q=1.25$} & \multicolumn{2}{c}{$q=1.50$} & \multicolumn{2}{c}{$q=1.75$}& \multicolumn{2}{c}{$q=2.00$} \\
	  \hline
	  16& $     1.3101\,\pm\,$ & $    0.0008$& $     2.5278\,\pm\,$ & $    0.0021$& $     4.2775\,\pm\,$ & $    0.0046$& $     6.6989\,\pm\,$ & $    0.0088$\\
	  32& $     1.3708\,\pm\,$ & $    0.0008$& $     2.7571\,\pm\,$ & $    0.0024$& $     4.8684\,\pm\,$ & $    0.0056$& $     7.9857\,\pm\,$ & $    0.0117$\\
	  64& $     1.4224\,\pm\,$ & $    0.0009$& $     2.9585\,\pm\,$ & $    0.0027$& $     5.4412\,\pm\,$ & $    0.0067$& $     9.3266\,\pm\,$ & $    0.0149$\\
	  128& $     1.4596\,\pm\,$ & $    0.0009$& $     3.1373\,\pm\,$ & $    0.0030$& $     5.9956\,\pm\,$ & $    0.0079$& $    10.6786\,\pm\,$ & $    0.0186$\\
	  256& $     1.4918\,\pm\,$ & $    0.0010$& $     3.2868\,\pm\,$ & $    0.0033$& $     6.4856\,\pm\,$ & $    0.0091$& $    12.0813\,\pm\,$ & $    0.0229$\\
	  512& $     1.5155\,\pm\,$ & $    0.0014$& $     3.4287\,\pm\,$ & $    0.0048$& $     6.9711\,\pm\,$ & $    0.0139$& $    13.4617\,\pm\,$ & $    0.0388$\\
	  1024& $     1.5364\,\pm\,$ & $    0.0023$& $     3.5411\,\pm\,$ & $    0.0085$& $     7.4506\,\pm\,$ & $    0.0258$& $    14.9718\,\pm\,$ & $    0.0742$\\
\hline
	\end{tabular}
      }
  \caption{\label{dynamic C1 data} 
      Estimates of $\tau_{\text{int},\mathcal{C}_1}$ from the Monte Carlo simulations at the critical point of the 2-dimensional random cluster model, as a function of $L$ and $q$,
    for $k=1$. The quoted error bar corresponds to one standard deviation.
  }
  \medskip
\end{table}

\begin{table}
\centering
      {\footnotesize
	\begin{tabular}{rr@{}lr@{}lr@{}lr@{}l}
	  \hline
	  $L$ & \multicolumn{2}{c}{$q=1.25$} & \multicolumn{2}{c}{$q=1.50$} & \multicolumn{2}{c}{$q=1.75$}& \multicolumn{2}{c}{$q=2.00$} \\
	  \hline
	  16& $     1.0547\,\pm\,$ & $    0.0007$& $     1.8490\,\pm\,$ & $    0.0015$& $     2.9362\,\pm\,$ & $    0.0031$& $     4.3973\,\pm\,$ & $    0.0058$\\
	  32& $     1.0744\,\pm\,$ & $    0.0007$& $     1.9293\,\pm\,$ & $    0.0017$& $     3.1495\,\pm\,$ & $    0.0036$& $     4.8826\,\pm\,$ & $    0.0071$\\
	  64& $     1.0920\,\pm\,$ & $    0.0007$& $     1.9968\,\pm\,$ & $    0.0018$& $     3.3571\,\pm\,$ & $    0.0041$& $     5.3808\,\pm\,$ & $    0.0086$\\
	  128& $     1.1020\,\pm\,$ & $    0.0007$& $     2.0563\,\pm\,$ & $    0.0020$& $     3.5553\,\pm\,$ & $    0.0047$& $     5.8582\,\pm\,$ & $    0.0102$\\
	  256& $     1.1120\,\pm\,$ & $    0.0008$& $     2.1031\,\pm\,$ & $    0.0021$& $     3.7110\,\pm\,$ & $    0.0052$& $     6.3568\,\pm\,$ & $    0.0120$\\
	  512& $     1.1195\,\pm\,$ & $    0.0010$& $     2.1560\,\pm\,$ & $    0.0030$& $     3.8788\,\pm\,$ & $    0.0078$& $     6.8290\,\pm\,$ & $    0.0197$\\
	  1024& $     1.1265\,\pm\,$ & $    0.0017$& $     2.1835\,\pm\,$ & $    0.0052$& $     4.0403\,\pm\,$ & $    0.0140$& $     7.3298\,\pm\,$ & $    0.0363$\\
\hline
	\end{tabular}
      }
  \caption{\label{dynamic C2 data} 
      Estimates of $\tau_{\text{int},\mathcal{C}_2}$ from the Monte Carlo simulations at the critical point of the 2-dimensional random cluster model, as a function of $L$ and $q$,
    for $k=1$. The quoted error bar corresponds to one standard deviation.
  }
  \medskip
\end{table}

\begin{table}
\centering
      {\footnotesize
	\begin{tabular}{rr@{}lr@{}lr@{}lr@{}l}
	  \hline
	  $L$ & \multicolumn{2}{c}{$q=1.25$} & \multicolumn{2}{c}{$q=1.50$} & \multicolumn{2}{c}{$q=1.75$}& \multicolumn{2}{c}{$q=2.00$} \\
	  \hline
	  16& $     1.0627\,\pm\,$ & $    0.0007$& $     1.9037\,\pm\,$ & $    0.0015$& $     3.0988\,\pm\,$ & $    0.0033$& $     4.7417\,\pm\,$ & $    0.0063$\\
	  32& $     1.0882\,\pm\,$ & $    0.0007$& $     2.0070\,\pm\,$ & $    0.0017$& $     3.3763\,\pm\,$ & $    0.0039$& $     5.3802\,\pm\,$ & $    0.0079$\\
	  64& $     1.1096\,\pm\,$ & $    0.0007$& $     2.0948\,\pm\,$ & $    0.0019$& $     3.6466\,\pm\,$ & $    0.0045$& $     6.0321\,\pm\,$ & $    0.0096$\\
	  128& $     1.1247\,\pm\,$ & $    0.0007$& $     2.1761\,\pm\,$ & $    0.0021$& $     3.9002\,\pm\,$ & $    0.0051$& $     6.6878\,\pm\,$ & $    0.0116$\\
	  256& $     1.1379\,\pm\,$ & $    0.0008$& $     2.2400\,\pm\,$ & $    0.0022$& $     4.1293\,\pm\,$ & $    0.0058$& $     7.3441\,\pm\,$ & $    0.0139$\\
	  512& $     1.1487\,\pm\,$ & $    0.0011$& $     2.3012\,\pm\,$ & $    0.0032$& $     4.3459\,\pm\,$ & $    0.0087$& $     7.9879\,\pm\,$ & $    0.0230$\\
	  1024& $     1.1573\,\pm\,$ & $    0.0018$& $     2.3518\,\pm\,$ & $    0.0056$& $     4.5466\,\pm\,$ & $    0.0158$& $     8.6848\,\pm\,$ & $    0.0430$\\
\hline
	\end{tabular}
      }
  \caption{\label{dynamic C3 data} 
       Estimates of $\tau_{\text{int},\mathcal{C}_3}$ from the Monte Carlo simulations at the critical point of the 2-dimensional random cluster model, as a function of $L$ and $q$,
    for $k=1$. The quoted error bar corresponds to one standard deviation.
  }
  \medskip
\end{table}

%
%
  \begin{sidewaystable}[ht]
    \centering
            {\footnotesize
              \begin{tabular}{|c|clr@{\ }c@{\ }c|c@{\ }l@{\ \ }l@{\ }rc@{\ }c@{\ }c|c|}
\cline{2-13}
   \multicolumn{1}{c}{\quad}
      & \multicolumn{5}{|c|}{Fits to $\xi/L = x^\star$}
      & \multicolumn{7}{|c|}{Fits to $\xi/L = x^\star + B L^{-p}$} &
   \multicolumn{1}{c}{\quad}  \\
\hline
   $q$ & $L_{\min}$ & \multicolumn{1}{c}{$x^\star$}
         & \multicolumn{1}{c}{$\chi^2$} & DF & CL(\%)
         & $L_{\min}$ & \multicolumn{1}{c}{$x^\star$}
         & \multicolumn{1}{c}{$B$} & \multicolumn{1}{c}{$p$}
         & \multicolumn{1}{c}{$\chi^2$} & DF & CL(\%) & $\Delta_1$ (exact) \\
                \hline
  1.25 & 256 &   0.91734(8) &    1.48 & 2 & 48 & 16 &   0.91735(7) &
0.900(243) &   1.81(10) &    2.53 & 4 & 64 &    1.80702\\
  1.50 & 128 &   0.90867(9) &    0.17 & 3 & 98 & 16 &   0.90857(10) &
 0.563(173) &   1.62(11) &    0.77 & 4 & 94 &    1.63551\\
  1.75 & 128 &   0.90496(12) &    5.19 & 3 & 16 & 16 &   0.90487(14) &
  0.582(228) &   1.62(14) &    6.93 & 4 & 14 &    1.47905\\
  2.00 & 512 &   0.90442(20) &    0.54 & 1 & 46 & 32 &   0.90387(74) &
  0.017(9) &   0.49(21) &    4.12 & 3 & 25 &    1.33333\\
  2.25 & 256 &   0.90786(17) &    1.30 & 2 & 52 & [16] &   0.90799(16)
&   0.465(184) &   1.55(15) &    8.19 & 4 & 8 &    1.19517\\
  2.50 & 64 &   0.91369(14) &    3.36 & 4 & 50 & 16 &   0.91335(21) &
 0.396(195) &   1.50(18) &    3.69 & 4 & 45 &    1.06191\\
\hline
              \end{tabular}
            }
\vspace*{5mm}
    \caption{
      Fits for $\xi/L$.
      The quoted error bar corresponds to one standard deviation.
      $\lmin$ in brackets indicates a poor fit (confidence level $< 10\%$).
    }
\label{table_xioverL_analysis}
  \end{sidewaystable}

\clearpage

%
%
  \begin{table}
    \centering
            {\small
              \begin{tabular}{|c|cllr@{\ }c@{\ }c|c@{\ \ }r@{\ \ }r|}
\cline{2-7}
   \multicolumn{1}{c}{\quad}
      & \multicolumn{6}{|c|}{Fits to $\chi = AL^{\gamma/\nu}$}
      & \multicolumn{3}{c}{\quad}  \\
\hline
   $q$ & $L_{\min}$ & \multicolumn{1}{c}{$\gamma/\nu$} & \multicolumn{1}{c}{$A$}
         & \multicolumn{1}{c}{$\chi^2$} & DF & CL(\%)
         & $\gamma/\nu$ (exact) & deviation & dev($\sigma$) \\
                \hline
  1.25 & 256 &   1.77773(9) &   1.0496(6) &    1.40 & 1 & 24 &
1.77764 &    0.00009 &    1.0\\
  1.50 & 128 &   1.76639(8) &   1.0645(5) &    0.30 & 2 & 86 &
1.76644 &   $-$0.00005 &   $-$0.6\\
  1.75 & 128 &   1.75759(12) &   1.0768(7) &    2.34 & 2 & 31 &
1.75738 &    0.00022 &    1.8\\
  2.00 & 128 &   1.74969(11) &   1.0937(6) &    2.54 & 2 & 28 &
1.75000 &   $-$0.00031 &   $-$2.9\\
  2.25 & 256 &   1.74428(26) &   1.1039(17) &    1.37 & 1 & 24 &
1.74404 &    0.00024 &    0.9\\
  2.50 & 256 &   1.73902(37) &   1.1209(24) &    0.01 & 1 & 94 &
1.73932 &   $-$0.00030 &   $-$0.8\\
  2.75 & 128 &   1.73575(28) &   1.1316(17) &    0.85 & 2 & 65 &
1.73576 & $-$0.00001 &  $-$0.03\\
  3.00 & 128 &   1.73438(28) &   1.1371(17) &    0.72 & 2 & 70 &
1.73333 &    0.00105 &    3.8\\
  3.25 & 128 &   1.73341(37) &   1.1460(23) &    0.30 & 2 & 86 &
1.73214 &    0.00127 &    3.4\\
  3.50 & 256 &   1.73304(99) &   1.1571(67) &    0.01 & 1 & 93 &
1.73246 &    0.00058 &    0.6\\
  3.75 & 128 &   1.73745(67) &   1.1431(40) &    1.46 & 2 & 48 &
1.73517 &    0.00229 &    3.4\\
  4.00 & 256 &   1.73968(152) &   1.1502(101) &    0.44 & 1 & 51 &
1.75000 &   $-$0.01032 &   $-$6.8\\
\hline
              \end{tabular}
            }
\vspace*{5mm}
    \caption{
      Fits of the susceptibility $\chi$ to a pure power law.
      The quoted error bar corresponds to one standard deviation.
      $\lmin$ in brackets indicates a poor fit (confidence level $< 10\%$).
      The final two columns show the deviation of the estimated $\gamma/\nu$
      from the known exact value, in absolute terms and in
      units of its standard deviation.
    }
\label{table_chi_analysis}
  \end{table}

\clearpage

%
%
  \begin{sidewaystable}[ht]
    \centering
            {\footnotesize
              \begin{tabular}{|c|clr@{\ }c@{\ }c|c@{\ }l@{\ \ }l@{\ }rc@{\ }c@{\ }c|c|}
\cline{2-13}
   \multicolumn{1}{c}{\quad}
      & \multicolumn{5}{|c|}{Fits to $\chi/L^{\gamma/\nu} = A$}
      & \multicolumn{7}{|c|}{Fits to $\chi/L^{\gamma/\nu} = A + B L^{-p}$} &
   \multicolumn{1}{c}{\quad}  \\
\hline
   $q$ & $L_{\min}$ & \multicolumn{1}{c}{$A$}
         & \multicolumn{1}{c}{$\chi^2$} & DF & CL(\%)
         & $L_{\min}$ & \multicolumn{1}{c}{$A$}
         & \multicolumn{1}{c}{$B$} & \multicolumn{1}{c}{$p$}
         & \multicolumn{1}{c}{$\chi^2$} & DF & CL(\%) & $\Delta_1$ (exact) \\
                \hline
  1.25 & 256 &   1.05017(5) &    2.38 & 2 & 30 & 16 &   1.05013(8) &
0.011(5) &   0.87(19) &    4.72 & 4 & 32 &    1.80702\\
  1.50 & 128 &   1.06417(6) &    0.70 & 3 & 87 & 16 &   1.06379(53) &
0.003(1) &   0.34(30) &    1.55 & 4 & 82 &    1.63551\\
  1.75 & 64 &   1.07811(7) &    7.16 & 4 & 13 & --- &   --- &      ---
&   --- &    --- & --- & --- &    1.47905\\
  2.00 & 512 &   1.09148(15) &    0.50 & 1 & 48 & [16] &   1.09195(8)
&  $-$0.231(316) &   1.88(50) &   15.63 & 4 &   0.4 &    1.33333\\
  2.25 & 256 &   1.10545(13) &    2.19 & 2 & 33 & [16] &   1.10565(15)
&  $-$0.078(43) &   1.22(21) &    7.84 & 4 & 10 &    1.19517\\
  2.50 & 128 &   1.11875(14) &    2.34 & 3 & 51 & 16 &   1.11912(26) &
 $-$0.062(22) &   0.94(14) &    3.96 & 4 & 41 &    1.06191\\
\hline
              \end{tabular}
            }
\vspace*{5mm}
    \caption{
      Fits for $\chi/L^{\gamma/\nu}$ with $\gamma/\nu$ set to its
      exact value \reff{eq.gammaovernu.exact}.
      The quoted error bar corresponds to one standard deviation.
      $\lmin$ in brackets indicates a poor fit (confidence level $< 10\%$).
    }
\label{table_chi_analysis_corrections}
  \end{sidewaystable}

\clearpage

%
%
  \begin{table}
    \centering
            {\small
              \begin{tabular}{|c|cllr@{\ }c@{\ }c|c@{\ \ }r@{\ \ }r|}
\cline{2-7}
   \multicolumn{1}{c}{\quad}
      & \multicolumn{6}{|c|}{Fits to $C_1 = AL^{d-\beta/\nu}$}
      & \multicolumn{3}{c}{\quad}  \\
\hline
   $q$ & $L_{\min}$ & \multicolumn{1}{c}{$\beta/\nu$} & \multicolumn{1}{c}{$A$}
         & \multicolumn{1}{c}{$\chi^2$} & DF & CL(\%)
         & $\beta/\nu$ (exact) & deviation & dev($\sigma$) \\
                \hline
  1.25 & 256 &   0.11111(6) &   0.9896(3) &    1.29 & 1 & 26 &
0.11118 &   $-$0.00007 &   $-$1.2\\
  1.50 & 16 &   0.11681(2) &   0.9958(1) &    2.01 & 5 & 85 &
0.11678 &    0.00003 &    1.6\\
  1.75 & 128 &   0.12115(8) &   1.0004(4) &    2.21 & 2 & 33 &
0.12131 &   $-$0.00016 &   $-$2.1\\
  2.00 & 128 &   0.12518(7) &   1.0080(4) &    2.04 & 2 & 36 &
0.12500 &    0.00018 &    2.7\\
  2.25 & 256 &   0.12787(17) &   1.0118(10) &    1.20 & 1 & 27 &
0.12798 &   $-$0.00011 &   $-$0.6\\
  2.50 & 128 &   0.13033(13) &   1.0177(7) &    1.48 & 2 & 48 &
0.13034 &   $-$0.00001 &  $-$0.10\\
  2.75 & 128 &   0.13230(18) &   1.0237(10) &    1.65 & 2 & 44 &
0.13212 &    0.00018 &    1.0\\
  3.00 & 64 &   0.13283(11) &   1.0245(5) &    1.24 & 3 & 74 &
0.13333 &   $-$0.00050 &   $-$4.6\\
  3.25 & 128 &   0.13366(24) &   1.0295(13) &    0.86 & 2 & 65 &
0.13393 &   $-$0.00027 &   $-$1.1\\
  3.50 & 64 &   0.13315(19) &   1.0289(9) &    2.39 & 3 & 50 &
0.13377 &   $-$0.00062 &   $-$3.3\\
  3.75 & 32 &   0.13174(15) &   1.0263(7) &    1.74 & 4 & 78 &
0.13242 &   $-$0.00067 &   $-$4.4\\
  4.00 & 256 &   0.13200(104) &   1.0357(62) &    0.54 & 1 & 46 &
0.12500 &    0.00700 &    6.8\\
\hline
              \end{tabular}
            }
\vspace*{5mm}
    \caption{
      Fits of the mean size $C_1$ of the largest cluster to a pure power law.
      The quoted error bar corresponds to one standard deviation.
      $\lmin$ in brackets indicates a poor fit (confidence level $< 10\%$).
      The final two columns show the deviation of the estimated $\beta/\nu$
      from the known exact value, in absolute terms and in
      units of its standard deviation.
    }
\label{table_C1_analysis}
  \end{table}

\clearpage

%
%
  \begin{sidewaystable}[ht]
    \hspace*{-8mm}
            {\footnotesize
              \begin{tabular}{|c|cllr@{\,}c@{\,}c|c@{\,}l@{\,\,}l@{\,}rc@{\,}c@{\,}c|c@{\ }r@{\,}r|}
\cline{2-14}
   \multicolumn{1}{c}{\quad}
      & \multicolumn{6}{|c|}{Fits to $C_H = AL^{\alpha/\nu}$}
      & \multicolumn{7}{|c|}{Fits to $C_H = AL^{\alpha/\nu} + B$} &
   \multicolumn{3}{c}{\quad}  \\
\hline
   $q$ & $L_{\min}$ & \multicolumn{1}{c}{$\alpha/\nu$} & \multicolumn{1}{c}{$A$}
         & \multicolumn{1}{c}{$\chi^2$} & DF & CL(\%)
         & $L_{\min}$ & \multicolumn{1}{c}{$\alpha/\nu$}
         & \multicolumn{1}{c}{$A$} & \multicolumn{1}{c}{$B$}
         & \multicolumn{1}{c}{$\chi^2$} & DF & CL(\%) & $\alpha/\nu$ (exact)
         & \multicolumn{1}{c}{dev} & dev($\sigma$) \\
                \hline
  1.25 & [256] &   0.0406(9) &   0.588(3) &   16.40 & 1 & $5 \times
10^{-3}$ & 16 &  $-$0.3675(49) &  $-$0.791(6) &   0.839(2) &    1.82 &
4 & 77 &   $-$0.35527 &   $-$0.012 &   $-$2.5\\
  1.50 & [256] &   0.0712(8) &   1.036(5) &   30.87 & 1 & $3 \times
10^{-6}$ & 128 &  $-$0.1977(186) &  $-$1.933(23) &   2.182(59) &
0.37 & 1 & 54 &   $-$0.22663 &    0.029 &    1.6\\
  1.75 & [256] &   0.1089(8) &   1.366(7) &   31.62 & 1 & $2 \times
10^{-6}$ & 128 &  $-$0.0965(135) &  $-$5.413(347) &   5.665(440) &
0.15 & 1 & 70 &   $-$0.10929 &    0.013 &    0.9\\
  2.00 & [256] &   0.1569(6) &   1.554(6) &   43.28 & 1 & $5 \times
10^{-9}$ & (see text) &   \multicolumn{1}{c}{---} &  \multicolumn{1}{c}{---} & \multicolumn{1}{c}{---} &   \multicolumn{1}{c}{---} & \multicolumn{1}{c}{---} & \multicolumn{1}{c|}{---} &
0($\log$) &    \multicolumn{1}{c}{---} &    \multicolumn{1}{c|}{---} \\
  2.25 & [256] &   0.2143(8) &   1.608(8) &   36.65 & 1 & $1 \times
10^{-7}$ & 128 &   0.1108(65) &   5.856(558) &  $-$5.554(640) &
0.45 & 1 & 50 &    0.10363 &    0.007 &    1.1\\
  2.50 & [256] &   0.2778(10) &   1.573(10) &    9.47 & 1 &   0.2 & 64
&   0.2025(32) &   3.421(109) &  $-$3.176(150) &    0.73 & 2 & 69 &
0.20357 &   $-$0.001 &   $-$0.3\\
\hline
              \end{tabular}
            }
\vspace*{5mm}
    \caption{
      Fits for the specific heat $C_H$.
      The quoted error bar corresponds to one standard deviation.
      $\lmin$ in brackets indicates a poor fit (confidence level $< 10\%$).
      The final two columns show the deviation of the estimated $\alpha/\nu$
      (from the fit $AL^{\alpha/\nu} + B$) from the known exact value,
      in absolute terms and in units of its standard deviation.
    }
\label{table_C_H_analysis}
  \end{sidewaystable}

\clearpage

%
%
\begin{table}[t]
\begin{center}
\begin{tabular}{rll}
\hline
     \multicolumn{1}{c}{$L$}   &  \multicolumn{1}{c}{$q=2$}
                               &   \multicolumn{1}{c}{$q=3.25$} \\
\hline
   16     &   2.143(3)   &   2.020(7)\    \\
   32     &   2.123(4)   &   2.013(9)\    \\
   64     &   2.100(4)   &   1.990(11)  \\
  128  &   2.080(4)      &   2.015(14)  \\
  256  &   2.069(5)      &   1.991(17)  \\
  512  &   2.047(7)      &   2.034(30)  \\
 1024 &   2.051(12)$\!$      &   1.953(56)  \\
\hline
\end{tabular}
\end{center}
\caption{
   Ratios
   $\tau_{{\rm int},\mathcal{E}'}(k=1) / \tau_{{\rm int},\mathcal{E}'}(k=2)$
   as a function of $L$
   for $q=2$ and $q=3.25$.
   Error bars are one standard deviation.
}
\label{table_kratios_examples}
\end{table}
%

%
%
\begin{table}
\begin{center}
\begin{tabular}{|ccc|c|ccc|}
\hline
  \multicolumn{7}{|c|}{Fits to
  $\tau_{{\rm int},\mathcal{E}'}(1) / \tau_{{\rm int},\mathcal{E}'}(k) = A$} \\
\hline
  $q$      & $k$   & $\lmin$  & $A$ &  $\chi^2$   & DF &  CL(\%)   \\
\hline
 2.00 & 2 & 512 &   2.048(6) & 0.09 & 1 & 76 \\
  2.25 & 2 & 256 &   2.036(5) & 0.51 & 2 & 77 \\
  2.50 & 2 & 128 &   2.022(5) & 0.02 & 3 & 100 \\
  2.75 & 2 & 128 &   2.006(6) & 0.87 & 3 & 83 \\
  3.00 & 2 & 32 &   2.015(5) & 3.63 & 5 & 60 \\
  3.25 & 2 & 32 &   2.005(6) & 5.40 & 5 & 37 \\
  3.50 & 2 & 16 &   2.015(5) & 6.74 & 6 & 35 \\
  3.75 & 2 & 16 &   2.014(6) & 1.68 & 6 & 95 \\
  4.00 & 2 & 16 &   2.011(7) & 2.75 & 6 & 84 \\
\hline
  3.00 & 3 & 256 &   3.011(16) & 0.66 & 2 & 72 \\
  3.25 & 3 & 64 &   3.012(11) & 3.18 & 4 & 53 \\
  3.50 & 3 & 64 &   3.014(13) & 3.67 & 4 & 45 \\
  3.75 & 3 & 16 &   3.034(9) & 2.32 & 6 & 89 \\
  4.00 & 3 & 64 &   2.988(19) & 1.66 & 4 & 80 \\
\hline
  4.00 & 4 & 64 &   4.002(25) & 3.01 & 4 & 56 \\
\hline
\end{tabular}
\end{center}
\caption{
   Estimates for the limiting ratios
   $\tau_{{\rm int},\mathcal{E}'}(1) / \tau_{{\rm int},\mathcal{E}'}(k)$
   as $L \to\infty$
   from fits to a constant $A$.
   Error bars are one standard deviation.
}
\label{table_kratios_fits}
\end{table}

\clearpage

\begin{table}
\begin{center}
\begin{tabular}{|ccc|cc|ccc|}
\hline
   \multicolumn{8}{|c|}{Fits to $\tau_{{\rm int},\E} = AL^z$} \\
\hline
  $q$  & $k$ & $\lmin$  &  $z_{{\rm int},\E}$  &  $A$  &
                                             $\chi^2$   & DF &  CL(\%)  \\
\hline\hline
  1.25 & 1 & [256] &   0.053(1) &   1.27(1) & 3.76 & 1 & 5   \\
\hline
  1.50 & 1 & [256] &   0.104(2) &   2.20(2) & 3.84 & 1 & 5   \\
\hline
  1.75 & 1 & 256 &   0.158(2) &   3.28(5) & 0.21 & 1 & 65   \\
\hline
  2.00 & 1 & 256 &   0.215(3) &   4.47(9) & 0.06 & 1 & 81   \\
  2.00 & 2 & [256] &   0.224(2) &   2.06(3) & 6.91 & 1 &   0.9   \\
\hline
  2.25 & 1 & 256 &   0.286(4) &   5.35(14) & 2.16 & 1 & 14   \\
\hline
\end{tabular}
\end{center}
\caption{
   Estimates for $z_{{\rm int},\E}$ from fits to
   $\tau_{{\rm int},\mathcal{E}'} = A L^z$.
   Error bars are one standard deviation.
   $\lmin$ in brackets indicates a poor fit (confidence level $< 10\%$).
}
\label{table_fits_ALz}
\end{table}

\begin{table}
\begin{center}
\begin{tabular}{|ccc|ccc|ccc|}
\hline
   \multicolumn{9}{|c|}{Fits to $\tau_{{\rm int},\E} = AL^z + B$} \\
\hline
  $q$  & $k$ & $\lmin$  &  $z_{{\rm int},\E}$  &  $A$  &  $B$  &
                                             $\chi^2$   & DF &  CL(\%)  \\
\hline\hline
  1.25 & 1 & 128 &  $-$0.213(35) &  $-$1.61(5) & 2.19(8) & 0.49 & 1 & 48  \\
\hline
  1.50 & 1 & 32 &  $-$0.034(7) &  $-$15.86(2.92) & 17.09(2.96) & 0.82 & 3 & 85  \\
\hline
  1.75 & 1 & 128 &  0.085(26) &  9.66(4.42) & $-$7.56(4.82) & 0.33 & 1 & 57  \\
\hline
  2.00 & 1 & 16 &   0.145(4) &  10.37(51) &  $-$8.44(57) & 2.32 & 4 & 68  \\
  2.00 & 2 & 32 &   0.141(5) &   5.48(32) &  $-$4.83(36) & 1.32 & 3 & 72  \\
\hline
  2.25 & 1 & 32 &   0.235(8) &   8.92(61) &  $-$6.68(86) & 1.62 & 3 & 65  \\
\hline
\end{tabular}
\end{center}
\caption{
   Estimates for $z_{{\rm int},\E}$ from fits to
   $\tau_{{\rm int},\mathcal{E}'} = A L^z + B$.
   Error bars are one standard deviation.
   $\lmin$ in brackets indicates a poor fit (confidence level $< 10\%$).
}
\label{table_fits_ALz+B}
\end{table}

\clearpage

\begin{table}
\begin{center}
\begin{tabular}{|c|cc|cc|c|cc|}
\cline{2-5}\cline{7-8}
   \multicolumn{1}{c}{\quad}
      & \multicolumn{2}{|c|}{Fits to $\tau_{{\rm int},\E} = AL^z$}
      & \multicolumn{2}{|c|}{Fits to $\tau_{{\rm int},\E} = AL^z + B$}
      & \multicolumn{1}{c|}{\quad} & $\alpha/\nu$ & $\beta/\nu$  \\
\cline{1-6}
   $q$ & $\lmin$ & $z$ & $\lmin$ & $z$ & Difference &
                                         (exact) & (exact) \\
\hline\hline
    1.25 & --- &   --- & 128 &  $-$0.213(35) &   --- &
$-$0.35527 &  0.11118 \\
    1.50 & --- &   --- & 32 &   $-$0.034(7) &   --- &
$-$0.22663 &  0.11678 \\
                1.75 & 256 &   0.158(2) & 128 &   0.085(26) &   0.074 &
$-$0.10929 &   0.12131 \\
                2.00 & 256 &   0.215(3) & 16--32 &   0.143(3) &
0.072 &   0.00000 &   0.12500 \\
                2.25 & 256 &   0.286(4) & 32 &   0.235(8) &   0.051 &
 0.10363 &   0.12798 \\
                2.50 & 256 &   0.353(3) & 32--64 &   0.307(7) &
0.046 &   0.20357 &   0.13034 \\
                2.75 & 256 &   0.424(8) & 16--128 &   0.408(5) &
0.016 &   0.30168 &   0.13212 \\
                3.00 & 128--256 &   0.511(3) & 16--32 &   0.497(3) &
0.013 &   0.40000 &   0.13333 \\
                3.25 & 128--256 &   0.587(5) & 32--64 &   0.572(7) &
0.015 &   0.50126 &   0.13393 \\
                3.50 & 16--128 &   0.693(1) & 16--64 &   0.689(4) &
0.004 &   0.61007 &   0.13377 \\
                3.75 & 32--128 &   0.784(2) & 16 &   0.796(4) &
$-$0.012 &   0.73760 &   0.13242 \\
                4.00 & 64--256 &   0.898(4) & 16--128 &   0.910(5) &
$-$0.012 &   1.00000 &   0.12500 \\

\hline
\end{tabular}
\end{center}
\caption{
   Summary of estimates for $z_{{\rm int},\E}$,
   averaged over all values of $k$
   for which the confidence level is $\ge 10\%$.
   Error bars are one standard deviation.
}
\label{table_fits_allk}
\end{table}

\clearpage

\begin{table}[ht]
  \centering
	  {\footnotesize
  \begin{tabular}{|llc|llc|llc|}
    \cline{4-9}
    \multicolumn{3}{c}{} & \multicolumn{3}{|c|}{$\tau/C_H = A\,\log L \,+\, B$}
                         &\multicolumn{3}{|c|}{$\tau/C_H = A\,L^p $} \\
    \hline
    \multicolumn{1}{|c}{$q$}  & $k$ & $L_{\min}$ &
       \multicolumn{1}{c}{$A$}  & \multicolumn{1}{c}{$B$}  &  CL(\%) &
       \multicolumn{1}{c}{$p$}  & \multicolumn{1}{c}{$A$}  &  CL(\%) \\ 
    \hline
  1.75 & 1 & 256 &   0.159(8) &   2.28(5) & 9 &   0.049(3) &   2.41(4) & 14\\
     \hline
  2.00 & 1 & 128 &   0.224(8) &   2.74(4) & 23 &   0.056(2) &   2.91(3) & 44\\
  2.00 & 2 & 128 &   0.125(3) &   1.23(1) & 8 &   0.064(1) &   1.34(1) & 30\\
  2.00 & 2 & 256 &   0.134(5) &   1.18(3) & 59 &   0.067(2) &   1.32(2) & 36\\
     \hline
  2.25 & 1 & 128 &   0.337(13) &   3.11(7) & 22 &   0.068(3) &   3.42(5) & 45\\
  2.25 & 1 & 256 &   0.373(25) &   2.89(14) & 86 &   0.072(5) &   3.33(9) & 69\\
\hline
  \end{tabular}
}
\vspace*{5mm}
\caption{
    Fits for the ratio $\tau_{{\rm int},\mathcal{E}'}/C_H$.
    The quoted error bar corresponds to one standard deviation.
}
   \label{table_tauint_CH_ratio}
\end{table}

\clearpage

\begin{table}
\begin{center}
\begin{tabular}{|ccc|cc|ccc|}
\hline
   \multicolumn{8}{|c|}{Fits to $\tau_{{\rm int},\mathcal{C}_2} = AL^z$} \\
\hline
  $q$  & $k$ & $\lmin$  &  $z_{{\rm int},\mathcal{C}_2}$  &  $A$  &
                                             $\chi^2$   & DF &  CL(\%)  \\
\hline\hline
        1.25 & 1 & 256 &   0.010(1) &   1.05(1) &    0.05 & 1 & 82\\
\hline
        1.50 & 1 & [256] &   0.030(2) &   1.78(2) &   10.00 & 1 & 0.2\\
\hline
        1.75 & 1 & 128 &   0.062(1) &   2.63(2) &    0.39 & 2 & 82\\
\hline
        2.00 & 1 & 256 &   0.103(3) &   3.59(7) &    0.01 & 1 & 91\\
        2.00 & 2 & [256] &   0.114(2) &   1.56(2) &    4.85 & 1 & 3\\
\hline
        2.25 & 1 & 128 &   0.174(3) &   4.08(6) &    0.40 & 2 & 82\\
\hline
\end{tabular}
\end{center}
\caption{
   Estimates for $z_{{\rm int},\mathcal{C}_2}$ from fits to
   $\tau_{{\rm int},\mathcal{C}_2} = A L^z$.
   Error bars are one standard deviation.
   $\lmin$ in brackets indicates a poor fit (confidence level $< 10\%$).
}
\label{table_fits_C2_ALz}
\end{table}

\begin{table}
\begin{center}
\begin{tabular}{|ccc|ccc|ccc|}
\hline
   \multicolumn{9}{|c|}{Fits to $\tau_{{\rm int},\mathcal{C}_2} = AL^z + B$} \\
\hline
  $q$  & $k$ & $\lmin$  &  $z_{{\rm int},\mathcal{C}_2}$  &  $A$  &  $B$  &
                                             $\chi^2$   & DF &  CL(\%)  \\
\hline\hline
  1.25 & 1 & 64 &    $-$0.189(89) &    $-$0.19(1) &     1.18(3) &
0.58 & 2 & 75\\
\hline
  1.50 & 1 & [128] &    $-$0.170(97) &    $-$1.03(4) &     2.51(22) &
  6.63 & 1 & 1\\
\hline
  1.75 & 1 & 128 &     0.047(67) &     3.77(6.81) &    $-$1.19(7.02) &
   0.34 & 1 & 56\\
\hline
  2.00 & 1 & 64 &     0.006(26) &   104.55(437.90) &
$-$102.03(438.13) &    0.85 & 2 & 65\\
  2.00 & 2 & 64 &     0.019(17) &    16.72(16.04) &   $-$15.66(16.12)
&    1.97 & 2 & 37\\
\hline
  2.25 & 1 & 16 &     0.118(8) &     8.25(80) &    $-$5.12(87) &
2.58 & 4 & 63\\
\hline
\end{tabular}
\end{center}
\caption{
   Estimates for $z_{{\rm int},\mathcal{C}_2}$ from fits to
   $\tau_{{\rm int},\mathcal{C}_2} = A L^z + B$.
   Error bars are one standard deviation.
   $\lmin$ in brackets indicates a poor fit (confidence level $< 10\%$).
}
\label{table_fits_C2_ALz+B}
\end{table}

\clearpage

\begin{table}
\begin{center}
\begin{tabular}{|c|ccc|ccc|}
\cline{2-7}
   \multicolumn{1}{c}{\quad}
    & \multicolumn{3}{|c|}{Fits to $\tau_{{\rm int},\mathcal{C}_2} = AL^z$}
    & \multicolumn{3}{|c|}{Fits to $\tau_{{\rm int},\mathcal{C}_2} = AL^z + B$}
             \\
\hline
   $q$ & $\lmin$ & $z$ & $z_{\mathcal{C}_2} - z_{\E}$
       & $\lmin$ & $z$ & $z_{\mathcal{C}_2} - z_{\E}$ \\
\hline\hline
   1.25 & 256 &   0.010(1) &  --- & 64 &  $-$0.189(89) & 0.024 \\
   1.50 & --- & --- & --- & --- & --- &  --- \\
   1.75 & 128 &   0.062(1) &   $-$0.096 & 128 &   0.047(67) &  $-$0.038 \\
   2.00 & 256 &   0.103(3) &   $-$0.112 & 64 &   0.016(14) &   $-$0.127 \\
   2.25 & 128 &   0.174(3) &   $-$0.112 & 16 &   0.122(4) &    $-$0.113 \\
   2.50 & 128 &   0.247(2) &   $-$0.106 & 16--128 &   0.217(7) &  $-$0.090 \\
   2.75 & 16--256 &   0.333(1) &  $-$0.091 & 16--128 &   0.322(7) & $-$0.086 \\
   3.00 & 16--64 &   0.429(1) &  $-$0.082 & 16 &   0.431(3) & $-$0.066 \\
   3.25 & 32--64 &   0.522(1) &  $-$0.065 & 16--32 &   0.543(4) & $-$0.029 \\
   3.50 & 64--256 &   0.637(4) &  $-$0.056 & 16--128 &   0.662(4) & $-$0.027 \\
   3.75 & 128 &   0.759(5) &  $-$0.025 & 16--128 &   0.777(5) & $-$0.019 \\
   4.00 & 128--256 &   0.892(6) &  $-$0.006 & 32--128 &   0.921(7) &  0.011 \\
\hline
\end{tabular}
\end{center}
\caption{
   Summary of estimates for $z_{{\rm int},\mathcal{C}_2}$,
   averaged over all values of $k$
   for which the confidence level is $\ge 10\%$.
   Error bars are one standard deviation.
   The last column of each group shows
   $z_{{\rm int},\mathcal{C}_2} - z_{{\rm int},\E}$
   from the corresponding fits.
}
\label{table_fits_C2_allk}
\end{table}

\clearpage


\addcontentsline{toc}{section}{Figures}

%
%

\clearpage

%
%

\clearpage

%
%

\clearpage

%
%

\clearpage

%
%

\clearpage

%
%

\clearpage

\end{document}